\documentclass[aps, prd, preprint, onecolumn, tightenlines, notitlepage, superscriptaddress, nofootinbib, preprintnumbers, floatfix,showkeys]{revtex4-1}

\usepackage{amstext}
\usepackage{amssymb}
\usepackage{amsmath,bm}
\usepackage{rotating,graphicx}
\graphicspath{{plots/}}
\usepackage{hyperref}
\usepackage{url}
\usepackage{color}
\usepackage{ulem}
\usepackage[utf8]{inputenc}
\pdfoutput=1
\usepackage[x11names]{xcolor}
\usepackage{textcomp}
\usepackage{booktabs}
\usepackage{epsfig,amsfonts,mathrsfs,amsmath,amssymb,graphicx,color,slashed,multirow}
\usepackage{amsmath,latexsym,amssymb,graphicx,slashed,hyperref,color,enumerate,url,cancel,gensymb}
\usepackage{textcomp}
\usepackage{framed} 
\usepackage{physics}
\usepackage{mathtools}
\usepackage{listings}
\usepackage{framed}

\definecolor{codegreen}{rgb}{0,0.6,0}
\definecolor{codegray}{rgb}{0.5,0.5,0.5}
\definecolor{codepurple}{rgb}{0.58,0,0.82}
\definecolor{backcolour}{rgb}{0.95,0.95,0.92}

\lstdefinestyle{mystyle}{
    backgroundcolor=\color{backcolour},   
    commentstyle=\color{codegreen},
    keywordstyle=\color{magenta},
    numberstyle=\tiny\color{codegray},
    stringstyle=\color{codepurple},
    basicstyle=\ttfamily\footnotesize,
    breakatwhitespace=false,         
    breaklines=true,                 
    captionpos=b,                    
    keepspaces=true,                 
    numbers=left,                    
    numbersep=5pt,                  
    showspaces=false,                
    showstringspaces=false,
    showtabs=false,                  
    tabsize=2
}

\renewenvironment{leftbar}[1][\hsize]
{%
    \MakeFramed{\hsize#1\advance\hsize-\width\FrameRestore}%
}
{\endMakeFramed}

\newcommand{\be}{\begin{equation}}
	\newcommand{\ee}{\end{equation}}
\newcommand{\bea}{\begin{equation} \begin{aligned}}
	\newcommand{\eea}{\end{aligned} \end{equation}}	
	
\hypersetup{colorlinks,citecolor= nicegreen,linkcolor= nicegreen}
\definecolor{nicered}{rgb}{0.7,0.1,0.1}
\definecolor{nicegreen}{rgb}{0.1,0.5,0.1}
\definecolor{mygreen}{rgb}{0,0.392,0}
\definecolor{mygreen}{rgb}{0,0,0.545}

\newcommand{\orcid}[1]{\href{https://orcid.org/#1}{#1}}

\newcommand {\red} 

\def \ie{{\it i.e. }}
\def \eg{{\it e.g. }}

\def\ang#1{{ \langle #1 \rangle}}

\newcommand{\di}{\mathrm{d}}

\AtBeginDocument{\hypersetup{citecolor=DarkOrange3,linkcolor=Green4,urlcolor=Green4}}
\begin{document}

\title{{\Large Unlocking the Inelastic Dark Matter Window with Vector Mediators}}

\author{Ana Luisa Foguel }\email{afoguel@usp.br}
\thanks{orcid \# \orcid{0000-0002-4130-1200}}
\author{Peter Reimitz}\email{peter@if.usp.br}
\thanks{orcid \# \orcid{0000-0002-4967-8344}}
\author{Renata Zukanovich Funchal}\email{zukanov@if.usp.br} 
\thanks{orcid \# \orcid{0000-0001-6749-0022}}
\affiliation{Departamento de F\'{\i}sica Matem\'atica, Instituto de F\'{\i}sica\\
Universidade de S\~ao Paulo, C. P. 66.318, 05315-970 S\~ao Paulo, Brazil\\[5mm] ~}

\preprint{}

\begin{abstract}
Despite the robust cosmological and astrophysical evidence confirming the existence of a non-baryonic matter component in the Universe, the underlying nature of Dark Matter (DM) remains a mystery. Among the several possible scenarios, light DM candidates thermally produced in the early Universe are especially interesting, since their abundance could be set via the standard freeze-out mechanism. Additionally, new light states can present a rich phenomenology and are attracting increasing attention due to recent experimental capabilities to probe dark sectors with feeble interactions. In particular, inelastic DM (iDM) candidates are an appealing option, since they can avoid cosmic microwave background (CMB) radiation
bounds as well as indirect and direct detection searches. Although such models have been intensively studied in the literature, the usual scenario is to consider a secluded dark photon mediator. In this work, we consider the case of iDM with general vector mediators and explore the consequences of such a choice in the relic density computation, as well as for the cosmological and experimental bounds. We examine models with couplings to baryon and lepton number and show new viable parameter regions for inelastic dark matter models. Especially, anomaly-free gauge groups with non-universal couplings to leptons open new windows of the parameter space for  thermal dark matter yet unexplored by experiments.  We also provide a numerical Python library to compute the relic densities for user-defined gauge charges.

\end{abstract}

\maketitle
\flushbottom
\newpage
\tableofcontents

\section{Introduction}
\label{sec:intro}

The particle nature of dark matter (DM) remains a pressing problem today.
For many decades the main experimental focus was to look for a TeV scale 
Weakly Interacting Massive Particle (WIMP), charged under the electroweak interaction that could {\it naturally} yield the cosmological observed relic abundance via the thermal freeze-out mechanism~\cite{Jungman:1995df}. Although the null results from quests in this direction seem now to disfavor this possibility~\cite{Arcadi:2017kky,Roszkowski:2017nbc}, the thermal freeze-out mechanism is still quite compelling. Even if DM is not a WIMP.

For the freeze-out mechanism to work, it is essential to have an efficient way to deplete the DM energy density in the early universe, usually by transferring it to the SM sector via a dark sector-SM mediator.  Thus it has been widely recognized that it is important to complement searches for the DM itself  with endeavours aiming at  other possible experimental signatures connected to the dark sector mediator. This also provides new  
opportunities for discovery and falsification and is the reason why an extensive program has emerged, in particular, around models that do not only explain today's DM abundance but can also be probed experimentally 
beyond direct and indirect detection, 
by exploiting beam-dump, fixed-target and collider experiments~\cite{Snowmass2013CosmicFrontierWorkingGroups1-4:2013wfj}. 

Dark sectors with (sub-)GeV mass particles have gained great attention in recent years~\cite{Knapen:2017xzo}. However, 
models of thermal DM in the (sub-)GeV range must sustain strong constraints from the measurement of the cosmic microwave background (CMB) radiation on s-wave DM annihilation~\cite{Planck:2018vyg}. 
Among the scenarios that can lead to negligible annihilation rates at the time of recombination 
we have asymmetric DM~\cite{Petraki:2013wwa,Zurek:2013wia},  p-wave DM~\cite{Diamanti:2013bia} and  the prominent  inelastic DM (iDM) model, originally introduced to attempt to solve the DAMA inconsistency~\cite{Tucker-Smith:2001myb}.

In the simplest implementation of iDM~\cite{Izaguirre:2015zva,Izaguirre:2017bqb},
the dark sector contains a pseudo-Dirac state with 
two Majorana fields $\chi_1$ and $\chi_2$ nearly degenerate in mass. The lightest state $\chi_1$, which is the stable DM species, couples to the SM only by interacting with a slightly heavier $\chi_2$. The relic DM abundance is basically set by the s-wave coannihilation ($\chi_1 + \chi_2 \leftrightarrow \rm SM$) process, instead of the suppressed symmetric s-wave iDM annihilation channel. Other processes, such as dark sector scattering ($\chi_2 + \chi_2 \to \chi_1 + \chi_1$) and decays ($\chi_2 \to \chi_1 + {\rm SM}$), are responsible for efficiently depleting the $\chi_2$ population after freeze-out, while scattering with SM particles ($\chi_2 + {\rm SM} \to \chi_1 + {\rm SM}$) not only contributes to depletion but also ensures thermal equilibrium between both sectors. Consequently, since the current population of $\chi_2$ states is negligible, this model does not only evade the CMB bound but also direct and indirect detection constraints~\footnote{
Note, however, that regeneration of the heavier dark state in galaxies at late times may give rise to observable gamma-rays in the 1 MeV to 100 MeV energy range that could be probed by future gamma-ray telescopes~\cite{Berlin:2023qco}.}.

Given the typically small couplings to the SM required to successfully account for the observed abundance of  $\Omega_{\chi_1} h^2 = 0.12$~\cite{Planck:2018vyg}, high-intensity accelerator experiments are necessary to produce anew the unstable dark $\chi_2$ particle. The decays of this particle are often only detectable in environments with low SM backgrounds. 
Due to the typically long lifetime of the $\chi_2$ dark state, most of the studies fall into the category of long-lived particle (LLP) searches. A characteristic feature of these iDM models is that production is driven by couplings to the SM via vector mediators, while the interaction and decay of the LLP are highly sensitive to the mass splitting between the nearly degenerate dark states.
So far, if not effectively coupling to the SM field strength via dipole operators~\cite{Jodlowski:2023ohn,Dienes:2023uve,Barducci:2024nvd}, one has considered exclusively as mediator 
the dark photon, a vector particle that kinetically mixes with the SM photon and is connected to a new secluded Abelian $U(1)$ symmetry.
Limits on this case have been investigated in the  
context of current collider~\cite{Izaguirre:2015zva,Berlin:2018jbm,
Duerr:2019dmv} and fixed target~\cite{Izaguirre:2017bqb,Tsai:2019buq,Batell:2021ooj,Mongillo:2023hbs} accelerator experiments.
Future sensitivities also have been predicted 
for several proposed LHC experiments~\cite{Bertuzzo:2022ozu}, the 
light DM LDMX experiment~\cite{Berlin:2018bsc,Berlin:2018pwi} as well as a possible Super Tau Charm Factory~\cite{Lu:2023cet}.
Most of the parameter space of this realization is currently excluded or will be probed in the near 
 future~\footnote{It is worth mentioning that removing some of the symmetry assumptions in the minimal iDM model can also relax the relic density constraint~\cite{Garcia:2024uwf}. 
 }.

Nevertheless, a wide range of $U(1)_{Q}$ extensions of the SM can be considered, unlocking new regions of viability for iDM. Especially, anomaly-free extensions, where the mediator couples non-universally  to the family lepton number $L_\alpha$, i.e. $Q=L_\mu -L_\tau$ or $Q=B-3L_\tau$, are of interest. 
As we will see, these types of models can effectively generate the correct DM relic abundance in different predictive regions of parameter space compared to the canonical dark photon mediator model in the minimal iDM setup, while also modifying the constraints set by current experimental limits and future predictions. This is mainly due to the additional, generally dominant, invisible $\chi_2\to \chi_1 \, \nu_\alpha \, \bar{\nu}_\alpha$ decays.

The focus of this paper is to study some of these $U(1)_{Q}$ extensions of the dark photon 
mediator iDM, which we will refer to as iDM$_Q$. We start in section~\ref{sec:model} by introducing the general model description, defining the chosen free parameters of the model and discussing the $Z_Q$ and $\chi_2$ decay rates, for a few $Q$ selections. In section~\ref{sec:relic}
we detail the  relic density calculation, highlighting the effect of the main free parameters and the 
differences with respect to the dark photon case.  The phenomenological implications of current and future accelerator experiments  are then analysed in section~\ref{sec:pheno}.
In section~\ref{sec:conclusion}, we summarize our findings and comment about future directions.
We provide a code \textsc{ReD-DeLiVeR} to compute the relic density for iDM/iDM$_Q$ as well as other simplified DM models and refer the reader to appendix~\ref{app:rdcode} for further details. 

\section{General Model and Parameter Choices}
\label{sec:model}

\subsection{Description of the General iDM$_{Q}$}
%
The iDM$_Q$ model we will discuss here is a simple extension of the one  
first proposed in \cite{Tucker-Smith:2001myb} 
and later explored in the context of collider searches~\cite{Izaguirre:2015zva,Berlin:2018jbm,Duerr:2019dmv}, 
fixed-target experiments~\cite{Izaguirre:2017bqb} and 
the anomalous magnetic moment of the muon~\cite{Mohlabeng:2019vrz}.
The dark sector, a pseudo-Dirac pair of two-component Weyl fermions, $\psi_1$ and $\psi_2$, 
is in our case connected  to an additional spontaneously broken $U(1)_Q$ Abelian gauge symmetry under which they have 
opposite charges. Their gauge interactions can be described by the following terms in the Lagrangian
\begin{align}
    \mathcal{L}\supset g_Q \, q_D \, Z_{Q\mu}(\psi_1^\dagger \bar{\sigma}^\mu \psi_1 - \psi_2^\dagger \bar{\sigma}^\mu \psi_2)  \, ,
    \label{eq:Lint0}
\end{align}
where $g_Q$ and $q_D$ represent the gauge coupling and the dark charge under the new $U(1)_Q$ symmetry, and $Z_Q$ is the associated gauge boson mediator.
 The fermion mass terms arise from
\begin{align}
    \mathcal{L}\supset -m_D \psi_1 \psi_2 - \frac12 (\delta_1 \, \psi_1^2 + \delta_2 \, \psi_2^2)+{\rm h.c.} \, ,
     \label{eq:Lmass}
\end{align}
where $m_D$ is a gauge-invariant Dirac mass and $\delta_{1,2}$ are $U(1)_Q$ breaking Majorana masses satisfying $\delta_{1,2}\ll m_D$. The mass eigenstates can be obtained after diagonalisation and correspond to a pseudo-Dirac pair
\begin{align}
    \chi_1 \simeq \frac{i}{\sqrt{2}}(\psi_1-\psi_2) ,\quad \chi_2 \simeq \frac{1}{\sqrt{2}}(\psi_1 + \psi_2)\, ,
   \label{eq:chi1chi2}  
\end{align}
with nearly degenerate masses
\begin{align}
    m_{1,2}\simeq m_D \mp \frac12 (\delta_1+\delta_2)\,,
   \label{eq:m1m2}  
\end{align}
such that the lightest one, $\chi_1$, will be the stable DM candidate.
We define the dimensionless and dimensional mass splitting parameters
\begin{align}\Delta&:=\frac{m_2-m_1}{m_1}=\frac{\delta_1+\delta_2}{m_1}< 1 \quad {\rm and} \notag  \\
    \delta&:=\delta_1+\delta_2 \, ,
     \label{eq:splittings}
\end{align}
respectively. The interaction term with the mediator $Z_Q$ 
given in eq.~(\ref{eq:Lint0}) is now off-diagonal and given by
\begin{align}
    \mathcal{L_{\rm int}^{\rm D}}=   i g_D Z_{Q\mu} \bar{\chi_2}\gamma^\mu \chi_1  + \rm h.c. \, ,\label{eq:Loffdiag}
\end{align}
up to terms of order $\mathcal{O} \qty(\delta_{1,2}/m_D)$, and where we defined $g_D \equiv g_Q \, q_D $. 

Due to the Abelian nature of the new gauge group, the mediator Lagrangian admits a kinetic mixing term with the photon \footnote{In general, one could include a kinetic mixing term with the hypercharge gauge boson. However, given that the coupling with the $Z$ boson current scales with the mass ratio $m_{Z_Q}^2/m_Z^2$ and our focus lies within the low-mass regime (\( m_{Z_Q} \lesssim 10 \) GeV), we can neglect such contributions without any loss of generality.}. After performing the kinetic mixing diagonalization, such term is rotated away into a new coupling connecting $Z_Q$ with the SM fields that is suppressed by the kinetic mixing parameter $\epsilon$~(see Ref.~\cite{Fabbrichesi:2020wbt} for a complete review). The full Lagrangian of the vector mediator is given by
\be
\mathcal{L} \supset -\frac14 Z_Q^{\mu \nu} Z_{Q \mu \nu} + \frac12 m_{Z_Q}^2 Z_Q^2 +    \mathcal{L_{\rm int}^{\rm D}} + \mathcal{L_{\rm int}^{\rm SM}}\, ,
\label{eq:Lmed}
\ee
where $Z_Q^{\mu \nu}$ is the mediator field strength tensor, $m_{Z_Q}$ is its mass and
\be
\mathcal{L_{\rm int}^{\rm SM}} = e \epsilon J^\mu_{\rm em} Z_{Q\mu} - g_Q J_Q^\mu Z_{Q\mu}  \, ,
\label{eq:Lint}
\ee
with $e=g \sin \theta_W$ the electric charge, $g$ and $g_Q$ the $SU(2)_L$ and $U(1)_Q$ coupling constants, respectively, and $\theta_W$ the SM weak mixing angle. As usual 
\be
 J^\mu_{\rm em} = \sum_{f} \bar{f} \gamma^\mu \, q^f_{\rm em} \, f\, , \label{eq:Jem}
 \ee
is the SM electromagnetic current, $q^f_{\rm em}$ is the fermion $f$ 
electric charge in units of $e$, and 
\be
 J^\mu_Q = \sum_{f}  q^f_{Q} \, \bar{f} \gamma^\mu \, f\, + \sum_{ \ell = e, \mu , \tau} q^{\nu_\ell}_{Q} \,\bar \nu_\ell \gamma^\mu P_L \nu_\ell,\label{eq:JQ}
 \ee
is the new vector current, with $q^{f(\nu_\ell)}_{Q}$ being the $Q$-charge of fermion $f$ (neutrino flavor $\nu_l$). The symmetry generator for these models can be written as
\be
Q = y_B B - y_e L_e - y_\mu L_\mu -y_\tau L_\tau \,, 
\label{eq:charges}
\ee
where $B$ is the baryon number and $L_e, L_\mu$  and $L_\tau$ are the lepton family number operators.

So far in the literature, the only vector portal considered for iDM is $Z_Q=Z_\gamma$, i.e. the so-called dark photon which is kinetically mixed with the photon and only couples to the SM sector via the $J_{\rm em}$ current~\cite{Izaguirre:2015zva,Izaguirre:2017bqb,Batell:2021ooj,Berlin:2018jbm,Tsai:2019buq,Duerr:2019dmv,Berlin:2023qco}. In principle, however, a wide range of mediators can be considered in connection with iDM$_Q$ models, depending on the choice of $U(1)_Q$. For these models one can have $e\epsilon \ll g_Q$, which is the case we investigate here.

Tab.~\ref{tab:models} presents a list of sample models we will examine in this work along with the corresponding SM fermion charges. We consider the  baryophilic anomaly-free $B-L$ and $B-3L_\tau$ models~\footnote{The condition of anomaly cancellation in such models requires the inclusion of extra right-handed neutrinos. In this work we assume that these particles are heavier than the typical scales and neglect them from here on. }, as well as the anomalous $B$ model. Additionally, we investigate the leptophilic anomaly-free $L_\mu-L_\tau$ model, which is well-known in the literature as a viable candidate to explain the muon $(g-2)$~\cite{Heeck:2011wj,Altmannshofer:2014pba,Altmannshofer:2016brv,Gninenko:2018tlp,Amaral:2021rzw,Huang:2021nkl}
as well as to relax the reported Hubble tension~\cite{Escudero:2019gzq}.

\begin{table}[htb]
\def\arraystretch{2.}
    \centering
    \begin{tabular}{|c|c|c|c|c|c c c c| }
    \hline
    \hline
        ~$y_B$~ & ~$y_e$~ & ~$y_\mu$~ & ~$y_\tau$~ & $Q$ & \multicolumn{4}{c|}{ $q^f_Q$}  \\
         & & & & & quarks & $e/\nu_e$ & $\mu/\nu_\mu$ & $\tau/\nu_\tau$\\
        \hline \hline
         1 & 1 & 1 & 1 &$B-L$ & $\frac{1}{3}$ & -1 & -1 & -1\\
         \hline
         1 & 0&0 & 3 & ~$B-3L_\tau$ & $\frac{1}{3}$ & 0 & 0 & -3\\
         \hline
         1 & 0&0 & 0 & $B$ & $\frac{1}{3}$ & 0 & 0 & 0\\
         \hline
          \hline
         0 & 0 & -1 & 1 & $L_\mu-L_\tau$ & 0 & 0 & 1 & -1 \\
         \hline \hline
    \end{tabular}
    \caption{Symmetry generators and fermion charges for the models considered in this work.}
    \label{tab:models}
\end{table}
%

\subsection{On the Choice and Role of Parameters}

Once $U(1)_Q$ is fixed, we have five free parameters to define the vector mediated iDM$_Q$ model~\footnote{ We will consider the limit where the tree-level kinetic mixing $\epsilon \to 0$. However, loop-generated kinetic mixing $\epsilon_{\rm loop}$ can also occur in some cases, such as in the $L_\mu-L_\tau$ model, which will be suppressed by at least $\epsilon_{\rm loop}/g_Q\lesssim e^2/(4\pi)^2 \simeq 6\cdot 10^{-4}$ compared to $g_Q$.}: the three new dark sector particle masses ($m_1, m_2,$ and $m_{Z_Q}$) the gauge coupling ($g_Q$) and the dark fermions charge ($q_D$). In our study, we prefer to work with $m_{Z_Q}$, the ratio $R=(m_{Z_Q}/{m_1})$, the dimensionless splitting  $\Delta$, and the constants $\alpha_Q = g_Q^2/(4\pi)$~\footnote{For the plots, we will often choose to fix the gauge coupling $g_Q$, as this approach is more common in the literature.} and $\alpha_D = g_D^2/(4\pi)$. Let us now discuss the validity range of each one of these parameters.

Regarding the mass of the mediator, we will consider the phenomenological interesting range $10^{-2} < m_{Z_Q}/{\rm GeV} < 10$, such that the dark sector states are light. We also want to focus in the regime where $m_{Z_Q} > m_1 + m_2 \sim 2 m_1$, which implies $R>2$, $\Delta \ll 1$. This hierarchy choice will ensure that the on-shell $Z_Q$ can decay into $\chi_1$ and $\chi_2$, which is crucial for the phenomenology and plays a major role to efficiently freeze-out the DM via the dominant coannihilation channel $\chi_1 \chi_2 \to Z_Q \to {\rm SM}$.
Therefore, our results are valid for $R > 2$. However, since larger $R$ values require higher gauge couplings to achieve the correct relic density, which often leads to stronger conflicts with experimental limits, we will mainly consider the case where $R = 3$. This choice also avoids the resonance region when $R \sim 2$. For a study of the ``forbidden DM" region $1<R<2$, we refer to~\cite{Fitzpatrick:2021cij}.

The dark fermion mass splitting $\Delta$ is the most crucial parameter in the iDM framework. In the limit where $\Delta \to 0$, we recover the pseudo-Dirac DM scenario, where direct detection bounds from elastic scattering apply. However, even for splittings as small as $\Delta \sim 10^{-6}$~\cite{Berlin:2018jbm,Duerr:2019dmv}, direct detection constraints can be evaded, as the kinetic energy of $\chi_1$ is insufficient to overcome the mass splitting to produce $\chi_2$. This effectively forbids inelastic upscattering, and elastic scattering is suppressed both by $\Delta$ and the DM velocity. Besides, the dark fermion mass splitting is often assumed to exceed twice the electron mass $m_e$, such that $\Delta \cdot m_1 \gtrsim 2 m_e \sim 1$ MeV. This is mainly because, below this threshold, $\chi_2$ decay channels are closed in the vanilla dark photon iDM scenario, which significantly impacts the depletion rate of the heavier dark fermion, as these decays are the dominant depletion process. Scattering processes are less efficient in depleting $\chi_2$, particularly for masses above the GeV scale~\cite{CarrilloGonzalez:2021lxm}. In contrast, this threshold does not exist in the iDM$_Q$ scenario if the mediator couples to any family lepton number $L_\alpha$, allowing $\chi_2$ to decay into SM neutrinos. As a result, the depletion of the heavier state remains efficient, enabling us to lower the values of $\Delta$ without concerning about cosmological or direct and indirect detection bounds. Additionally, the inclusion of neutrinos will affect both the cosmological evolution history of the dark fermion candidates, as well as the phenomenology of such models, since the decays into neutrinos will contribute to invisible signatures, influencing the experimental constraints. 

For the upper limit on $\Delta$, large mass splittings could, in principle, be considered in the iDM$_Q$ scenario. However, since the coannihilation cross section is inversely proportional to $\Delta$, increasing the splitting weakens the cross section. To prevent an overabundance of $\chi_1$, the coannihilation process must remain efficient, requiring an increase in the gauge coupling $g_Q$ to produce the correct DM abundance. Unfortunately, the region of large couplings is already excluded by various experimental searches, as discussed in section~\ref{sec:pheno}. Therefore, we restrict ourselves to $\Delta \leq 0.4$ to avoid these experimental bounds.

Finally, the coupling constant $\alpha_{Q} \lesssim 1$ can be chosen freely, with an upper limit set by the naive perturbative premise $g_{Q} \lesssim \sqrt{4\pi}$. We also work in the regime where $\alpha_D \gg \alpha_Q$, motivated by the fact that $g_Q$ is already tightly constrained by multiple experimental bounds. Additionally, in this regime, the dark mediator decays primarily and promptly into dark fermion states, as we will show in the next section, leading to interesting phenomenological consequences. Note that this choice implies that the dark fermion charges are significantly larger than the SM charges, which is feasible as long as $\alpha_{D}$ remains within the perturbative regime~\cite{Batell:2021snh,Berlin:2018bsc}.

Based on the points discussed above, in the main text of the paper we will focus on the following benchmark parameter region
\begin{align}
    & R=3 \, ; \quad \Delta=0.1,0.4 \, ; \quad \alpha_D=0.1 \, ; \\
    & 0.01<m_{Z_Q}/{\rm GeV}<10 \, ; \quad \alpha_Q\lesssim 1 \, ,\notag
\end{align}
while consistently emphasizing how variations in these parameters impact the results. Before we discuss the relic density calculation, let us first examine the decay rates of the unstable dark sector particles to better understand some of the above considerations.

\subsection{Branching Ratios and Decay Modes}
\label{sec:BR}

In order to correctly determine the DM relic abundance, as well as to compute the experimental bounds in the parameter space of a particular iDM$_{Q}$, one needs to calculate the decay widths and branching ratios of the additional dark particles of the model. The mediator $Z_Q$ width into SM fermions can be written as
\begin{align}
\Gamma(Z_{Q} \to  f \bar f) 
= C^f \frac{\alpha_Q (q_Q^f)^2 }{3} m_{Z_Q}\left( 1 + 2\frac{m_f^2}{m^2_{Z_Q}}\right)\sqrt{1-\frac{4 \, m^2_f}{m^2_{Z_Q}}}\,,
    \label{eq:GZff}
\end{align}
where $C^f=(3,1,1/2)$, respectively, for quarks, charged leptons and neutrinos and $m_f$ is the corresponding fermion mass. For the decay width into hadronic final states we use the vector meson dominance approach, as described in Ref.~\cite{Foguel:2022ppx}. The mediator can also decay into the dark sector fermions, with its width given by
\begin{align}
\Gamma(Z_{Q} \to  \chi_1 \chi_2) 
=  \frac{\alpha_D }{3} m_{Z_Q}  \qty(1- \frac{\Delta^2}{R^2})^{3/2}  \qty( 1 + \frac{(\Delta +2)^2}{2 R^2}) \sqrt{1 -\frac{(\Delta+2)^2}{R^2} } \,.
    \label{eq:GZXX}
\end{align}

\begin{figure}[bht]
\begin{center}
\includegraphics[width=0.49\textwidth]{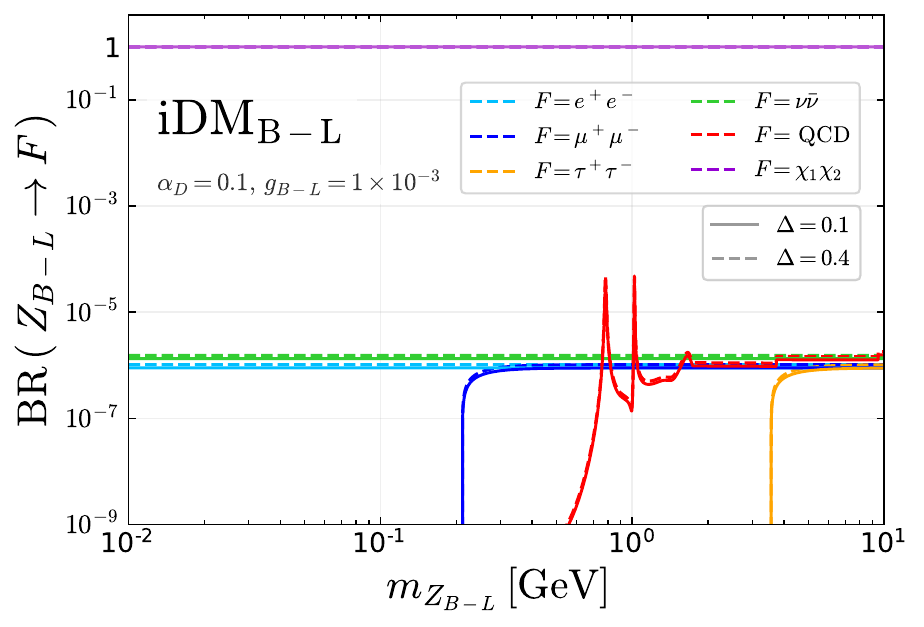}
\includegraphics[width=0.49\textwidth]{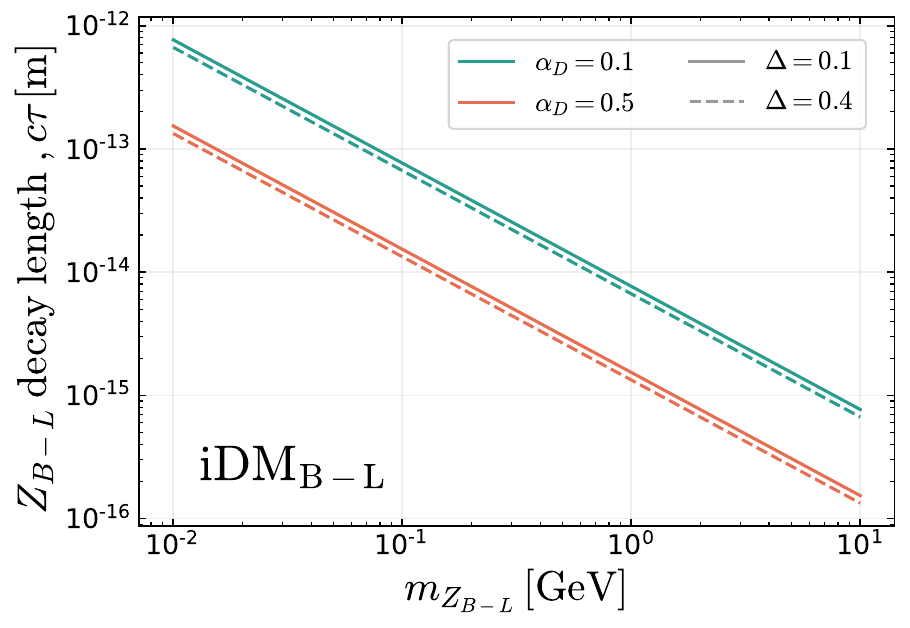}
\end{center}
\vglue -0.8 cm
\caption{\label{fig:zqmed} The left panel shows the $Z_{B-L}$ branching ratio into dark fermions (purple), neutrinos (green), electrons (light blue), muons (dark blue), taus (yellow) and hadrons+quarks (red), for the iDM$_{B-L}$  with $R=3$, $\alpha_D=0.1$, $g_{B-L}=1\times 10^{-3}$, and $\Delta = 0.1$ (solid lines) or $\Delta = 0.4$ (dashed lines). The right panel shows the rest frame decay length of the mediator for $\alpha_D =0.1$ (green) and $\alpha_D =0.5$ (orange), indicating the promptly behavior of $Z_{B-L}$. }
\end{figure}

The panels of figure~\ref{fig:zqmed} illustrate the general behavior for the iDM$_{B-L}$. In the left panel we show the $Z_{B-L}$ branching ratio across various final states, depicted by the color-coded labels, for the parameter values $\alpha_D=0.1$, $g_{B-L} = 1\times 10^{-3}$, and $R=3$. It also shows the comparison of two different values of $\Delta$, although the difference, as expected from eq.~(\ref{eq:GZXX}), is negligible. One can see that the dominant decay channel is the one into  dark fermions (purple line), such that we can safely assume $BR(Z_{B-L}\to \chi_1 \chi_2)\sim 1$ for all values of $m_{Z_{B-L}}$ considered in the following analysis. This behavior is a consequence of the hierarchy $\alpha_D \gg \alpha_{B-L}$, which arises from the strong experimental bounds on the kinetic mixing and $U(1)_{B-L}$ gauge couplings~\cite{BaBar:2017tiz,NA62:2019meo,Banerjee:2019pds}, implying that $\Gamma(Z_{B-L}\to {\rm SM})/ \Gamma(Z_{B-L}\to \chi_1\chi_2) \approx (\alpha_{B-L}/\alpha_D) \ll 1$. 
The rest frame decay length of the mediator for two choices of $\alpha_D$ and $\Delta$ as shown in the right panel of figure~\ref{fig:zqmed}, confirm that the $Z_{B-L}\to \chi_1 \chi_2$ happens promptly.
This result is valid for any choice of $\alpha_{B-L} < \alpha_D$. Similarly, the same conclusions apply for different mediator models.

Let us now move on to describe the decays of the heavier dark fermion $\chi_2$. Due to the presence of the $\Delta$ mass splitting, $\chi_2$ can undergo three-body decays into $\chi_1$ and a pair of SM fermions $\chi_2 \to \chi_1 \, \bar f f$, as well as n-body decays with final hadronic states $\chi_2 \to \chi_1  +  \rm{hadrons} $, via the emission of an off-shell $Z_Q^*$ boson. In the limit where $m_{Z_Q} \gg m_2 > m_1$ and $m_f \to 0$, the decay width of $\chi_2$ into fermions can be approximated as
\be \label{eq:chi2decay}
\Gamma(\chi_2 \to \chi_1 \, \bar f f) \simeq \frac{4 \, \alpha_Q \, \alpha_D \, \Delta^5 \, m_{Z_Q}}{15\pi R^5} \, .
\ee
In our study, we refrain from using this approximate expression and instead compute the full analytical three-body decay width for fermions, as well as for the dominant hadronic channels. These computations were also verified by comparing them with numerical outputs from MadGraph. For further details regarding the three-body computations and the consistency checks, we refer to appendix~\ref{app:chi2dec}.

\begin{figure}[h]
\begin{center}
\includegraphics[width=0.48\textwidth]{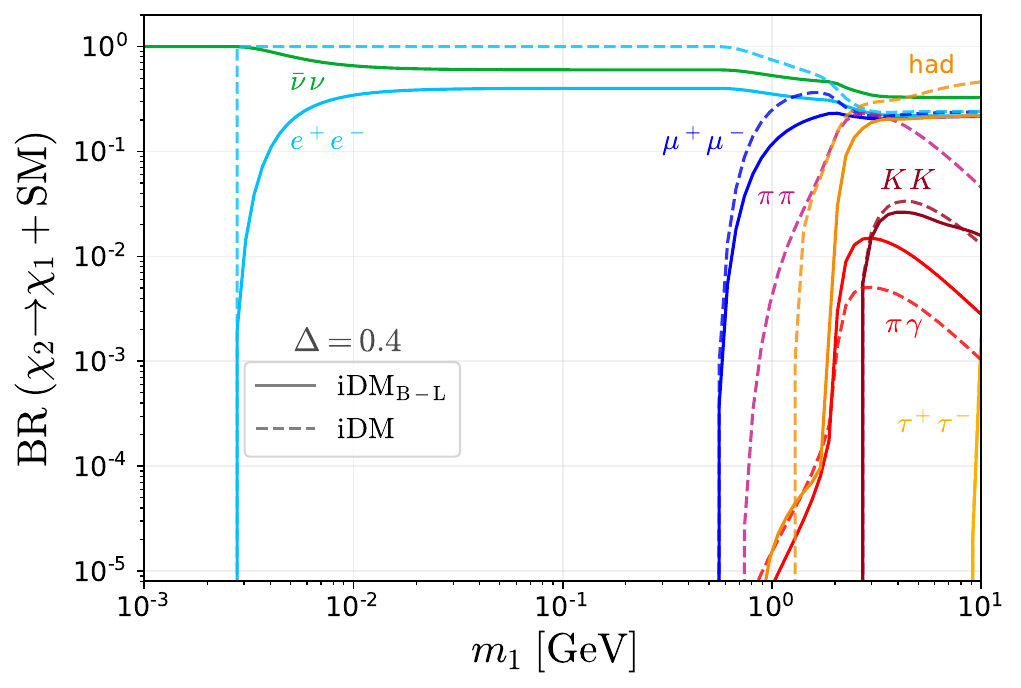}
\includegraphics[width=0.485\textwidth]{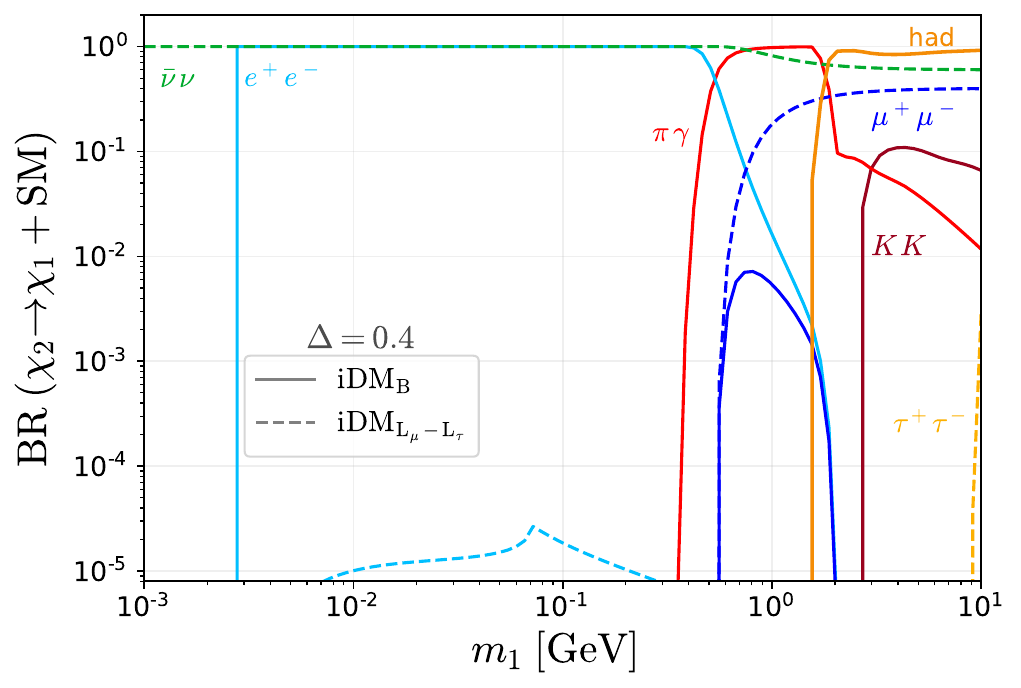}
\end{center}
\vglue -0.8 cm
\caption{\label{fig:brchi2d04} 
We show on the left (right) panel the branching ratio of $\chi_2 \to \chi_1 +\rm{SM}$ as a function of $m_{1}$ for the iDM$_{B-L}$  and the vanilla dark photon iDM (iDM$_B$ and iDM$_{L_\mu - L_\tau}$). Here 
$\rm SM$ stands for neutrinos (green), electrons (light blue), muons (dark blue), taus (yellow), $\pi^+ \pi^-$ (magenta), $\pi^0 \gamma$ (red), $KK = K^+ K^- + \bar K^0 K^0$ (brown) and other hadrons (orange), for the iDM$_{B-L}$ (solid) and iDM (dashed). The right panel shows the same, but for iDM$_B$ (solid) and iDM$_{L_\mu-L_\tau}$ (dashed). 
We have fixed $\Delta =0.4$.
}
\end{figure}

In the left (right) panel of figure~\ref{fig:brchi2d04} we show the branching ratio of $\chi_2$ decays into various channels, distinguished by the colored labels, for the iDM$_{B-L}$ (iDM$_B$ and iDM$_{L_\mu - L_\tau}$)  for $\Delta =0.4$. The orange lines labeled `had' represent an estimate of additional hadronic contributions~\cite{Jodlowski:2019ycu} originating from other final states different from $\pi^+ \pi^-$, $\pi^0 \gamma$ and $KK$.
We show the dark photon iDM model as a reference.

Some additional comments regarding the $\chi_2$ branching ratios are in order. Firstly, for the case of the iDM$_B$ (iDM$_{L_\mu-L_\tau}$), the coupling to leptons (electrons) is loop-generated via the induced kinetic-mixing with the photon~\cite{Amaral:2020tga,Tulin:2014tya}.
Secondly, the channel $\chi_2 \to \chi_1 \pi^+ \pi^-$ is exclusive to the iDM, as $B$-like models lack couplings to the vector meson $\rho$ current~\cite{Foguel:2022ppx}, and the iDM$_{L_\mu-L_\tau}$  exhibits negligible decays into hadrons. 
As mentioned before, neutrinos are the main SM fermions produced in the decay in all models coupled to lepton number. 
Lastly, we omitted the display of branching ratios for $\Delta=0.1$ since the results are similar to the one presented here, albeit with all curves shifted to larger values of $m_1$. This shift arises due to the fact that to compensate 
for the smaller $\Delta$ one needs a larger 
$m_1$ to reach the same mass threshold for a specific channel.

\begin{figure}[h]
\begin{center}
\includegraphics[width=0.7\textwidth]{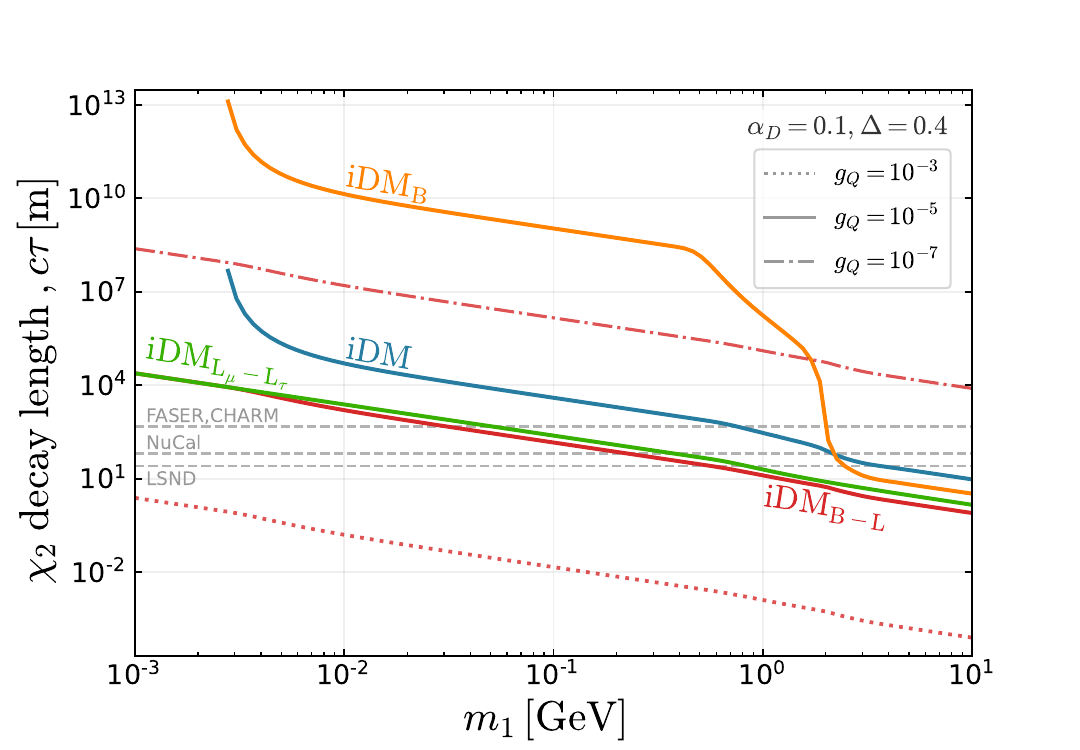}
\end{center}
\vglue -0.8 cm
\caption{\label{fig:declchi2} 
Decay length of $\chi_2$ in its rest frame for several models with the  parameter values fixed at $\Delta=0.4$, $R= 3$, $\alpha_D = 0.1$, $g_Q = 10^{-5}$ (solid lines): iDM$_{B-L}$ (red), iDM$_{L_\mu - L_\tau}$ (green), dark photon iDM (blue) and iDM$_B$  (orange). For iDM$_{B-L}$, we also show in dotted (dash-dotted) the curve for fixed $g_Q = 10^{-3}$ ($g_Q = 10^{-7}$). The horizontal dashed gray lines indicate the distances from the target to the detectors (decay region) for the FASER, CHARM, NuCal and LSND experiments.}
\end{figure}

Finally, figure~\ref{fig:declchi2} shows the rest frame $\chi_2$ decay length as a function of $m_1$ for the fixed values $\Delta=0.4$, $R= 3$, $\alpha_D = 0.1$ and $g_Q= 10^{-5}$ 
(solid lines), and considering different iDM$_Q$ as labeled by the colors in the plot. For the iDM$_{B-L}$, we also show the decay length for the choices $g_Q= 10^{-3}$
(dotted line) and $g_Q= 10^{-7}$
(dash-dotted line). The gray dashed horizontal lines correspond to the length of the accessible decay region for  some  experiments: FASER/CHARM (480 m), NuCal (64 m) and LSND (25.85 m). See section~\ref{sec:pheno} for more details on these experiments.

As seen in the figure, the heavier dark fermion exhibits macroscopic decay lengths for several regions of the parameter space, suggesting potential decay signatures in different experimental settings. Additionally, the plot indicates that changing the mediator model can lead to drastic changes in the typical $\chi_2$ decay length, which will impact later the associated experimental bounds. For instance, when models  couple to neutrinos (iDM$_{B-L}$ and iDM$_{L_\mu - L_\tau}$), the inclusion of such couplings increases the decay width, resulting in shorter lifetimes and consequently smaller decay lengths. However, it is important to note that the neutrino decay channel contributes to an invisible signal, thus affecting the balance between visible and invisible searches.  We have 
shown the curves only for a rather large mass splitting ($\Delta=0.4$). If we decrease this splitting all the decay length curves shift upward. This also means $\chi_2$  may behave as invisible for some accelerator searches in this case.  These aspects will be further discussed in section~\ref{sec:pheno}.

\section{Relic Density and Thermal Target}
\label{sec:relic}

The dark fermion $\chi_1$ stands as the  naturally stable dark matter candidate. Moreover, it acts as a light thermal relic, allowing for the computation of the DM abundance via thermal freeze-out from the SM bath. This computation enables us to derive the thermal target curve, which represents the collection of points in the model parameter space that reproduce the correct DM relic abundance, i.e. give $\Omega_{\chi_1} h^2 = 0.12$~\cite{Planck:2018vyg}, and serves as a reference when comparing current 
experimental bound and future sensitivities for different iDM$_Q$.

In order to describe the evolution of the DM abundance in the early universe and determine today's relic density one needs to solve a system of coupled Boltzmann equations for number densities $n_i$ of particles $\chi_i$. In terms of the comoving number densities $Y_{1,2} \equiv n_{1,2}/s$, where $s$ is the entropy, and by parameterizing the time evolution as $x \equiv m_2/T$, with $T$ the SM bath temperature,  these Boltzmann equations can be written as
\begin{equation} \label{eq:boltziDM}
\begin{aligned}
    \dv{Y_{1,2}}{x} = \frac{s}{H x}  \, \bigg[ & - \ang{\sigma v}_{12 \to ff}  \big( \, Y_1 Y_2 - Y_1^{\rm eq}Y_2^{\rm eq} \, \big) \pm 2 \, \ang{\sigma v}_{22 \to 11} \qty( (Y_2)^2 - \qty( Y_1 \frac{Y_2^{\rm eq}}{Y_1^{\rm eq}})^2 ) \\
    & \pm \frac{1}{s} \; \qty( \Gamma_{2f \to 1f} + \ang{\Gamma}_{2 \to 1 ff}  ) \qty( Y_2 - Y_1 \frac{Y_2^{\rm eq}}{Y_1^{\rm eq}} ) \bigg] \, ,
\end{aligned}
\end{equation}
where the Hubble expansion rate $H$ is given by $H(T)= \sqrt{8 \pi^3 g_\rho(T)/90} \, T^2 M_{Pl}^{-1}$, with $M_{Pl}$ the Planck mass and $g_\rho$ counts the relativistic degrees of freedom that contribute to the energy density at temperature $T$, $\ang{\sigma v}_{ij \to kl}$ represents the thermally averaged cross-section for the $ij \to kl$ process times the velocity,  $\Gamma_{2f \to 1f}$ corresponds to the $\chi_2 f \to \chi_1 f$ scattering 
rate  and $\ang{\Gamma}_{2 \to 1 ff}$ is the thermally averaged $\chi_2$ decay rate. The equilibrium yield $Y_i^{\rm eq}$ is given by~\cite{Steigman:2012nb}

\be \label{eq:eqyield}
Y_i^{\rm eq} \equiv \frac{n_i^{\rm eq}}{s} \, , \: {\rm with} \: \begin{dcases}
    \, n_i^{\rm eq} = \frac{3 \zeta(3) }{4 \pi^2} \, g_i \, T^3  & \, {\rm for} \quad T \gg m_i \, , \\[10pt]
    \, n_i^{\rm eq} = g_i  \qty(\frac{ m_i \, T }{2 \pi})^{3/2} \exp \qty(-\frac{m_i}{T})  & \, {\rm for} \quad  T \ll m_i \, ,
\end{dcases}
\ee
where $g_i$ is the number of degrees of freedom of species $i$ (for the iDM$_Q$ case $g_{1,2}=2$), $s = 2 \pi^2  g_s(T) T^3/45$ represents the entropy and $g_s(T)$ counts the effective number of entropic relativistic degrees of freedom at temperature $T$~\footnote{ Let us emphasize that $g_s= g_\rho$, unless there are relativistic particles that have already decoupled from the photons. Given the SM particle content, this will be the case only for temperatures below $T \sim m_e$.}. Each term in eq.~\eqref{eq:boltziDM} represents a relevant number-changing process that contributes to the abundance calculation. For the case of $2 \to 2$ processes, the thermally averaged cross-section can be obtained with the general formula~\cite{Gondolo:1990dk,Chu:2011be} 
\be \label{eq:avgxsec}
\langle \sigma v\rangle_{ij \to kl} = \frac{ \gamma({ij \leftrightarrow kl})}{n_i^{\rm eq} n_j^{\rm eq}} \, ,
\ee
where the reaction density $\gamma$ is defined as
\be\label{eq:lilgam}
\gamma({ij \leftrightarrow kl}) = \frac{T}{64 \pi^4} \int_{s_{\rm min}}^{\infty} {\rm d}s   \sqrt{s} \, \hat \sigma_{ij \leftrightarrow kl}(s) K_1\left(\frac{\sqrt{s}}{T}\right) \, ,
\ee
with $s_{\rm min} = {\rm max}[(m_i+m_j)^2, (m_k+m_l)^2]$ and $K_n$ corresponding to the modified Bessel function of the second type of order $n$. The reduced cross-section $\hat \sigma$ is given by
\be \label{eq:redxsec}
\hat \sigma_{ij \leftrightarrow kl}(s) = \frac{g_i g_j}{c_{ij}} \frac{2 \lambda(s, m_i^2, m_j^2)}{s} \sigma_{ij \to kl}(s) \, ,
\ee
in which $\lambda(x,y,z) := (x-y-z)^2 - 4yz$ is the K\"{a}ll\'{e}n-function,  $c_{ij}$ is a symmetry factor that equals to $2$ if the particles $i$ and $j$ are identical and $\sigma_{ij \to kl}$ is the cross-section for two states $i$ and $j$ going into the states $k$ and $l$. Note that the reduced cross-section, as well as the reaction density, do not depend on the particular choice of the process direction $ij \leftrightarrow kl$. 

The general eq.~\eqref{eq:lilgam} was derived under the assumption that the initial particles follow Maxwell-Boltzmann statistics. Nevertheless, this result remains valid for other statistical distributions as long as the system is in the non-relativistic regime, which holds for~$T< 3 \, m_{i,j}$. In this regime, where the temperature is much lower than the energy, both the Fermi-Dirac and Bose-Einstein distribution functions reduce to the Maxwell-Boltzmann form,  $f_{i,j} \sim e^{-E/T}$. For the same reason, the Bose enhancement and Pauli blocking factors can be neglected, since $(1 \pm f_{i,j}) \sim 1$. In our case, this condition is satisfied in the coannihilations and $\chi_2-\chi_1$ scattering processes, as the dark fermions undergo freeze-out non-relativistically.

However, this does not apply to dark fermion-SM fermion scatterings, as the initial state SM fermions can be relativistic at freeze-out. Therefore, to compute the kinetic rate $\Gamma_{2f \to 1f}$, we will follow the 
prescriptions outlined in Refs.~\cite{Gondolo:2012vh,Bertoni:2014mva,Berlin:2023qco} and write
\be
\Gamma_{2f \to 1f} = - \frac{g_f}{6 m_1 T} \int_0^{\infty} \frac{d^3p}{(2 \pi)^3} f_d (1-f_d) \frac{p}{\sqrt{p^2 + m_f^2}} \int_{-4 p^2}^{0} dt  \, t \, \frac{d \sigma_{\chi f}}{dt} \, ,
\ee
where $g_f$ represents the statistical degrees of freedom (for example, in the case of scattering with electrons, $g_e= 2$ spin states, and for the case of quarks we need  $g_q= 6$ to account for spin and color), $p$ is the momentum of the initial SM fermion, $f_d= (e^{E/T} +1)^{-1}$ is the Fermi-Dirac distribution function for fermions with energy $E$, $m_f$ is the fermion mass, $t$ is the usual Mandelstam variable and $\sigma_{\chi f}$ is the $\chi_2 f \to \chi_1 f$ scattering cross-section. For the $U(1)_Q$ models considered here the fermion $f$ will usually represent electrons, neutrinos or quarks~\footnote{As the $\chi_2$-$f$ scattering only has a subdominant effect on the relic abundance calculation, for simplicity, only light $u,d$ quarks are used as an approximation instead of scatterings with pions and other mesons.}.

Finally, in the case of decays, we can express the thermally averaged decay rate of particle $i$ into the set of particles $\{j\}$ as
\begin{align}
   \ang{\Gamma}_{i \to \{j\}} = \Gamma_{i\to \{j\}} \frac{K_1(m_i/T)}{K_2(m_i/T)}~,
\end{align}
which holds for any n-body final-state configuration if there are no further restrictions, \eg Pauli blocking in certain environments. The ratio of Bessel functions represents the temperature dependence of the decay. At high temperatures, the decaying particle is boosted, leading to slower decay rates due to time dilation. Consequently, in this limit, the ratio approximates the inverse boost factor.

\begin{figure}[h!]
\begin{center}
\includegraphics[width=1.\textwidth]{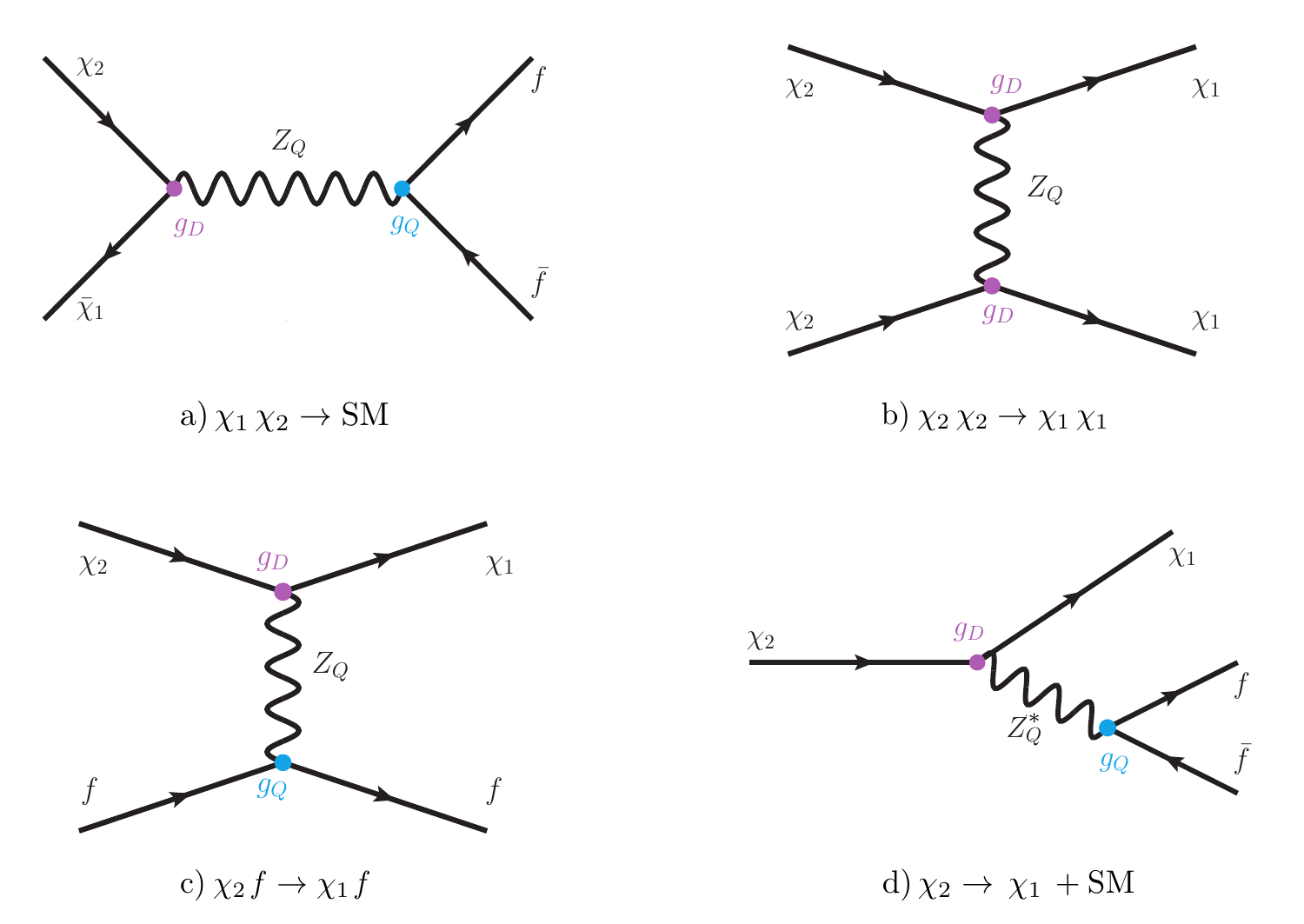}
\end{center}
\vglue -0.8 cm
\caption{\label{fig:relic_diagrams} Feynman diagrams for the processes that determine the relic abundance of $\chi_1$ dark matter in iDM$_Q$: coannihilation -- $\sigma_{12ff}$ (upper left), $\chi_2 - \chi_1$ scattering --  $\sigma_{2211}$ (upper right), $\chi_2$-fermion inelastic scattering -- $\sigma_{2f1f}$ (lower left), and $\chi_2$ decays -- $\Gamma_{2\to1ff}$ (lower right). }
\end{figure}

In figure~\ref{fig:relic_diagrams} we illustrate the Feynman diagrams for the relevant tree-level processes that enter the Boltzmann equations in iDM$_Q$. Additional in-depth studies of the possible processes involved in the thermal evolution of iDM can be found in Ref.~\cite{CarrilloGonzalez:2021lxm}. Note that, given the mass hierarchy $m_{Z_Q} > m_1 + m_2$, the self-annihilation channel $\chi_1 \chi_1 \to Z_Q Z_Q$ is kinematically forbidden for non-relativistic iDM$_Q$, and, hence, the coannihilation process $\chi_1 \chi_2 \to {\rm SM}$ dominates the freeze-out. This hierarchy choice is essential because the self-annihilation thermal average cross-section is velocity-independent ($s$-wave), leading to continuous energy injection into the plasma which also remains during the recombination era. Such injection distorts the CMB power spectrum, imposing a stringent bound on the dark matter mass $m_{\text{DM}} \gtrsim 10$ GeV~\cite{Planck:2018vyg,Galli:2011rz,Leane:2018kjk}. Conversely, although the coannihilation thermal average cross-section $\ang{\sigma v}_{12 \to ff}$ also exhibits $s$-wave dominance, this channel has the feature that, at later times, the heavier state $\chi_2$ has already decayed. This results in a low abundance of $\chi_2$ during recombination, which inhibits $\chi_1 \chi_2 \to {\rm SM}$ processes, evading the CMB bounds.

To determine the DM relic density at freeze-out, as well as to track the heavier dark fermion population suppression, one should solve the system of coupled Boltzmann Eqs.~\eqref{eq:boltziDM} considering all  four channels pictured in figure~\ref{fig:relic_diagrams}. However, as we will show below, in most regions of the parameter space it is sufficient to sum the two Boltzmann equations and solve for the number density of dark species $n \equiv n_1 + n_2$. Since scatterings and decays conserve the total number of $\chi_1 + \chi_2$, the only process that survives is the coannihilation channel. Therefore, by using this approximation we end up with a much simpler single equation for $Y^{\rm eff}= n/s$, given by
\begin{equation} \label{eq:boltzcoann}
\begin{aligned}
    \dv{Y^{\rm eff}}{x} = \frac{s}{H x}  \, \bigg[ & - 2 \, \ang{\sigma v}_{\rm eff}  \big( \, (Y^{\rm eff})^2 - (Y^{\rm eq})^2 \, \big) \bigg] \, ,
\end{aligned}
\end{equation}
where
\be
\ang{\sigma v}_{\rm eff}  =  \ang{\sigma v}_{12 \to ff}   \frac{n_1^{\rm eq} n_2^{\rm eq}}{(n^{\rm eq})^2} \, ,
\ee
and we have used the coannihilation approximation $n_1 n_2^{\rm eq} \sim n_2 n_1^{\rm eq}$~\cite{Griest:1990kh} . Since the coannihilation dominates the freeze-out, it is useful to show the parametric dependence of this cross-section, which can be expressed by the following formula
\be
\sigma_{12ff} (s)=\frac{12\pi s^2}{(s-m_{Z_Q}^2)^2 + m_{Z_Q}^2 \Gamma_{Z_Q}^2}\frac{\Gamma_{Z_Q\to \,{\rm SM} }(s) \, \Gamma_{Z_Q\to \chi_1\chi_2}(s)}{\lambda(s,m_1^2,m_2^2)} \, ,
\ee
where $\Gamma_{Z_Q}$ is the mediator total decay width. The partial decay widths, $\Gamma_{Z_Q\to \,{\rm SM}}(s)$ into  SM states (see eq.~\eqref{eq:GZff} for fermions), and $\Gamma_{Z_Q\to \chi_1\chi_2}(s)$ into dark fermions (see eq.~\eqref{eq:GZXX}), are computed with the replacement $m_{Z_Q}^2 \to s$. The leading order term in the relative velocity expansion of the thermally-averaged cross-section appears at $v^0$ ($s$-wave term) and has the following parametric dependence in the non-relativistic limit
\be \label{eq:sigvcoann}
 \ang{\sigma v}_{12 \to ff} \simeq \frac{ (q_Q^f g_Q)^2 \alpha_D (m_1 +m_2)^2}{[ (m_1 + m_2)^2-m_{Z_Q}^2]^2 + m_{Z_Q}^2 \Gamma_{Z_Q}^2} = \frac{ 4 \pi (q_Q^f)^2 \alpha_Q \alpha_D (\Delta + 2)^2 R^{2}}{ m_{Z_Q}^2[ (\Delta + 2)^2-R^2]^2 + R^4 \Gamma_{Z_Q}^2} \,.
\ee

\begin{figure}[h!]
\begin{center}
\includegraphics[width=0.49\textwidth]{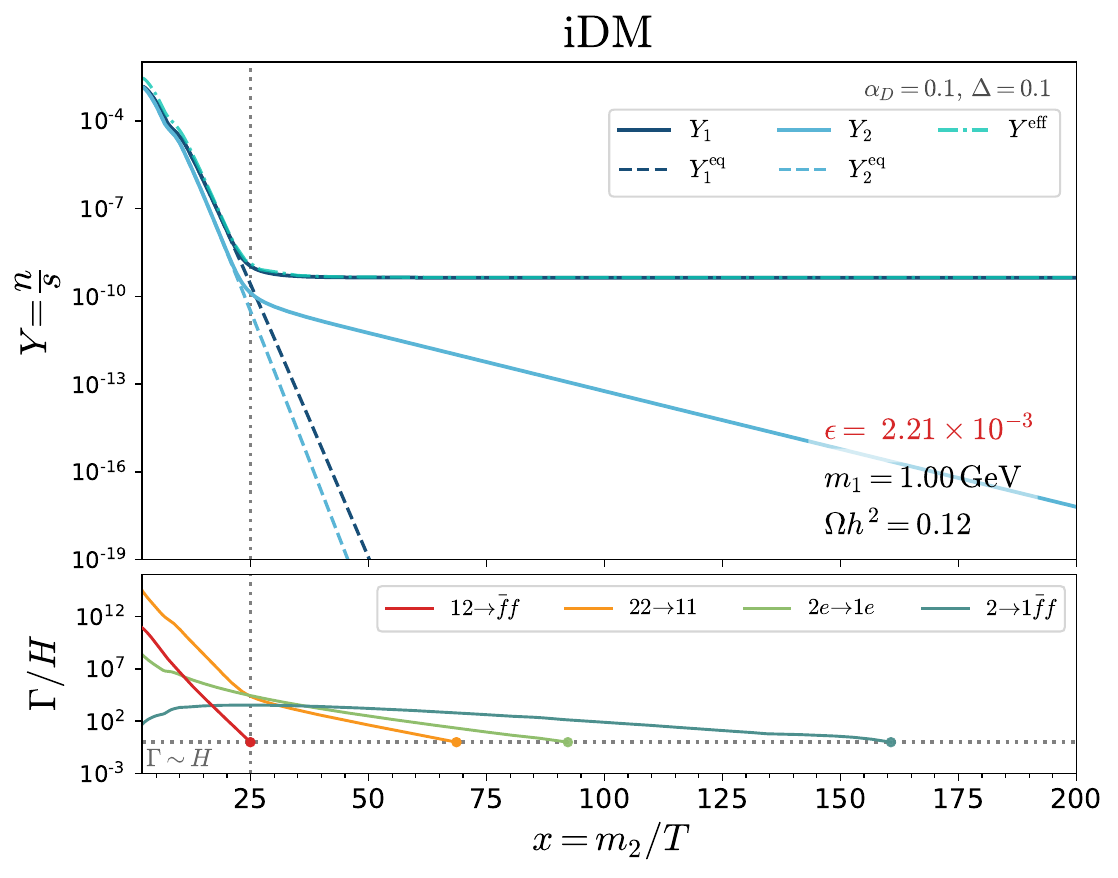}
\includegraphics[width=0.49\textwidth]{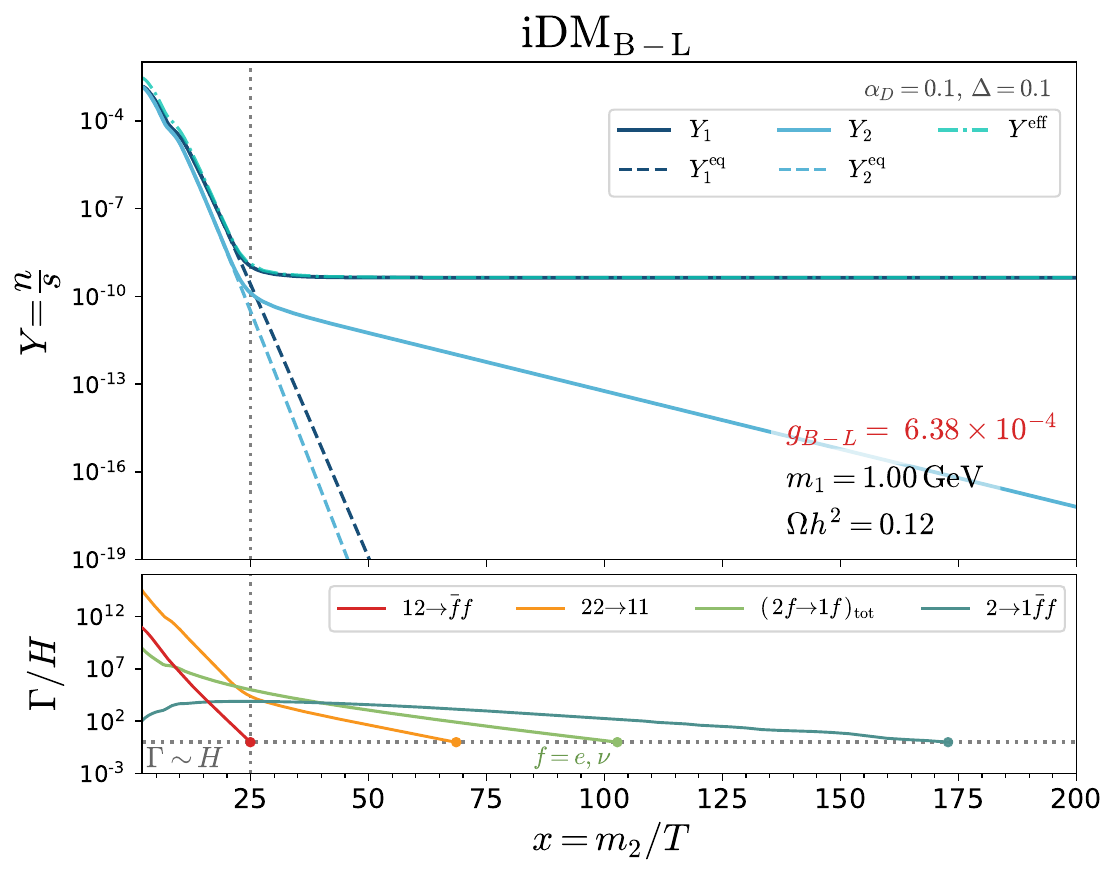}
\includegraphics[width=0.49\textwidth]{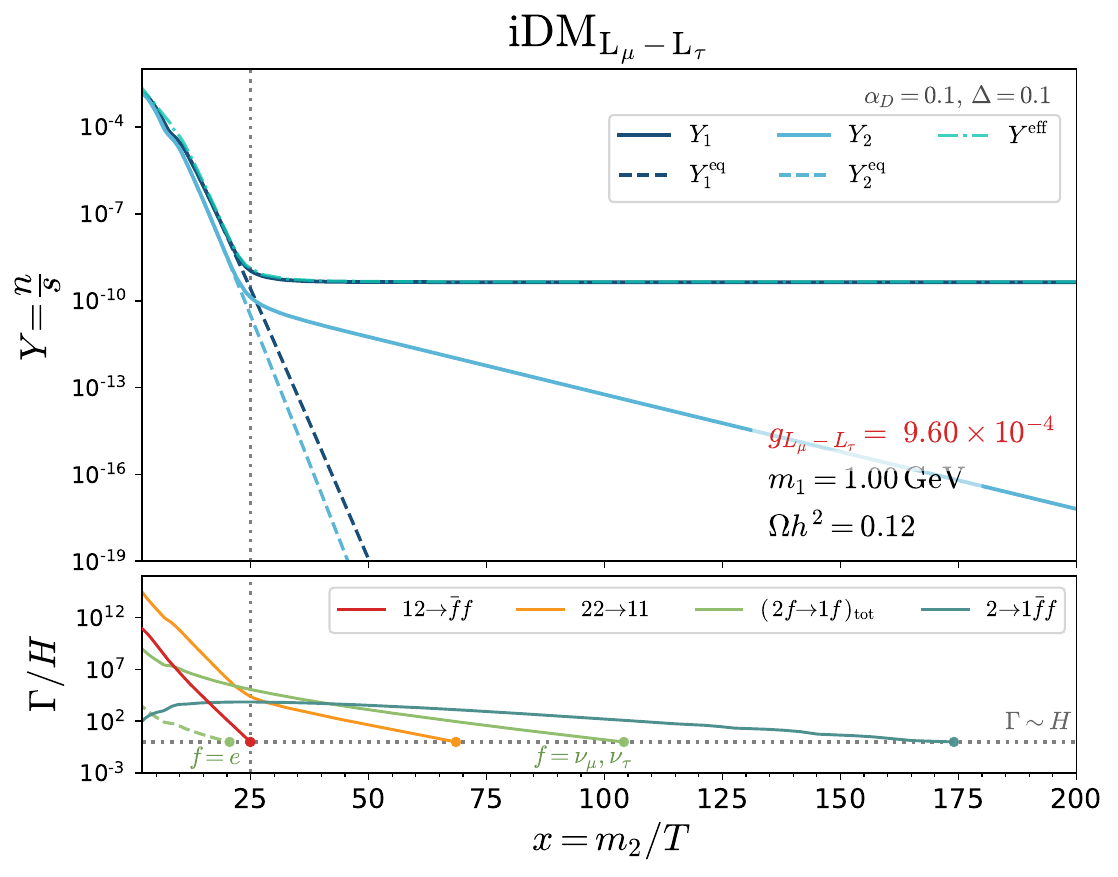}
\includegraphics[width=0.49\textwidth]{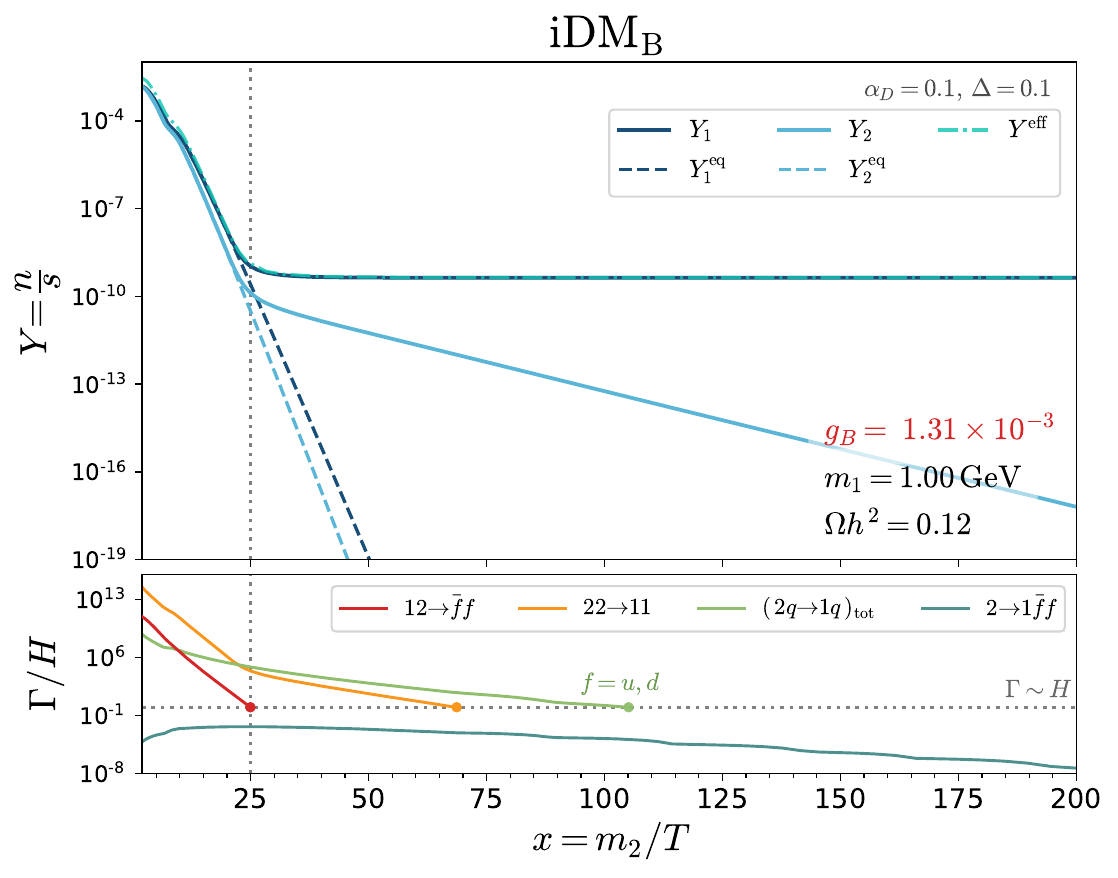}
\end{center}
\vglue -0.8 cm
\caption{\label{fig:fo_curve_models} The freeze-out scenarios for different iDM$_Q$: dark photon iDM (upper left), iDM$_{B-L}$ (upper right), iDM$_{L_\mu- L_\tau}$ (lower left) and iDM$_B$ (lower right). 
The upper panels of each plot show the comoving number density evolution of $\chi_1$ (dark blue) and $\chi_2$ (light blue) as a function of $x = m_2/T$. The dashed lines represent their respective equilibrium yields. The dash-dotted cyan lines correspond to the $Y^{\rm eff}$ solution of the Boltzmann equation with the coannihilation approximation. Each plot indicates the value of $g_Q$ chosen to reproduced the correct DM relic density. In the lower panels of each plot, we show the rates for coannihilation (red), $\chi_2 - \chi_1$ scattering (orange), $\chi_2 -$fermion scattering (green) and $\chi_2$ decays (blue), all normalized by the Hubble expansion rate. The colored dots that intersect the horizontal line $\Gamma \sim H$ signalize the decoupling temperature of each process, allowing one to track the cosmic history. For iDM$_{B-L}$, the fermion scattering channel involves electrons and neutrinos; the corresponding rate shown in the plots corresponds to a representative fermion, \ie either an electron or a neutrino flavor. Finally, the vertical dotted line marks the moment of coannihilation decoupling, which is a good approximation for the freeze-out temperature.}
\end{figure}

Another important ingredient to consider when analyzing the thermal history is the definition of the channel rate, which should be compared to Hubble with $\Gamma(T_{\rm fo})/H(T_{\rm fo})\overset{!}{=}1$ in order to find the freeze-out temperature $T_{\rm fo}$ of each process. In terms of such rates, we rewrite eq.~\eqref{eq:boltziDM} for the DM as
\begin{equation} \label{eq:boltzrates}
\begin{aligned}
    \dv{\ln Y_1}{\ln x} =   - \frac{\Gamma_{12}}{H}  \qty( \, 1  - \frac{Y_1^{\rm eq}Y_2^{\rm eq}}{Y_1 Y_2 } \, ) + \frac{\Gamma_{22}}{H} 
     \qty( 1 - \qty[\frac{Y_1 Y_2^{\rm eq}}{Y_2  Y_1^{\rm eq}}]^2 ) + \qty( \frac{\Gamma_{2f}}{H}+ \frac{ \Gamma_{2}}{H}) \qty(1 - \frac{ Y_1 Y_2^{\rm eq}}{Y_2 Y_1^{\rm eq}} )  \, ,
\end{aligned}
\end{equation}
where
\begin{gather} \label{eq:chrates}
    \begin{aligned}
   \Gamma_{12} & \equiv \ang{\sigma v}_{12 \to ff} n_2 
    \qquad \quad & \Gamma_{2f} &\equiv \Gamma_{2f \to 1f} \, \frac{n_2}{n_1} 
    \\ 
    \Gamma_{22} &\equiv 2 \ \ang{\sigma v}_{22 \to 11} n_2  \, \frac{n_2}{n_1} 
    \qquad \quad & \Gamma_{2} &\equiv \ang{\Gamma}_{2 \to 1 ff}  \,  \frac{n_2}{n_1} \, .
    \end{aligned}
\end{gather}
To discuss the main features of iDM$_Q$, we present the
time evolution of the dark sector 
population and the corresponding freeze-out curves, for $\alpha_D = 0.1$ and $m_1 = 1$ GeV, in figure~\ref{fig:fo_curve_models}. 
In each plot, the value of $g_{Q}$ is fixed to achieve the correct dark matter relic density $\Omega_{\chi_1} h^2 = 0.12$. In the upper panels, the solid dark (light) blue curves represents the evolution of $\chi_{1}$ ($\chi_{2}$) number density, while the dashed lines corresponds to their respective equilibrium yields, given by eq.~\eqref{eq:eqyield}. The dot-dashed cyan lines give, in each case, the total dark sector yield $Y^{\rm eff}$ obtained by solving the single Boltzmann equation with the coannihilation approximation (eq.~\eqref{eq:boltzcoann}). The lower panels illustrate the channel rates of eq.~\eqref{eq:chrates} normalized by the Hubble parameter, with different colors representing the various channels, as indicated by the labels. The horizontal dotted line corresponds to $\Gamma \sim H$, and the colored dots mark the moments when each rate crosses this line, indicating the temperatures at which each process decouples during cosmological history. In the case of $\chi_2 -$fermion scatterings, the included fermion species are specified in green.
The vertical dotted line crossing both panels indicates the $x$ value at which $\Gamma_{12}$ decouples, highlighting the dominance of the coannihilation channel in the iDM/iDM$_Q$ freeze-out. After such decoupling, the number density of $\chi_1$ stabilizes, while the $\chi_2$ one continues to decrease due to scatterings and decays. We omit the curve for the iDM$_{B-3L\tau}$ model, as it is similar to the iDM$_{B-L}$ case, with the only difference being a small decrease in the $g_Q$ value that correctly reproduces the observed DM density.

Another comment regarding the freeze-out curves is the appearance of wiggle features at the beginning of the number density yields. This pattern represents the transition when the temperature drops below $\delta = m_1 \Delta$. At higher temperatures, both directions of the considered processes are very efficient. However, after this threshold, the annihilation of $\chi_2$ becomes more favorable, leading to a separation in the equilibrium yield curves proportional to $n_2^{\rm eq}/n_1^{\rm eq} \sim e^{-\delta/T}$. For a detailed discussion on the effects of varying the mass splitting parameter $\Delta$ on the freeze-out curves, along with a comparison of the thermally-averaged cross-sections for different channels, we refer to appendix~\ref{app:rddetails}. In addition to the individual freeze-out curves, we also present the thermal target curves in the dark sector parameter space for different choices of gauge couplings and mass ratio $R$ values in~appendix~\ref{app:rddetails}.


Comparing the {\it vanilla} dark photon iDM with the iDM$_{B-L}$ in figure~\ref{fig:fo_curve_models}, one can see that they present similar cosmic histories. However, in the iDM$_{B-L}$, dark scatterings and decays decouple later, primarily due to the presence of neutrinos. The iDM$_{L_\mu-L_\tau}$ exhibits similar features. The main difference compared to iDM$_{B-L}$ is that, as the coupling to electrons is loop suppressed in this model, the green solid line indicates the scattering rate with muon and tau neutrinos. 
Scatterings with electrons decouple much earlier as shown by the dashed green line.
In the case of the iDM$_B$, one can see that the decay rate is never efficient, given that the $\chi_2$ decays into leptons are highly loop-suppressed. Furthermore, the fermions represented in the $\chi_2-$fermion scattering rate (green line) correspond to the $u$ and $d$ quarks. Apart from the cosmic histories, one of the key difference in the proposed iDM$_Q$ scenarios is the shift in the value of the gauge coupling $g_Q$ that correctly reproduces the observed DM abundance. As we will see in the next section, this shift opens up new allowed regions in the parameter space of these models.

All the results displayed in this section can be reproduced via the python code \textsc{ReD-DeLiVeR}, which we made publicly available at GitHub~\cite{ReD-DeLiVeR}. This package is designed to solve the Boltzmann equations numerically and evaluate the relic density curves and thermal targets for any iDM$_Q$ scenario as well as for other simplified DM models. It is closely related to the \textsc{DeLiVeR} code~\cite{DeLiVeR}, which is a package capable of computing light vector mediator decays within user-defined $U(1)_Q$ models including hadronic channels. More details about the \textsc{ReD-DeLiVeR} code can be found in appendix~\ref{app:rdcode}.

\section{Phenomenological Implications}
\label{sec:pheno}

\subsection{Experimental Sensitivities}

Now we turn to the analysis of the cosmological motivated region of the parameter space available for iDM$_Q$, where the relic abundance satisfies $\Omega_{\chi_1}h^2 \lesssim 0.12$, and which can be probed by current and future high intensity accelerator experiments. Many of the constraints considered in our analysis for the baryophilic iDM$_Q$ models were inspired by the study in~\cite{Batell:2021snh} whereas the iDM$_{L_\mu-L_\tau}$ analysis has similarities with searches for the $L_\mu-L_\tau$ minimal DM model~\cite{Foldenauer:2018zrz}. For additional details regarding the recasting methods and specifications on the bound computations, we refer to appendix~\ref{app:bounds}.

We will first consider several types of facilities where the dark state $\chi_2$ can be produced via the prompt decay $Z_Q \to \chi_1 \chi_2$, and may subsequently decay either visibly or semi-visibly. Let us start by listing these facilities, describing the main production channels of the mediator $Z_Q$, highlighting the iDM$_Q$ models they can constrain, and identifying the current detectors and experiments that possess relevant data, as well as those expected to provide pertinent data in the near future:

\vspace{3pt}
 
\paragraph{$e^+e^-$ colliders:} here the main production process is $e^+e^-\to \gamma Z_Q$, which requires the mediator to couple to electrons either directly as in iDM, iDM$_{B-L}$ or loop-induced as in iDM$_B$ and iDM$_{L_\mu-L_\tau}$. So, in principle, $e^+e^-$ experiments can place limits on all of these models if they have conducted dark photon  ($Z_\gamma$)  searches. The experiments that have available data 
are: (i) the BaBar experiment  at the PEP-II asymmetric-energy $e^+e^-$ B-factory of the SLAC National Accelerator Laboratory~\cite{BaBar:2001yhh} that collected data from 1999 to 2008 at the center of mass energy $\sqrt{s} \simeq 10$~GeV; and (ii) the BES II experiment at the Beijing Electron Positron Collider that started in 1991 and is still collecting data at $\sqrt{s} \simeq (3-5.6)$~GeV~\cite{BES:2001vqx}. 
The Belle II detector~\cite{Belle-II:2018jsg} started to collect data produced by $e^+e^-$ collisions at $\sqrt{s} \simeq 10$~GeV at SuperKEKB in Japan in 2018. We expect to have results from this detector that can probe iDM$_Q$ soon.

\vspace{3pt}

\paragraph{$\ell^-$ beam dump:} here $Z_Q$
can be produced by $\ell^- X \to \ell^- X Z_Q$ ($\ell^-$ Bremsstrahlung), where $X$ is a target, as long as it couples to electrons ($\ell = e$) or muons  ($\ell = \mu$). 
So just as explained before, in principle, all iDM$_Q$ can be constrained. There are two experiments that provide useful data in this category: (i) the NA64 experiment at the CERN Super Proton Synchrotron (SPS), which has been collecting data since 2016 and can operate with a 100 GeV $e^- $ (NA64-$e$) or a 160 GeV $\mu^-$ (NA64-$\mu$) incident beam on an active target~\cite{Kirsanov:2019inh}; and (ii) the old 20 GeV $e^-$ beam dump E137 experiment at SLAC~\cite{Bjorken:1988as}. There is also a proposal for a future Light Dark Matter eXperiment (LDMX) that could operate with a 4 to 8 GeV $e^-$ beam at SLAC or 4 to 16 GeV $e^-$ beam at CERN~\cite{LDMX:2018cma}.

\vspace{3pt}

\paragraph{proton beam dump:} 
 these are, in general, neutrino experiments operating in the beam dump mode at different proton energies. The main idea is that the $\pi^0$'s and $\eta$'s produced by a high intensity proton beam hitting a thick targets can decay as  $\pi^0, \eta \to \gamma Z_Q$ followed by   $Z_Q \to \chi_1 \chi_2$, in models coupled to $B$. For $m_{Z_Q} \gtrsim 0.5$ GeV, proton Bremsstrahlung becomes the main production channel, before the Drell-Yan process starts to prevail.
 The available data here comes from a number of experiments: (i) the NuCal experiment~\cite{Blumlein:1990ay}, which operated with a 70 GeV proton beam hitting an iron target; (ii) the CERN-Hamburg-Amsterdam-Rome-Moscow (CHARM) experiment~\cite{CERN-Hamburg-Amsterdam-Rome-Moscow:1980ppk}, which used the 400~GeV proton beam from the CERN SPS hitting a copper target;
 (iii) the MiniBooNE experiment (BooNE is an acronym for the Booster Neutrino Experiment) at the Fermi National Accelerator Laboratory (Fermilab) using the 8~GeV Booster proton beam delivered to a steel target~\cite{MiniBooNE:2017nqe,MiniBooNEDM:2018cxm};
(iv) the Liquid Scintillator Neutrino Detector (LSND) experiment~\cite{LSND:2001akn} at Los Alamos, which employed an 800 MeV proton beam directed at a water-copper target; and (v) the NA62 experiment~\cite{NA62:2017rwk}, which is the only one that is not a neutrino experiment but rather a rare kaon decay facility. The detector started operating in 2015 and is still collecting data. It utilizes the 400 GeV proton beam from the CERN SPS, which strikes a beryllium target to produce a 75 GeV secondary beam of positive particles, of which 6\% are charged kaons.

\vspace{3pt}

\paragraph{hadron colliders:}

The relatively new ForwArd Search ExpeRiment (FASER) detector~\cite{FASER:2022hcn} is an experiment dedicated to the search of light, extremely weakly-interacting particles at CERN’s Large Hadron Collider (LHC). It started operating in 2022 and is expected to run for 4 years. Located on the beam collision axis around 480 m from the ATLAS detector, it receives beams from the LHC proton-proton collisions with $\sqrt{s}=$ 13.6 TeV. A future upgraded version of this facility, FASER-2, is also a possibility for the future~\cite{FASER:2018eoc,Feng:2022inv}.
Additionally, the Collider Detector at Fermilab (CDF), an experiment located at the Tevatron accelerator, can place limits on the production of dark particles through monojet searches, \ie $p \bar p \to j +$ missing transverse energy at $\sqrt{s}=1.96$ TeV~\cite{CDF:2012tfi}, when the mediator couples to $B$. Besides, the Compact Muon Solenoid (CMS) experiment at the LHC can use the production channel $pp \to 4 \mu$ at $\sqrt{s}=13$ TeV to constrain models that couple to $L_\mu-L_\tau$~\cite{CMS:2018yxg}. Finally, the LHCb detector at the LHC can also probe a free region of the iDM$_B$ parameter space through an inclusive dark photon search with muon final states~\cite{LHCb:2017trq}.

The experiments of the above facilities can perform two 
types of searches:

\subsubsection{Invisible Searches}

In invisible searches, it is assumed that once $Z_Q$ is produced, it promptly decays via $Z_Q \to \chi_1 \chi_2$, with $\chi_2$ either decaying invisibly into neutrinos within the detector or decaying outside the detector’s sensitive volume. Another possible scenario involves $Z_Q$ decaying directly into neutrinos, although this case becomes relevant only for large $g_Q$ couplings.

The main bounds from the BaBar experiment comes from their mono-photon event search, using 53 fb$^{-1}$ $e^+ e^-$ scattering data, consistent with the production $e^+ e^- \to \gamma Z_\gamma$ followed by the decay $Z_\gamma \to {\rm invisible}$
for $m_{Z_\gamma} \leq 8$ GeV~\cite{BaBar:2017tiz}.
This analysis excludes a region of the parameter space ($m_{Z_\gamma}$, $\epsilon$), given in figure~5 of  Ref.~\cite{BaBar:2017tiz}, that can be recast 
to provide bounds on iDM$_Q$.
BaBar also has searched for invisible decays of $\Upsilon$(1S) meson  using a sample of $91.4 \times 10^6$ $\Upsilon$(3S), which can produce $\Upsilon$(1S) via the decay
$\Upsilon$(3S)$\to \pi^+ \pi^- \Upsilon$(1S). 
They were able to place a limit of $\rm{BR}(\Upsilon({\rm 1S}) \to \rm{invisible}) < 3.0 \times 10^{-4}$~\cite{BaBar:2009gco}. Similarly, 
the BES II Collaboration also explored invisible meson decay, constraining $\rm{BR}(J/\psi \to \rm{invisible}) < 7.2 \times 10^{-4}$~\cite{BES:2007sxr} at 90\% confidence level. These limits can be translated into a bound on $g_Q$ if the mediator couples to $B$, which is significant around $m_{Z_Q} \simeq m_{\Upsilon(\rm 1S)}$ and $m_{Z_Q} \simeq m_{J/\psi}$, respectively. In the future, the Belle II experiment is expected to enhance the sensitivity of searches for invisible decays of the $\Upsilon$(1S) meson, reaching $\rm{BR}(\Upsilon({\rm 1S}) \to \rm{invisible}) \lesssim 1.3 \times 10^{-5}$~\cite{Belle-II:2018jsg}. Likewise, the BES III experiment is projected to improve the sensitivity of $J/\psi$ invisible decay searches, with an estimated limit of $\rm{BR}(J/\psi \to \rm{invisible}) \lesssim 3 \times 10^{-5}$~\cite{BESIII:2020nme}.
Besides, a muon-philic search has been carried out recently by BES III~\cite{BESIII:2023jji} in the process $J/\Psi \to \mu^+\mu^- Z_Q$ where $Z_Q$ decays invisibly. We expect those results to lie in the range of BaBar and Belle II which will be discussed below.

Regarding the NA64 detector at CERN, the experiment conducted two different invisible searches for each beam dump mode configuration. The NA64-$e$ Collaboration performed a missing energy search for $Z_\gamma \to {\rm invisible}$~\cite{NA64:2016oww,Banerjee:2019pds,NA64:2023ehh}, using $2.75 \times 10^9$ $e^-$ on target, for $m_{Z_\gamma}< 100$ MeV. They 
have found no evidence of $Z_\gamma$, allowing them to place limits on the plane ($m_{Z_\gamma}$, $\epsilon$), as showed in figure 3 of Ref.~\cite{NA64:2023ehh}, which can be recast to the iDM$_Q$ parameter space. The NA64-$\mu$ experiment used $1.98 \times 10^{10}$ $\mu^-$ with 160 GeV momentum  direct onto a target to place limits on a new vector boson coupled to $L_\mu - L_\tau$ \cite{NA64:2024klw}. This was done by looking for events with a single scattered muon with momentum $< 80$ GeV in the final state, accompanied by missing energy. No event was observed so they derived bounds 
 valid for $m_{Z_{L_\mu-L_\tau}}< 2.4$ GeV assuming the {\it vanilla} $L_\mu-L_\tau$ model. Their bounds in the plane ($m_{Z_{L_\mu-L_\tau}}$, $g_{L_\mu-L_\tau}$) given in figure 4 of Ref.~\cite{NA64:2024klw} can be recast to iDM$_{L_\mu-L_\tau}$. 

The NA62 experiment used a total of $4.12 \times 10^{8}$ 
$\pi^0$'s, produced through the decay chain $K^+ \to \pi^+ \pi^0$,
$\pi^0 \to \gamma Z_\gamma$, to search for $Z_\gamma \to {\rm invisible}$. No signal was observed in their data, allowing them to improve upon previous searches in the mass range of 60-110 MeV~\cite{NA62:2019meo}. We recast the limits presented in figure 7 of their paper to obtain constraints for our models.

Finally, the CDF experiment looked for events with a single energetic jet and large missing transverse energy using an integrated luminosity of 6.7 fb$^{-1}$~\cite{CDF:2012tfi}, finding no discrepancies compared to the SM prediction. In this work, we recast the monojet bounds from the CDF search in the low DM mass regime, as presented in figure 4 of Ref.~\cite{Shoemaker:2011vi}, which are relevant for the iDM$_B$ scenario.

In the future, Belle II can set more stringent limits by searching for $e^+e^-\to \gamma Z_Q \to \gamma \, \chi_2 \chi_1$ using missing momentum and displaced $\chi_2$ decay searches~\cite{Duerr:2019dmv}. We project the sensitivity of Belle II to the iDM$_Q$ parameter space with $Z_Q$ decaying invisibly by recasting the predictions from figure 209 of Ref.~\cite{Belle-II:2018jsg}, which correspond to 20 fb$^{-1}$ of integrated luminosity. Additionally, we verified that our rescaled result for Belle II is consistent with a MadGraph simulation (see appendix~\ref{app:bounds}).

Similarly, the proposed LDMX experiment would place stringent limits on the iDM parameter space using missing energy searches~\cite{LDMX:2018cma}, which can be translated to iDM$_Q$ models. We recast the projected sensitivity curves considering a 10\% radiation length tungsten target and an 8 GeV beam, assuming $10^{16}$ $e^-$ on target, as presented in Ref.~\cite{Berlin:2018bsc}.

\subsubsection{Semi-visible Searches \label{subsub:semivis}}
In semi-visible searches, one assumes that after $Z_Q$ is produced, it promptly decays as $Z_Q \to \chi_1 \chi_2$  following by the semi-visible decay $\chi_2 \to \chi_1 \, \ell^+ \ell^-$ inside the detector~\footnote{Note that another possible signature is the direct visible decay $Z_Q \to \ell^+ \ell^-$; however, in most regions of the parameter space, this channel is negligible, as it only becomes relevant for $g_Q \sim \mathcal{O}(1)$. }.

The BaBar experiment can set constraints on $\mu$-philic models where the mediator couples to muons,  such as in iDM$_{L_\mu-L_\tau}$, through their 4-muon channel analysis, \textit{i.e.}, $e^+ e^- \to \mu^+ \mu^- Z_Q$, followed by $Z_Q \to \chi_1 (\chi_2 \to \chi_1 \mu^+ \mu^-$), using 514 fb$^{-1}$~\cite{BaBar:2016sci} to constrain  $g_{L_\mu - L_\tau}$ for $0.2 \lesssim (m_{Z_{L_\mu-L_\tau}}/{\rm GeV})  \lesssim 8$.

The data from the old 20 GeV $e^-$ beam dump E137 experiment at SLAC~\cite{Bjorken:1988as} was used to place limits on iDM in Ref.~\cite{Mohlabeng:2019vrz}. Their analysis is based on the non-zero probability that $\chi_{1,2}$ travel to the detector and scatter off $e^-$ targets. They also consider the scenario where the heavier dark fermion decays inside the detector as $\chi_2 \to \chi_1 e^+ e^-$. In this work, we considered the decay signatures and recast them for iDM$_Q$ models. However, it is important to emphasize that there are some caveats regarding a robust derivation of iDM constraints from Ref.~\cite{Bjorken:1988as}. As highlighted in Refs.~\cite{Berlin:2018pwi,Berlin:2018jbm,Tsai:2019buq}, the appropriate energy threshold for deriving the bounds (ranging from 1 to 3 GeV) remains uncertain, and the original analysis used a visual display that lacks precision by today's standards.

There are a few proton beam dump experiments with data that can be used to constrain models through the semi-visible decay $\chi_2 \to \chi_1 e^+ e^-$. The NuCal experiment collected $1.7\times 10^{18}$ protons on target (POT) in beam dump mode, producing 1.96 $\pi^0$'s and 0.22 $\eta$'s per POT. Similarly, the CHARM experiment performed a search for heavy neutral leptons $N$ decaying to $N\to \nu e^+ e^-$ in their detector~\cite{CHARM:1985nku} with a total of $2.4\times 10^{18}$ POT and 2.4 $\pi^0$'s and 0.26 $\eta$'s produced per POT. The authors of Ref.~\cite{Tsai:2019buq} derived bounds from the NuCal and CHARM data for the iDM scenario. They considered the distance from the target to the decay region to be 64 m (NuCal) and 480 m (CHARM), with fiducial particle decay lengths of 23 m (NuCal) and 35 m (CHARM). We recast their bounds from figure 1 and 2 of their paper to place limits on iDM\(_Q\).

The Liquid Scintillator Neutrino Detector (LSND) was a neutrino experiment conducted at Los Alamos from 1993 to 1998~\cite{LSND:1996jxj}. With its high-luminosity, relatively low-momentum 800 MeV proton beam directed onto the target, the experiment was capable of producing dark vector bosons primarily through secondary $\pi^0$ decays, as other production modes were negligible. LSND observed 55 background-subtracted events from their neutrino-electron elastic scattering analysis~\cite{LSND:2001akn}. This signal was used in Ref.~\cite{Izaguirre:2017bqb} to compute bounds on iDM for $Z_{\gamma} \lesssim 100$ MeV, by considering that the LSND data could be consistent with the iDM semi-visible signal, $Z_\gamma \to \chi_1 (\chi_2 \to \chi_1 e^+ e^-)$, if the angular separation of the $e^+e^-$ pair satisfied $\theta_{e^+ e^-}< 12^\circ$ and the combined energy $E_{e^+}+ E_{e^-}< 50$ MeV. To compute the expected iDM yield, they assumed a total production of \(10^{22}\) $\pi^0$'s, a 19\% detection efficiency for e$^\pm$, and simulated the detector as having a length of approximately 10 m and a radius of about 3 m, positioned roughly 30 m from the beam stop. We followed their approach to derive similar limits for iDM$_{B-L}$, iDM$_B$, and iDM$_{L_\mu-L\tau}$.

The CMS detector at LHC has searched for the process $pp \to \mu^+\mu^- Z_{L_\mu-L_\tau}$, followed by the decay $Z_{L_\mu-L_\tau}\to \mu^+\mu^-$, using an integrated luminosity of 77.3 fb$^{-1}$ at $\sqrt{s}=13$ TeV~\cite{CMS:2018yxg}. This analysis yielded limits on the coupling $g_{L\mu - L_\tau}$ and the mass $m_{Z_{L_\mu-L_\tau}}$ in the mass range
5-70 GeV. Here, we recast these limits to constraint the iDM$_{L_\mu-L_\tau}$ model in the high mass region. Similarly, the LHCb detector conducted an inclusive search for dark photons decaying into muons~\cite{LHCb:2017trq}. Although the derived bounds are less stringent compared to the other limits presented here, there remains a small region in the high mass regime of the iDM$_B$ model where these constraints become significant, assuming a model-dependent loop-generated kinetic mixing coupling the mediator to muons. We considered the LHCb muon search limits for a $B$ boson mediator~\cite{Ilten:2018crw,Foguel:2022ppx} to constrain this region in the iDM$_B$ model. This is because, in the high coupling/mass regime of iDM$_B$, the dark sector states are more likely to decay visibly (see figure~\ref{fig:PinvB}).

Regarding the FASER (FASER-2) experiments, we have calculated their sensitivity to iDM$_B$, iDM$_{B-L}$ and iDM$_{B-3L_\tau}$ considering the  following settings: a total integrated luminosity of 250 fb$^{-1}$ (3 ab$^{-1}$), a distance of 480 m (620 m) from the interaction point to the detector decay volume, and particle decay lengths of 1.5 m (10 m) before the tracker. We use the meson and Bremsstrahlung spectra as provided by the FORESEE~\cite{Kling:2021fwx} code. For the Drell-Yan process, such as $q\bar{q}\to Z_{Q} \to \chi_1\chi_2$, we generate events with Madgraph 5~\cite{Alwall:2014hca} and \textsc{Pythia8}~\cite{Sjostrand:2014zea} using the parton distribution functions NNPDF 3.1 NNLO~\cite{NNPDF:2017mvq} and our own iDM$_Q$ FeynRules model file. Although for most of the considered mass range the visible decays of the heavier state $\chi_2$ are predominantly into electrons, when the kinematic threshold condition $m_{Z_Q}\gtrsim 6 \, m_e/\Delta$ is satisfied, we  additionally expect some hadronic decays for masses greater than 1 GeV, as depicted in figure~\ref{fig:brchi2d04}. For that reason, not only for the $B-L$ model but for all the following baryophilic models, the FASER experiment is most sensitive to $\chi_2$ decays for $m_{Z_{B-L}}$ between $1-10$~GeV, where the visible branching ratio exceeds that of the invisible decays to neutrinos. In principle, the $B-3L_\tau$ model with $\Delta=0.1$ is testable above $m_{B-3L_\tau}\gtrsim 3m_{\pi^0}/\Delta \approx 4.05$~GeV where the $\chi_2\to \chi_1 \pi^0 \gamma$ decay channel opens up. However, no sensitivity to hadronic decays can be achieved. In the iDM$_B$ scenario, where $\chi_2$ decays into neutrinos are absent, the only possible decay for $\chi_2$ is the loop-induced decay into electrons. Even this suppressed decay can set mild limits around $g_B<10^{-3}$ on the B model for most of the considered parameter range. The sensitivity grows with the onset of hadronic decays in the few GeV range.

In addition to the above predicted sensitivities, we can establish a limit line for the iDM$_B$ model. The FASER experiment has carried out a search for axion-like particles decaying into a pair of photons, using data corresponding to an integrated luminosity of 57 fb$^{-1}$ produced by LHC proton-proton collisions at  $\sqrt{s}=$13.6 TeV, probing a mass range between 50 and 500 MeV~\cite{CERN-FASER-CONF-2024-001,FASER:2024bbl}. The signature of a signal event is characterized by no observed activity in the veto scintillators, along with a significant energy deposit in the calorimeter consistent with a high-energy photon pair. They observed a single event, consistent with their background expectation of $ 0.42 \pm 0.38$ from neutrino interactions. Their data can be employed to set limits on iDM$_Q$ baryophilic models, where the heavier dark state can decay as $\chi_2\to \chi_1 \pi^0 \gamma \to \chi_1 3\gamma$. For this search, one can make
use of the full 4~m length of the detector volume comprising all magnets and the tracking stations. In our analysis, we assume the same luminosity, expected background events, energy cuts, and efficiencies as in Ref.~\cite{CERN-FASER-CONF-2024-001,FASER:2024bbl} to establish 90\% CL exclusion limits for the iDM$_B$ model with $\Delta=0.4$.

\subsubsection{Other Constraints} 

Besides the bounds from invisible and semi-visible searches, we also consider additional constraints. Below, we list the relevant ones, dividing them by the iDM$_Q$ model to which they were included.

\vspace{6pt}

\underline{\textbf{iDM$_{B-3L_{\tau}}$}}
\vspace{5pt}

Neutrino neutral current non-standard interactions (NSI)~\cite{Wolfenstein:1977ue,Guzzo:1991hi} may 
affect neutrino propagation in matter, modifying the standard coherent forward scattering~\cite{Wolfenstein:1977ue,Mikheyev:1985zog}, which in turn could impact solar, atmospheric, reactor, and long-baseline accelerator neutrino oscillation experimental data. Coherent elastic neutrino-nucleus scattering, first observed by the COHERENT experiment, is another sensitive probe of new vector neutral current interactions. The authors of Refs.~\cite{Han:2019zkz, Coloma:2020gfv} were able to use the latest neutrino experimental data to constraint the mass $m_{Z_{B-3 L_\tau}}$ and the coupling $g_{B-3L_\tau}$ of the $B-3L_\tau$ mediator model. Since these constraints depend only on the interaction between neutrinos and matter, rather than the nature of dark matter, we directly adopt the results from~\cite{Coloma:2020gfv} for our model iDM\(_{B-3L_\tau}\).

\vspace{6pt}

\underline{\textbf{iDM$_{B}$}}
\vspace{5pt}

Due to the anomalous nature of the $U(1)_B$ mediator model, anomaly-related constraints coming from missing energy searches in rare $Z$ decays and in flavor-changing meson decays, such as $K\to \pi Z_B$ and $B\to K Z_B$, need to be considered. We use the bounds presented in Refs.~\cite{Dror:2017ehi, Dror:2017nsg} for this purpose.

\vspace{6pt}

\underline{\textbf{iDM$_{Q}$}}
\vspace{5pt}

The MiniBooNE-DM collaboration~\cite{MiniBooNE:2017nqe,MiniBooNEDM:2018cxm} performed dedicated searches for DM produced by a dark photon, using $1.86 \times 10^{20}$ protons delivered by the Fermilab 8 GeV Booster proton beam operating in beam dump mode. The detector's sensitive volume is located 490 m from the steel beam dump region, where dark photons can be produced via $\pi^0$, $\eta$ decays, or proton Bremsstrahlung. After production, the dark photon decays promptly into DM particles that can be probed by the MiniBooNE through scattering signatures with electrons or nucleons. Although this limit would, in principle, be weaker compared to other proton beam-dump decay searches, it should be considered in models that do not couple to electrons and muons. Therefore, especially for the iDM$_B$ and iDM$_{B-3L_\tau}$ models, this bound can be very relevant, leading us to focus exclusively on the nucleon scattering analysis. In this case, the final signal is characterized by a pattern of hits consistent with a track from a single proton or neutron with a few hundred MeV of kinetic energy~\cite{MiniBooNE:2017nqe,MiniBooNEDM:2018cxm}. The experiment observed a total of $1465 \pm 38$ events, consistent with the estimated background, allowing them to place bounds in the vector portal parameter space, as presented in figure 7 of \cite{MiniBooNE:2017nqe} and figure 23 of \cite{MiniBooNEDM:2018cxm}. Here, we consider a similar analysis for a signature where, after the production and decay of the mediator, the DM candidate $\chi_1$ up-scatters off target nucleons via the mediation of the dark vector boson. However, it is important to comment that another signature could contribute in the iDM$_Q$ scenario:  if $\chi_2$ is not long-lived enough, it can decay into $\chi_1$, which could then up-scatters in the detector. Alternatively, if $\chi_2$ is sufficiently long-lived, it can also down-scatter in the detector. To account for such cases, a more dedicated simulation would be required, which we leave for future study. As a result, the MiniBooNE bound we present here is conservative. For the recasting of the dark photon MiniBooNE study to other mediators, we used the production flux from Ref.~\cite{deNiverville:2016rqh}.

A joint analysis of Big Bang Nucleosynthesis (BBN) and the CMB provides model-dependent limits on the dark sector by constraining the effective number of neutrino species, $N_{\rm eff}$~\cite{Giovanetti:2021izc,Berlin:2023qco,Ghosh:2024cxi}. As emphasized in~\cite{Giovanetti:2021izc}, CMB-only constraints cannot break the degeneracy between DM induced entropy injection, which heats up photons relative to neutrinos after neutrino decoupling, effectively lowering $N_{\rm eff}$, and the existence of additional inert relativistic degrees of freedom. BBN, however, can help breaking this degeneracy, as primordial elemental abundances are sensitive in different ways to radiation degrees of freedom and the neutrino temperature. The $N_{\rm eff}$ limit presented in our plots of $m_{Z_Q} \lesssim 23.7$ MeV~\footnote{This value applies for a Dirac DM particle coupled to a dark photon kinetically mixed with the SM photon, as discussed in~\cite{Giovanetti:2021izc}.} represents only a rough estimate of the lower boundary in our parameter space. A more rigorous determination of this constraint would require in-depth calculations with modern numerical tools, such as those provided in~\cite{Escudero:2018mvt,EscuderoAbenza:2020cmq,Pitrou:2018cgg,Pitrou:2020etk,Burns:2023sgx}. For our purposes, the value serves as an approximate guideline for the left side of the DM mass window.

There are also other experiments and searches that would constrain the iDM$_Q$ parameter space, but their limits are weaker compared to the ones presented before, such that their curves would be hidden in the plots. For example, searches conducted by the LEP-I experiments at CERN using $Z$ decay data, such as the ALEPH constraints based on $Z \to \tau \tau$ decay searches place limits on $g_{B-3L_\tau}$ for $m_{Z_{B-3 L_\tau}}< 70$ GeV~\cite{Ma:1998dp}. Additionally, LEP can place model-independent bounds
on the kinetic mixing arising from the mixing with the $Z$ boson~\cite{Hook:2010tw}. These bounds would apply to the loop-generated kinetic mixing in the iDM$_Q$ models. For instance, we found that for the iDM$_B$ model, such constraints exclude couplings $g_{B} \gtrsim 14$, for $\epsilon \sim g_B e/(4 \pi)^2$. We also have limits from \( K^+ \to \pi^+ + \text{inv} \) set by Brookhaven E949~\cite{BNL-E949:2009dza}, which we verified to be weaker. Turning to astrophysical bounds, white dwarfs could also impose significant constraints in models featuring $L_\mu-L_\tau$ mediators, as discussed in~\cite{Foldenauer:2024cdp,Dreiner:2013tja}. However, these bounds are weaker than those from the NA64$\mu$ search. Let us remind the reader that these are just a few non-exhaustive examples of searches that could also constrain iDM$_Q$ models. For instance, different approaches for constraining the iDM parameter space can be found in Refs.~\cite{Gustafson:2024aom,Acevedo:2024ttq}.

\begin{figure}[h!]
\begin{center}
\includegraphics[width=0.45\textwidth]{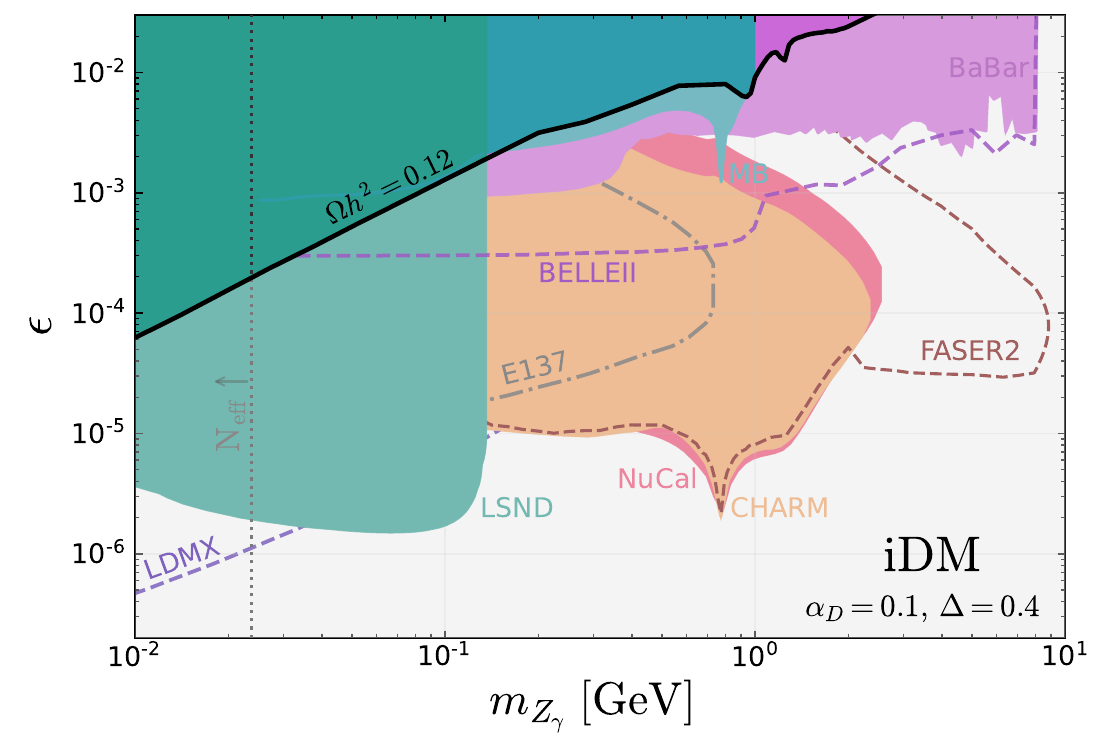}
\includegraphics[width=0.45\textwidth]{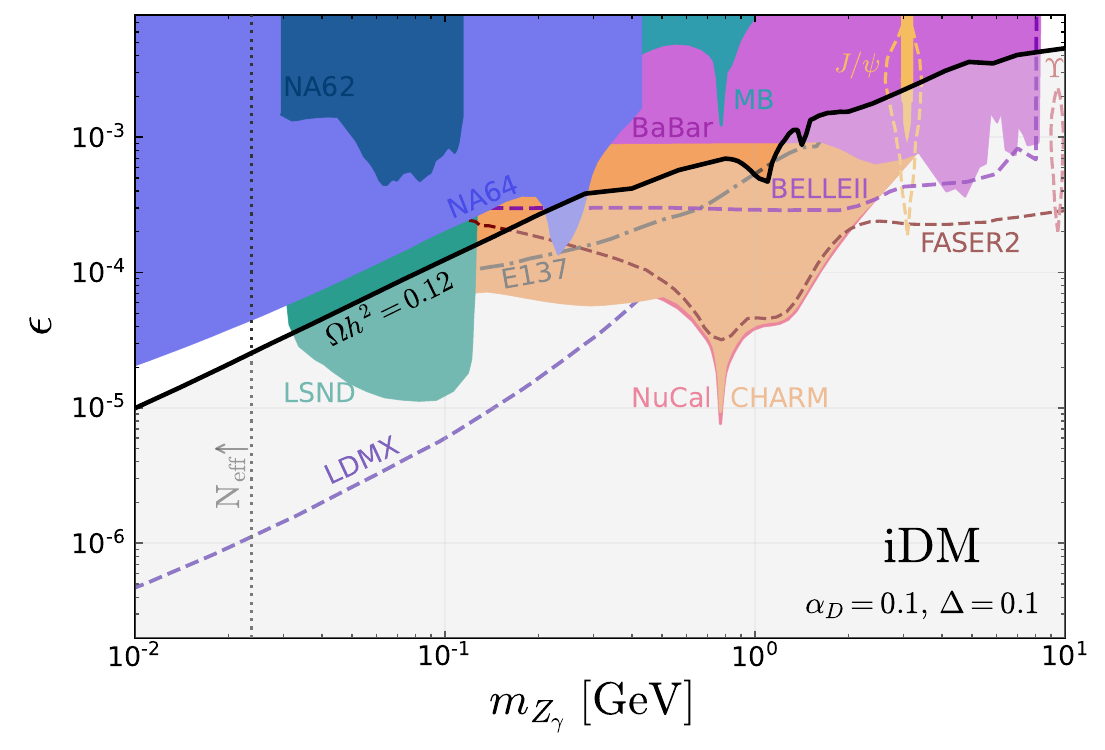}
\includegraphics[width=0.45\textwidth]{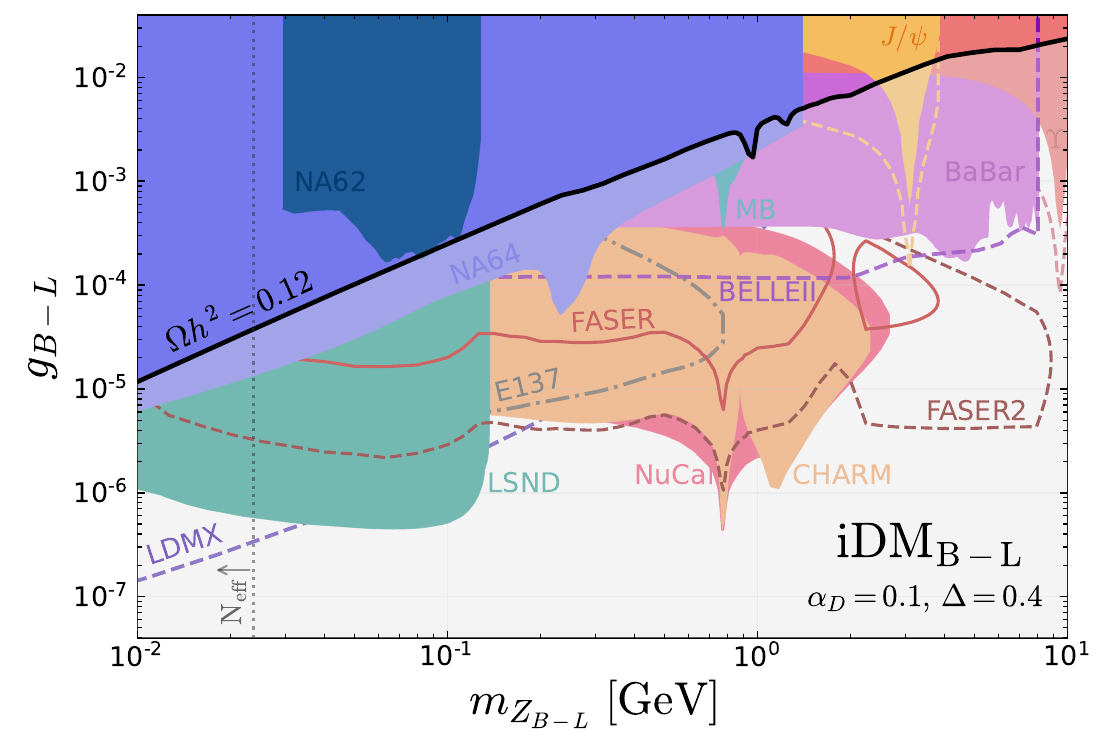}
\includegraphics[width=0.45\textwidth]{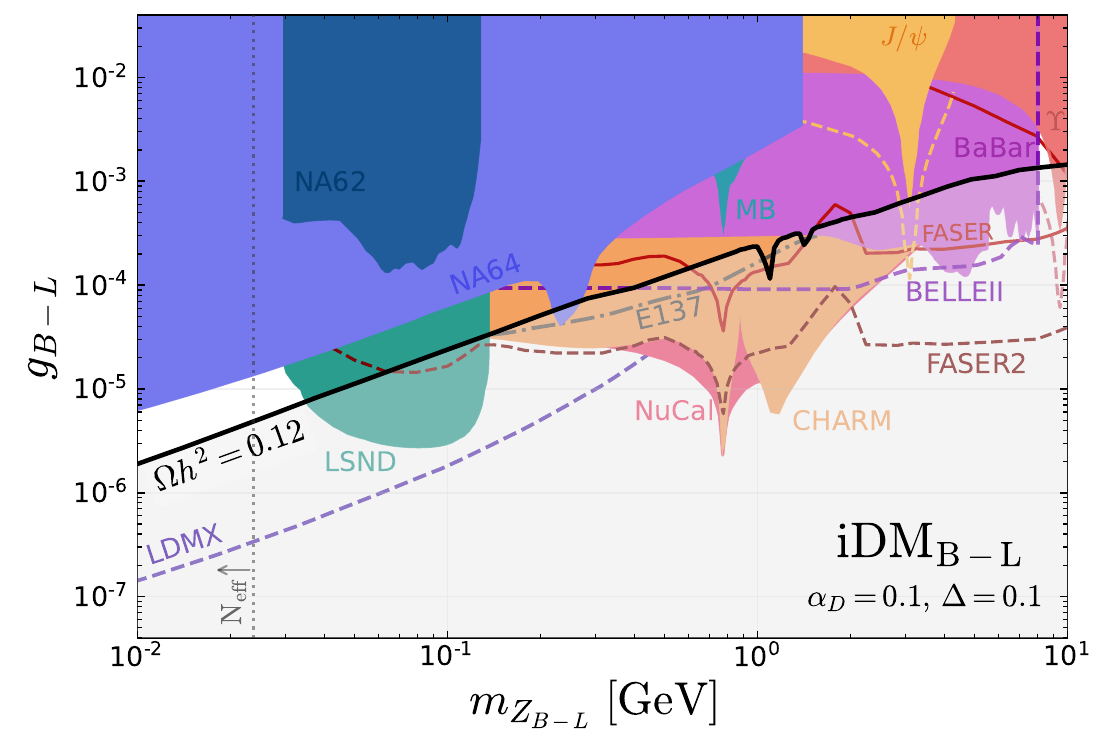}
\includegraphics[width=0.45\textwidth]{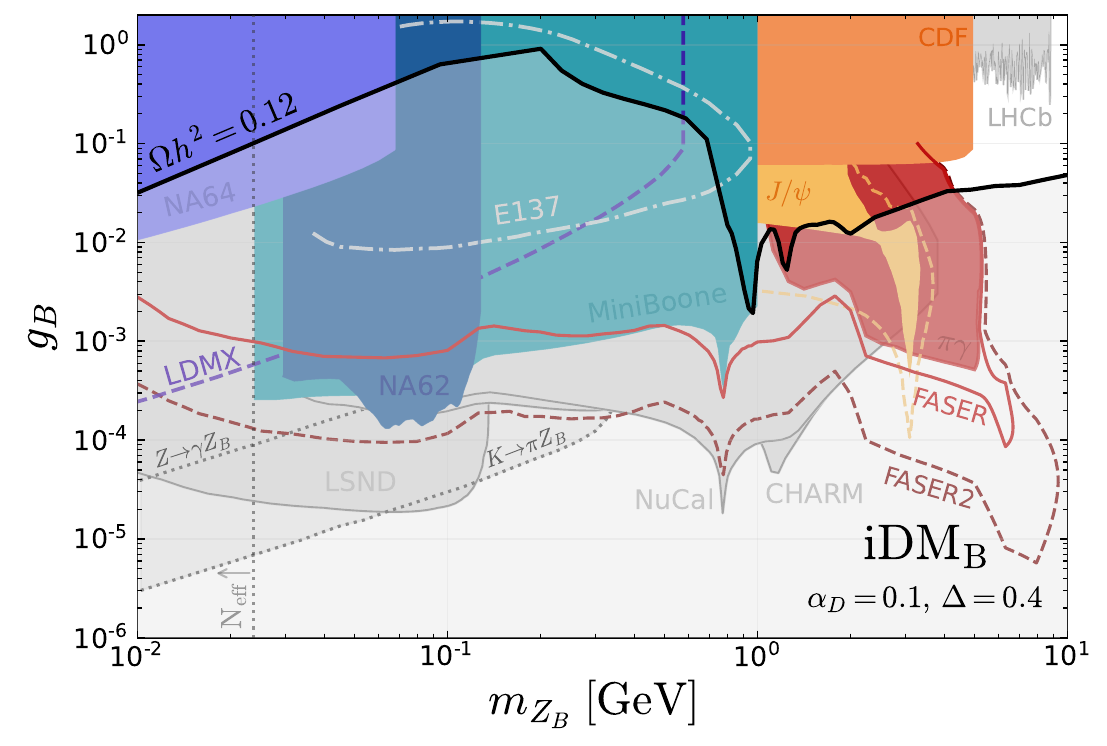}
\includegraphics[width=0.45\textwidth]{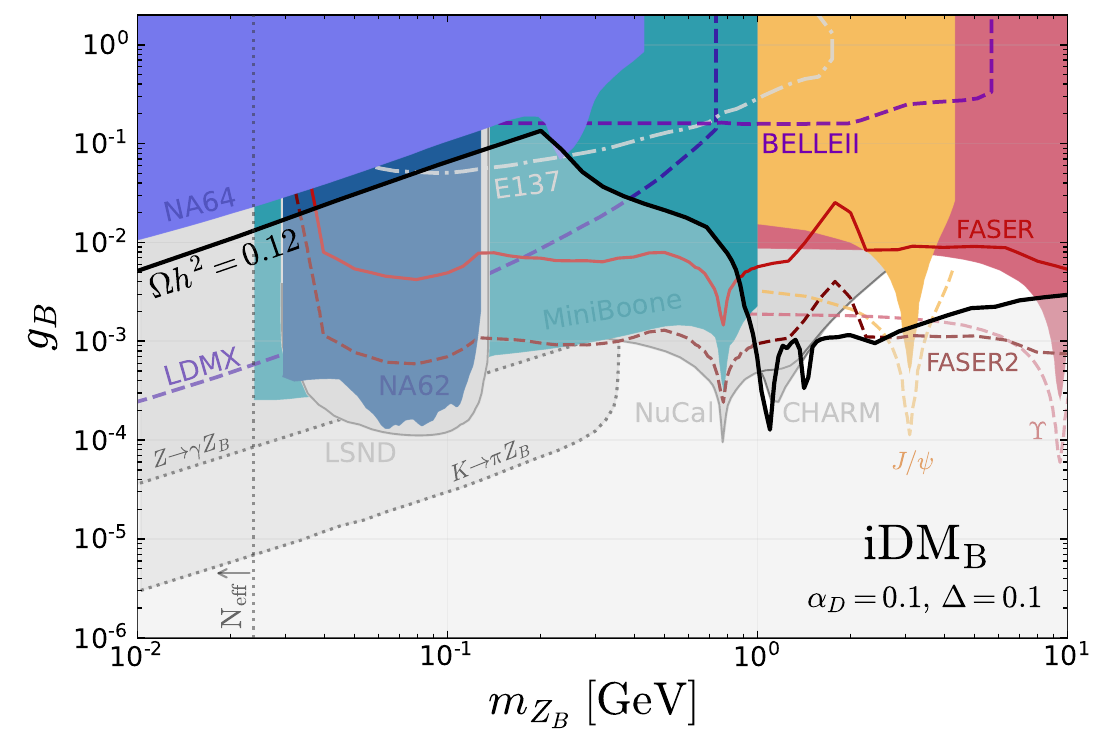}
\end{center}
\vglue -0.8 cm
\caption{Experimental limits and future sensitivities to baryophilic iDM$_Q$ models with universal couplings to the lepton sector. In all cases, we set $R=3$ and $\alpha_D=0.1$, while the left (right) panels correspond to $\Delta=0.4$ ($\Delta=0.1$). The gray shaded region below the freeze-out thermal target curve of $\Omega h^2=0.12$ (black line) indicates the overproduction of thermal DM. For details about the experiments, we refer to the main text.\label{fig:limits1}
}
\end{figure}
\begin{figure}[h!]
\begin{center}
\includegraphics[width=0.48\textwidth]
{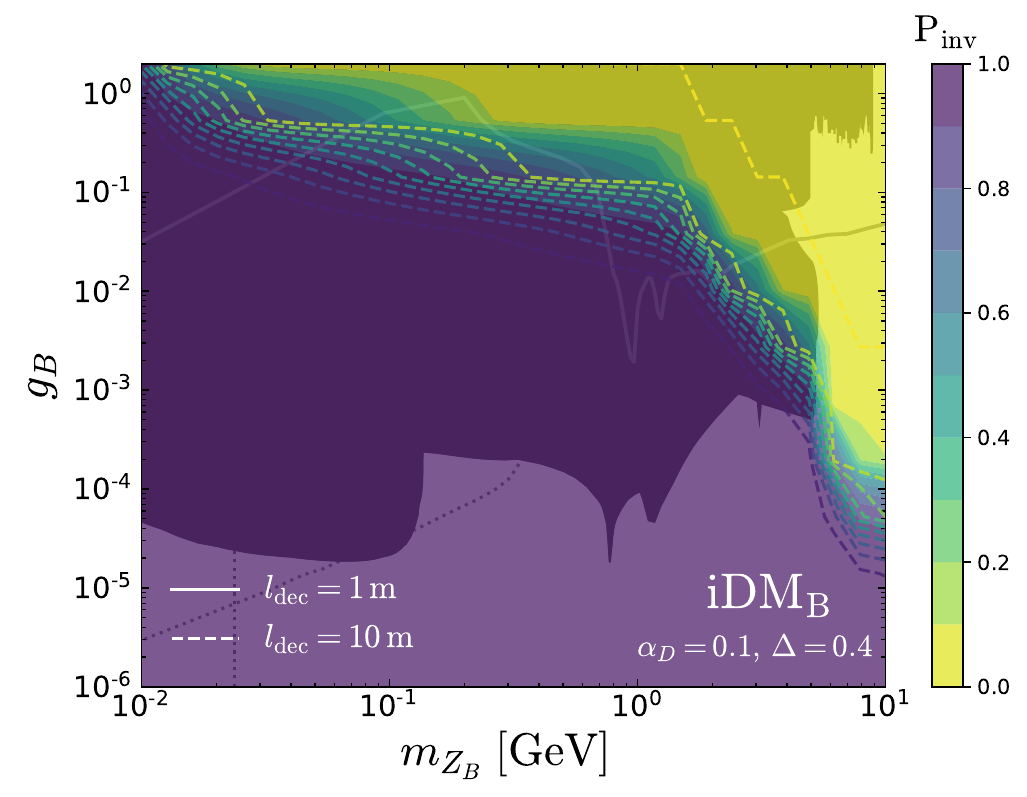}
\includegraphics[width=0.48\textwidth]
{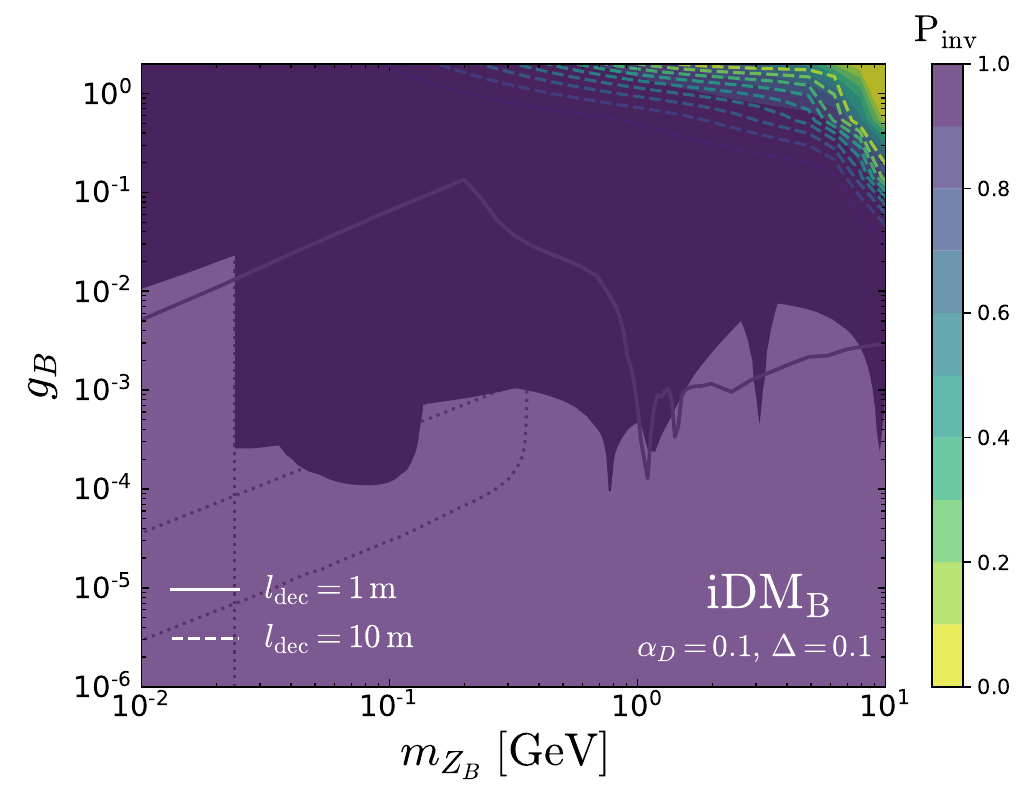}
\end{center}
\vglue -0.8 cm
\caption{\label{fig:PinvB} Probability of invisible decays $P_{\rm inv}$ for the iDM$_B$ model considering two mass splittings, $\Delta=0.4$ (left panel) and $\Delta=0.1$ (right panel), and for two different decay distances, $l_{\rm dec} = 1$~m (solid contours) and $l_{\rm dec} = 10$~m (dashed contours). The shaded regions in the background correspond to the iDM$_B$ limits presented in figure~\ref{fig:limits1}.}
\end{figure}

\subsection{Results}

\begin{figure}[h!]
\begin{center}
\includegraphics[width=0.45\textwidth]
{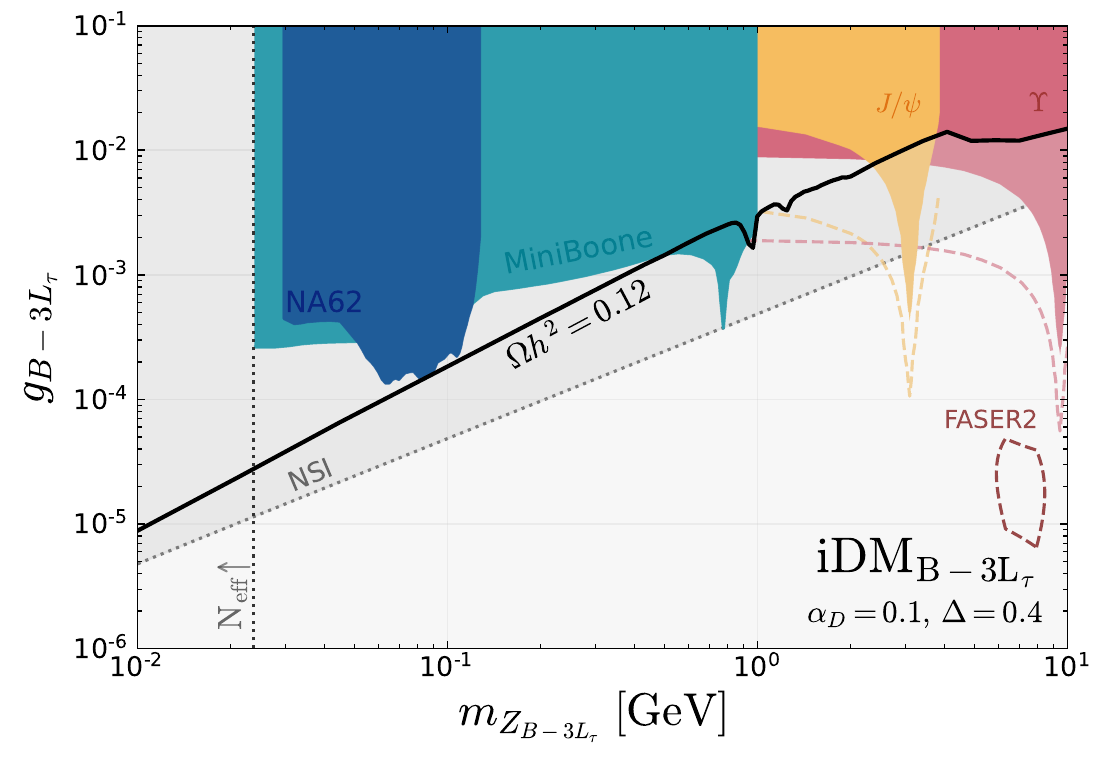}
\includegraphics[width=0.45\textwidth]
{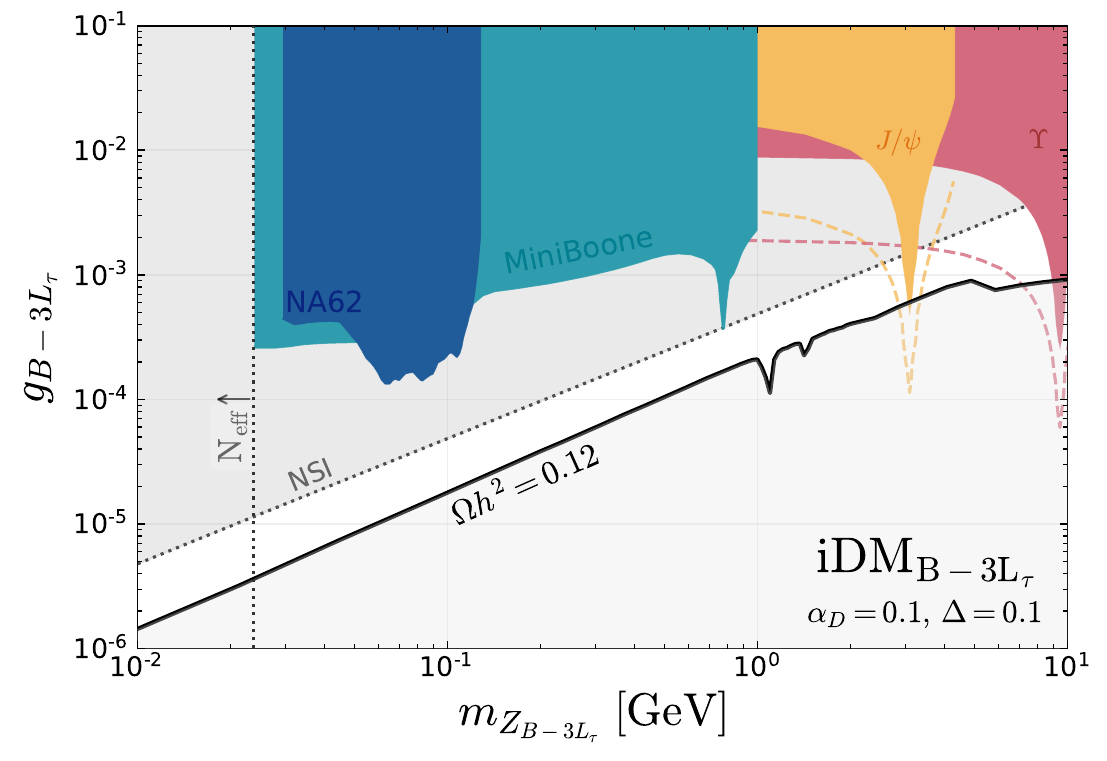}
\includegraphics[width=0.45\textwidth]{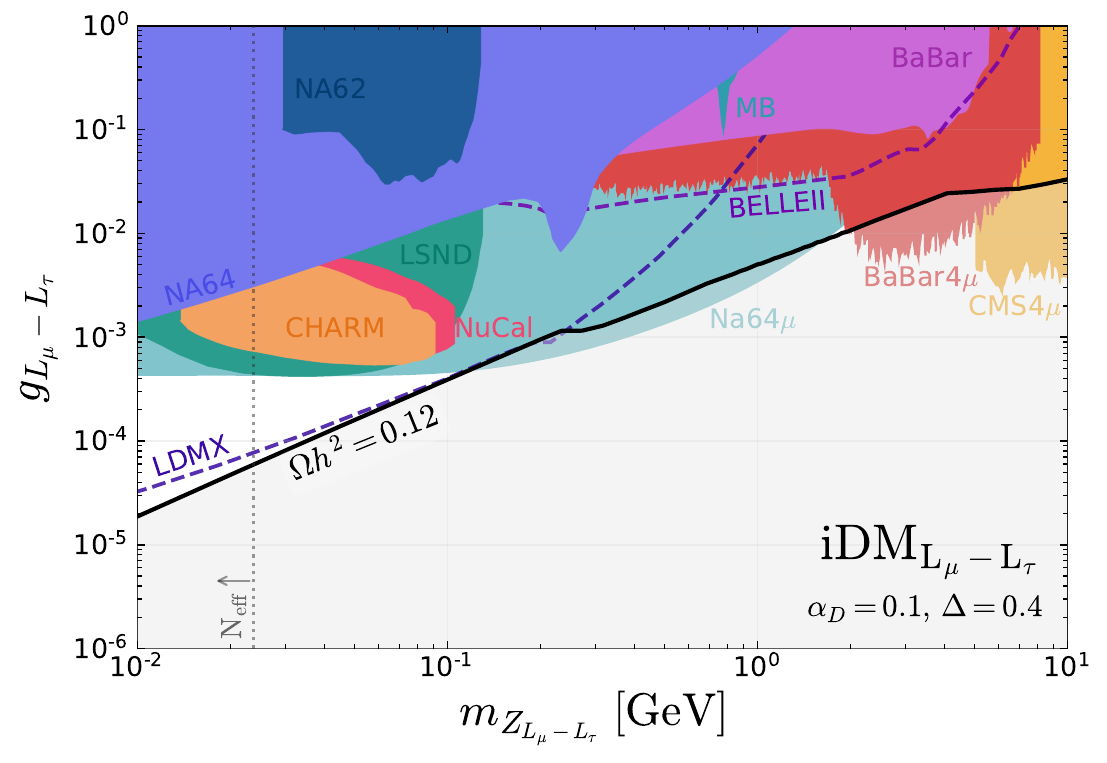}
\includegraphics[width=0.45\textwidth]{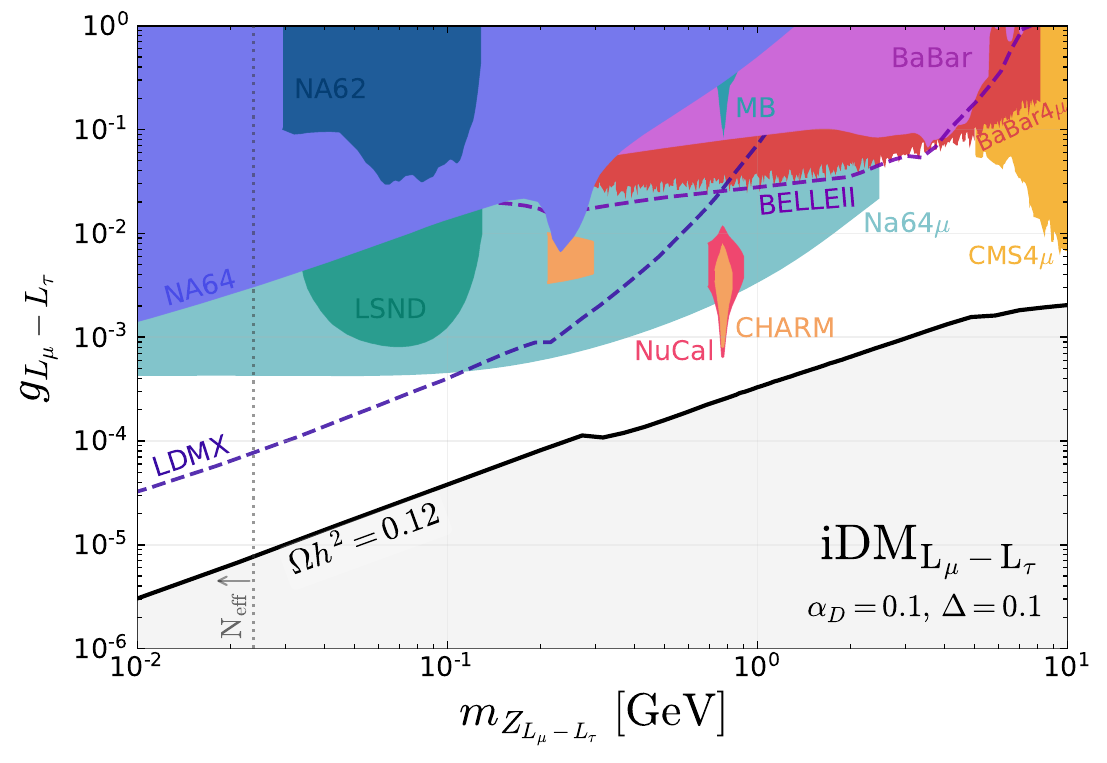}
\end{center}
\vglue -0.8 cm
\caption{\label{fig:limits2} Same as in figure~\ref{fig:limits1}, but for iDM$_Q$ models that couple non-universally to lepton family number.
}
\end{figure}

The results for several iDM$_Q$ models can be grouped into two categories. Models with universal couplings to lepton number are listed in figure~\ref{fig:limits1}, while models where the mediator couples non-universally to lepton family numbers are shown in figure~\ref{fig:limits2}. For each model, we present the allowed regions of the $g_Q \times m_{Z_Q}$ parameter space for two different mass splitting values, $\Delta = 0.4$ (left panel) and $\Delta = 0.1$ (right panel), always fixing $\alpha_D=0.1$ and $R=3$, as an illustration.

In all plots, the black solid line represents the thermal target curve, \textit{i.e.}, the points that reproduced the observed DM relic density of $\Omega_{\chi_1}h^2 = 0.12$ under the assumption of freeze-out DM production. Regions below this line correspond to overproduction of thermal inelastic DM, making them incompatible with a viable thermal DM window~\footnote{Note, however, that alternative production mechanisms, such as freeze-in, could still be considered~\cite{An:2020tcg,Heeba:2023bik}.}.The regions filled with solid colors represent the bounds set by current experimental data, while the colored lines indicate the projected sensitivity of current (solid) or future (dashed) experiments. We use a dot-dashed line for the limits from E137 due to concerns regarding the reliability of this bound. Additionally, bounds from anomaly constraints (iDM$_B$), NSI (iDM$_{B-3L_\tau}$) and $N_{\rm eff}$ are shown as gray dotted lines. From the plots, we observe that smaller $\Delta$ values shift the relic target curve towards lower $g_Q$ values, and the experimental limits appear to weaken. As a result, more parameter space becomes available for iDM$_Q$ models to satisfy the relic abundance constraint $\Omega_{\chi_1}h^2\lesssim 0.12$.

Let us start by discussing the models presented in figure~\ref{fig:limits1}. In the two top plots, we show the limits in the $\epsilon \times m_{Z_\gamma}$ plane for the {\it vanilla} iDM as a reference. It is evident that nearly all the parameter region where $\chi_1$ could be considered a viable thermal DM candidate has already been ruled out. For $\Delta = 0.1$, only two small regions around $m_{Z_\gamma} \sim 3 \times 10^{-2}$ GeV and $m_{Z_\gamma} \sim 10$ GeV remain phenomenologically available. In contrast, for $\Delta=0.4$, the entire parameter space has already been excluded.

The middle panels in figure~\ref{fig:limits1} show the results for the iDM$_{B-L}$ model. Similar to the \textit{vanilla} iDM scenario,  much of the parameter range is excluded by experimental data, with the exception of small regions around $m_{Z_{B-L}} \sim 3 \times 10^{-2}$ GeV and $m_{Z_\gamma} \sim 8$ GeV when $\Delta =  0.1$. The main experiments that rule out most of the relevant parameter space include NA64 and LSND for lower masses, NuCal/CHARM and BaBar for intermediate masses, and BaBar and FASER for higher masses. Upcoming experiments such as LDMX could close the lower mass window, while FASER is expected to explore the remaining relic target parameter space by the end of LHC Run 3 for $m_{Z_{B-L}}>1$~GeV.

The shape of the NuCal/CHARM and FASER limits mostly follows the characteristics of the $Z_Q$ production modes. Decays of $\pi^0$ and $\eta$ dominate until their respective masses are insufficient to produce the dark mediator. Beyond this threshold, proton Bremsstrahlung production becomes significant, peaking around 0.78~GeV, where the dark vector is resonantly radiated through mixing with the omega meson. Above 1~GeV, the dark mediator is predominantly produced via Drell-Yan processes. In future studies, we aim to update this analysis with more recent Bremsstrahlung flux predictions, as calculated in~\cite{Foroughi-Abari:2024xlj}.

Finally, the limits for iDM$_B$ are presented in the two bottom plots. Due to the absence of tree-level couplings to leptons, the iDM$_B$ case exhibits distinctive features compared to the other models. For instance, the thermal target curve requires significantly higher values of the gauge coupling  $g_B$ for smaller $m_{Z_B}$ masses. This necessity arises because the gauge coupling must compensate for the weakening of the co-annihilation cross-section resulting from the loop-suppressed leptonic couplings. The curve begins to decline around $m_{Z_B} \sim 10^{-1}$~GeV, when the $\pi \gamma$ channel opens up, and continues to drop as additional hadronic channels become available. Besides, the semi-visible searches from LSND, CHARM, NuCal, and LHCb depend on leptonic decay signatures, which, in the case of iDM$_B$, relate to the value of the loop-induced kinetic mixing. For this reason, these bounds are model-dependent, and their limits are represented as gray regions, indicating that a more refined analysis is necessary to make robust statements. In contrast, the other limits shown in colored regions depend solely on baryophilic couplings or invisible searches, implying that they are directly valid. The dark red filled region labeled as $\pi \gamma$ in the $\Delta=0.4$ plot represents the FASER bound based on the aforementioned search for axion-like particles~\cite{CERN-FASER-CONF-2024-001}.

Regarding the available regions of the parameter space, we find that, in the iDM$_B$ scenario, there are still unexplored regions even for higher mass splittings, such as $\Delta =0.4$, where $m_{Z_{B}} \gtrsim 5$ GeV remains possible.  Although this region corresponds to larger values of the gauge coupling, experimental bounds are limited there since invisible searches no longer apply. This fact is supported by figure~\ref{fig:PinvB}, which illustrates the probability of invisible decays, $P_{\rm inv}$, in the $g_B \times m_{Z_B}$ plane for two decay distances: $l_{\rm dec} = 1$~m (solid contours) and $l_{\rm dec} = 10$~m (dashed contours)~\footnote{This is an estimate that does not account for the particles boost. The effect of the boost would reduce the yellow regions, i.e., increase the lifetime of the dark particles. This explains why the CDF bound, despite being an invisible limit, still applies to some parts of the yellow region.}. The left panel corresponds to $\Delta = 0.4$, while the right panel considers $\Delta = 0.1$. The shaded regions in the background represent the experimental bounds that appear in figure~\ref{fig:limits1}. For $\Delta = 0.4$,  the figure reveals a specific region in the parameter space where both dark particles decay before $l_{\rm dec}$, indicating that standard invisible limits do not apply. Conversely, for $\Delta = 0.1$, the $\chi_2$ fermion is long-lived over most of the parameter space, which means invisible bounds, such as those from Upsilon decays, are still relevant. Returning to figure~\ref{fig:limits1}, we can see that for $\Delta  = 0.1$ mediator masses within the range from 2 to 8 GeV can still be realized. In the future, the FASER-2 experiment will be able to cover some of these unexplored windows.

It is important to note that, for the iDM$_B$ case, the condition for invisible decays corresponds to the dark particles escaping the detector. However, for models that also couple to neutrinos, decays into these particles also contribute to the estimation of the invisible probability.

\begin{figure}[h!]
\begin{center}
\includegraphics[width=0.6\textwidth]
{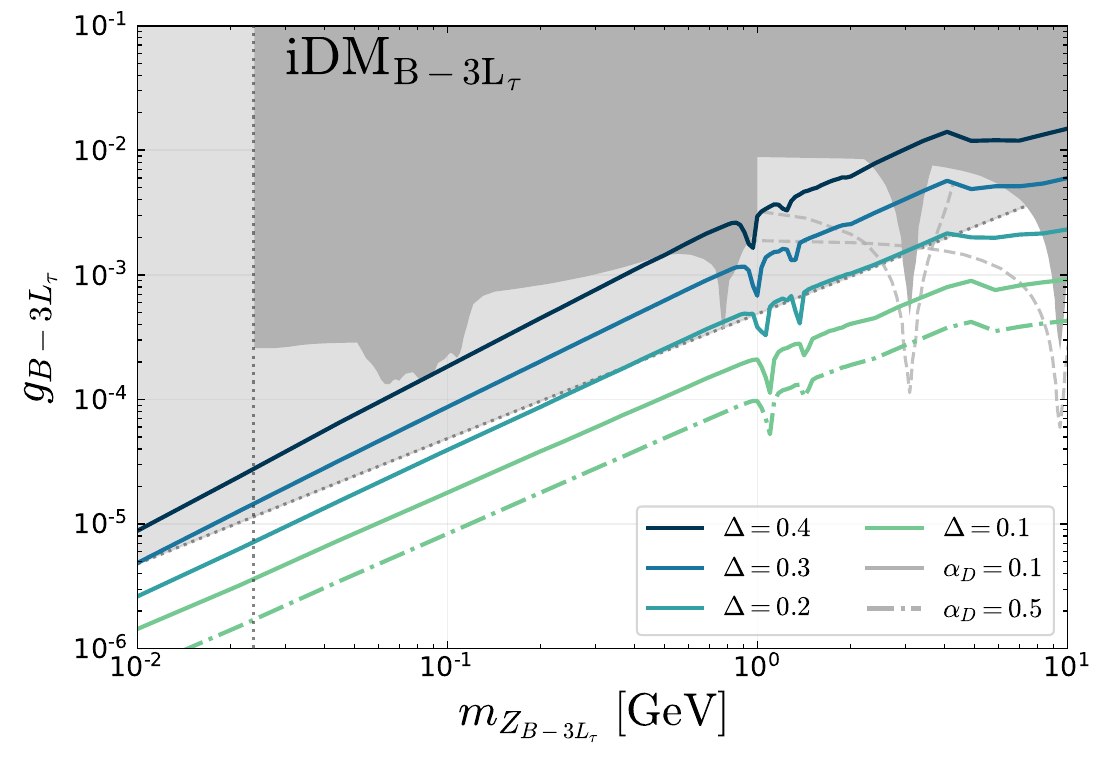}
\end{center}
\vglue -0.8 cm
\caption{\label{fig:relicB3lt} Comparison of thermal target curves for iDM$_{B-3L_\tau}$ with varying $\Delta$ values (indicated by colors) and $\alpha_D = 0.1$ (solid lines) or $\alpha_D = 0.5$ (dash-dotted lines).
}
\end{figure}

Now, we turn our attention to the models displayed in figure~\ref{fig:limits2}. On the top panels we show the parameter space for the iDM$_{B-3L_\tau}$ model. Although the inelastic DM thermal window is closed for $\Delta = 0.4$, it remains basically completely open for $\Delta = 0.1$, except for a small region excluded by invisible decays of $J/\Psi$ and $\Upsilon$. To our knowledge, there are currently no experimental searches, present or planned, that could probe this open region. The invisible search limits shown in the plot are not significantly dependent on the $\Delta$ value, as the branching ratio for $\chi_2$ decays into neutrinos is approximately  $ 100 \%$ in this scenario. Furthermore, the NSI bounds are independent of the DM. Consequently, in figure~\ref{fig:relicB3lt} we illustrate the iDM$_{B-3L_\tau}$ parameter space along with various $\Delta$ and $\alpha_D$ choices for the thermal target curves, set against a relatively fixed background of available limits. We found that for $\Delta \lesssim 0.25$ viable regions of the parameter space begin to appear. Moreover, increasing $\alpha_D$ results in lower $g_Q$ values, which also opens up new available regions.

In the bottom panels of figure~\ref{fig:limits2}, we present the results for the iDM$_{L_\mu-L_\tau}$ model. For $\Delta =0.4$, there is a viable region for $m_{Z_{L_\mu - L_\tau}} \lesssim 10^{-1}$ that can potentially be probed in the future by LDMX. In the case of $\Delta =0.1$, the mass window for iDM$_{L_\mu-L_\tau}$ remains completely open. Once again, we did not identify any future searches that could probe such regions within this scenario.

Note that we have focused on the standard values $\Delta=0.1,0.4$ as found in the literature. However, as discussed previously, the cosmological evolution of the proposed iDM$_Q$ models does not significantly changes with respect to decreasing values of $\Delta$, while some of the visible experimental bounds can weaken considerably. This flexibility of iDM$_Q$ in that regard motivates future searches for non-minimal DM models at the intensity frontier. In particular, future experiments like SHiP~\cite{SHiP:2020noy} will potentially be able to probe the baryophilic iDM$_Q$ models discussed here.

Overall, we conclude that modifications to the portal couplings in iDM beyond the {\it vanilla} dark photon case have significant effects on the phenomenology of such DM scenarios. The iDM$_Q$ models presented here demonstrate that the thermal target window is far from being closed, highlighting the importance of further investigation in future experimental facilities. 

\section{Conclusions}
\label{sec:conclusion}

Dark Matter accounts for approximately 85\% of all matter in the Universe. Nevertheless, despite three decades of tenacious scientific efforts—including numerous dedicated direct and indirect detection experiments—the pursuit for the DM particle has thus far been unsuccessful. Although we have observed a wide range of gravitational effects caused by DM, such as its influence on galaxy motion, the expansion of the Universe, and the growth of cosmic structures, it remains a mystery why we have still not discovered it in the laboratory.

The inelastic DM scenario is a clever theoretical proposal capable of naturally evading bounds from direct and indirect detection, as well as cosmological constraints, while still yielding the correct thermal relic observed abundance.
In this scenario, two pseudo-Dirac fermions, $\chi_1$ 
and $\chi_2$, nearly degenerate in mass, couple directly to a dark photon, which is kinetically mixed with the SM photon. The massive dark photon mediator required for this scheme can be produced in accelerators through its photon-like couplings to the SM particles. Unfortunately, in the phenomenologically interesting MeV–GeV mass range, most of the model's parameter space has been already ruled out by experiments.

In this work, we have extended the {\it vanilla} inelastic DM model to include generic $Q$ charges for the spontaneously broken $U(1)_Q$ symmetry, which connects the dark fermions to the corresponding dark vector boson $Z_Q$. We refer to these models as iDM$_Q$. Considering such scenarios, we have investigated a few realizations for the charge $Q$, such as $Q=B, B-L, B-3 L_\tau$, and $L_\mu-L_\tau$, under the assumption that the tree-level kinetic mixing is negligible compared to the gauge coupling $g_Q$. Our models depend on five parameter: $m_{Z_Q}$, the mediator-to-DM mass ratio $R$, 
the dimensionless mass splitting $\Delta$, and the two couplings $\alpha_Q$ and $\alpha_D$. For illustration, we have considered scenarios with fixed $R=3$ and $\alpha_D = 0.1$.

To examine the regions of the iDM$_Q$ parameter space that reproduce the correct observed DM abundance, the thermal target curve was computed by solving the complete set of Boltzmann equations for each iDM$_Q$ scenario. We studied how the rates for coannihilation, $\chi_2-\chi_1$ scattering, $\chi_2$-fermion scattering, and $\chi_2$ decays influence the relic abundance and alter the value of $g_Q$ required to achieve the correct DM yield. Following this, a careful examination of the current accelerator constraints was performed, considering $e^+e^-$ colliders, $e$- and $\mu$-beam dumps, proton beam dumps and hadron collider experiments that impact the iDM$_Q$ model's parameter space. Additionally, the future sensitivity of ongoing experiments, such as Belle II and FASER, as well as the proposed FASER-2 and LDMX, was analyzed. Indirect bounds on iDM$_Q$ were also investigated, including constraints from neutrino oscillations, coherent elastic neutrino-nucleus scattering, anomaly constraints due to non-conserved currents, and N$_{\rm eff}$.

Our main results are presented in Figs.~\ref{fig:limits1} and 
~\ref{fig:limits2} for iDM$_Q$ with universal and non-universal 
couplings to leptons, respectively, considering two values of the mass splitting, $\Delta = 0.4$ and $0.1$. On the one hand, for the iDM$_Q$ models depicted in figure~\ref{fig:limits1}, a very limited region of the parameter space where $\chi_1$ could still suit as a DM candidate remains unexplored by experiments. However, upcoming proposed experiments would be able to constraint such regions in the near future. On the other hand, iDM$_{B-3L_\tau}$ for $\Delta \lesssim 0.25$, and iDM$_{L_\mu-L_\tau}$ for $\Delta \lesssim 0.4$ appear to be viable DM models across the entire dark vector mass window, \textit{i.e.}, from 
$\sim 25$ MeV to 10 GeV. This shows that this class of $Q$-charged cosmological viable inelastic dark matter models can unlock new windows of the parameter space not yet explored by experiments, presenting interesting opportunities for future investigation.

\begin{acknowledgements}
PR would like to thank Felix Kling for assistance with the calculation of the FASER limits. The authors thank Patrick Foldenauer for valuable comments on our paper. ALF is supported by Funda\c{c}\~ao de Amparo \`a Pesquisa do Estado de S\~ao Paulo (FAPESP) under the contracts 2022/04263-5, and 2024/06544-7. PR acknowledges support by FAPESP under the contract 2020/10004-7. RZF is partially supported by FAPESP and Conselho Nacional de Ci\^encia e Tecnologia (CNPq).
\end{acknowledgements}%

\appendix

\section{Three-body Decays of $\chi_2$}
\label{app:chi2dec}

In order to correctly account for the three-body decays of the heavier dark fermion into SM fermions $\chi_2 \to \chi_1 \, \bar f f$ or hadronic states $\chi_2 \to \chi_1  +  \rm{hadrons}$, we computed the full three-body decay amplitude. This appendix elaborates on the details of this computation.

Consider the decay of a particle with mass $M$ and momentum $p$ into $n$ particle states with masses $m_i$, momenta $p_i$ and energies $E_i$, where $i=1,\dots,n$. The differential decay width $ \di \Gamma$ can be expressed as
\be \label{eq:dGamma}
\di \Gamma = \frac{1}{2M} \frac{1}{S} \, \overline{\abs{\mathcal{M}}^2} \, \di \Phi_n \, ,
\ee
where $S$ is a symmetry factor accounting for identical particles in the final state, $\mathcal{M}$ represents the matrix element of the decay process, with the overline indicating an average over the initial spin states, and $\di \Phi_n$ is the final state phase-space element, given by
\be 
 \di\Phi_n = \delta^4(p - \Sigma \, p_i) (2 \pi)^4 \; \prod_{i=1}^{n} \frac{\di^3 p_i}{(2 \pi)^3} \frac{1}{2 E_i} \, .
\ee

Let us now consider the case of three-body decays, \textit{i.e.}, when $n=3$. From energy-momentum conservation, we have $p = p_1 + p_2 + p_3$. In order to simplify the problem, we define three Lorentz invariants, $s_i = (p - p_i)^2$, such that

\be
\begin{aligned}
 & s_1 \equiv m_{23}^2  = (p - p_1)^2 = (p_2+p_3)^2 \\
 & s_2 \equiv m_{31}^2  = (p - p_2)^2 = (p_3+p_1)^2 \\
 & s_3 \equiv m_{12}^2  = (p - p_3)^2 = (p_1+p_2)^2  \, , 
\end{aligned}
 \ee
where $m_{ij}$ represents the invariant mass of the subsystem composed by particles $i$ and $j$ and the following relation holds
\be
s_1 + s_2 + s_3 = M^2 + m_1^2 + m_2^2 + m_3^2 \,.
\ee

The three-body phase-space $\Phi_3$ can be expressed as
\be 
\begin{aligned}
   \Phi_3 & = \int d\Phi_3 = \int  \delta^3( \vec p - \Sigma \, \vec p_i) \, \delta(E - \Sigma E_i) \; \frac{\di^3 p_1 \; \di^3 p_2 \; \di^3 p_3}{(2 \pi)^5 } \frac{1}{8 E_1 E_2 E_3} \, \\
   & = \frac{1}{8 (2 \pi)^5 } \int \delta(E - \Sigma E_i) \, \frac{\di^3 p_1 \; \di^3 p_2 }{E_1 E_2 E_3 } \, ,
\end{aligned}
\ee
where in the second line we considered the center-of-mass (CM) frame, where $\vec p = 0$, to eliminate one of the momenta integrals through the Dirac delta. Now, using the relation $\di^3 p_i = \abs{\vec p_i}^2  \di\abs{\vec p_i} \di\Omega_i = \abs{\vec p_i} E_i \di E_i \di\Omega_i $, where $\di \Omega = \di\mathrm{cos}\theta \; \di\phi $ is the solid angle element, we can rewrite the phase-space as
\be 
\begin{aligned}
   \Phi_3 & =  \frac{1}{8 (2 \pi)^5 } \int \delta(E - \Sigma E_i) \, \frac{\abs{\vec p_1}  \abs{\vec p_2} \; \di E_1 \di E_2 \; \di \Omega_1  (\di\mathrm{cos}\theta_{12} \; \di\phi_{12}) }{ E_3 } \, .
\end{aligned}
\ee
To simplify further, we use the relation
\be 
\begin{aligned}
   E_3^2 & = m_3^2 + \vec{p_3}^2 \\
   & = m_3^2 +  (\vec p_1 + \vec p_2)^2 = m_3^2 + \abs{\vec p_1}^2 + \abs{\vec p_2}^2 + 2 \cos{\theta_{12}} \abs{\vec p_1} \abs{\vec p_2} \,,
\end{aligned}
\ee
so that by differentiating both sides, we find
\be 
\begin{aligned}
  \di\mathrm{cos}\theta_{12} =  \frac{E_3 \,  \di E_3}{\abs{\vec p_1} \abs{\vec p_2}} \, .
\end{aligned}
\ee
Replacing this into the phase-space formula gives
\be 
\begin{aligned}
   \Phi_3 & =  \frac{1}{8 (2 \pi)^5 } \int \delta(E - \Sigma E_i) \,  \di E_1 \di E_2  \di E_3 \; \di \Omega_1   \di\phi_{12} \, ,
\end{aligned}
\ee
where the remaining Dirac delta can be used to eliminate the $\di E_3$ energy integral. Finally, we want to express the phase-space integrals in terms of the Lorentz invariants. In the CM frame, they can be written in terms of the energies as
\be \label{eq:siEi}
\begin{aligned}
s_i & = p^2 + p_i^2 - 2 \, p \cdot p_i \\
& = M^2 + m_i^2 - 2 M E_i \, ,
\end{aligned}
\ee
which leads to the expression $\di s_i = - 2 M \di E_i$. Therefore, we obtain the final formula
\be 
\begin{aligned}
   \Phi_3 & =  \frac{1}{(2 \pi)^5 }  \frac{1}{32 M^2} \int   \di s_1 \int \di s_2  \; \int \di \Omega_1   \di\phi_{12} \, .
\end{aligned}
\ee
If the squared amplitude is independent of the angular variables,  the expression for the decay width simplifies to
\be \label{eq:GammaF}
 \Gamma = \frac{1}{32 M^3} \frac{1}{(2 \pi)^3}\frac{1}{S} \, \int_{s_1^-}^{s_1^+} \di s_1 \int_{s_2^-}^{s_2^+} \di s_2 \; \overline{\abs{\mathcal{M}}^2} \; .
\ee
To complete the computation, the limits of integration $s_1^\pm$ and $s_2^\pm$ in the above decay width expression need to be determined. Since $s_1$ and $s_2$ are invariants, these limits can be calculated in any reference frame. In the CM frame, as seen from eq.~\eqref{eq:siEi}, the maximum value of $s_1$ occurs when $E_1 = \sqrt{m_1^2 + p_1^2}$ is minimized. This corresponds to $E_1^{\rm min} = m_1$, which implies $s_1^{+} = (M- m_1)^2$. For the minimum value of $s_1$, we consider the rest frame R23 of the (2,3) subsystem, where $ \vec{\hat{p_2}} = -  \vec{\hat{p_3}}$. In this case, $s_1 = (\hat E_2+ \hat E_3)^2 \geq (m_2+m_3)^2$. Therefore, we have 
\be \label{eq:s1ilim}
\begin{aligned}
& s_1^{-} = (m_2+m_3)^2 \\ 
& s_1^{+} = (M- m_1)^2 \, .
\end{aligned}
\ee

Now, to find the integration limits of $s_2$ for a fixed value of $s_1$, we need to consider again the reference frame R23. Let us start by expanding the expression for $s_2$
\be \label{eq:s2eq}
\begin{aligned}
   s_2 & = (p_1+ p_3)^2 \\
       & = m_1^2 +m_3^2 + 2 \, (\hat E_1 \hat E_3 - |\overrightarrow{\hat p_1}| |\overrightarrow{\hat p_3}| \cos{\theta_{13}} )   \, .
\end{aligned}
\ee
To proceed further, we want to express the momenta and energies that appear in this equation in terms of the fixed variables and $s_1$. Since in the R23 frame $\vec{\hat p} = - \vec{\hat{p_1}}$, we can write
\be 
\begin{aligned}
   s_1 = (\hat E - \hat E_1)^2 = \qty( \sqrt{M^2 +|\overrightarrow{\hat p_1}|^2} - \sqrt{m_1^2 + |\overrightarrow{\hat p_1}|^2 } )^2 \, ,
\end{aligned}
\ee
that can be solved for $\vec{p_1}$
\be \label{eq:p1eq}
|\overrightarrow{\hat p_1}| = \frac{1}{2 \sqrt{s_1}} \sqrt{\lambda(s_1, M^2,m_1^2)} \, ,
\ee
where $\lambda$ is the K\"{a}ll\'{e}n-function, previously defined in eq.~\eqref{eq:redxsec}. Similarly, we obtain for the other momenta
\be \label{eq:p23eq}
|\overrightarrow{\hat p_2}| =  |\overrightarrow{\hat p_3}| = \frac{1}{2 \sqrt{s_1}} \sqrt{\lambda(s_1, m_2^2,m_3^2)} \, .
\ee
Now, for the energies in the R23 frame, we can write
\be 
\begin{aligned}
\hat E_3^2  &= |\overrightarrow{\hat p_3}|^2 + m_3^2 = |\overrightarrow{\hat p_2}|^2 + m_3^2 = (\hat E_3^2 - m_2^2) + m_3^2 \\
& = (\sqrt{s_1} - \hat E_3)^2 + m_3^2 - m_2^2 \, ,
\end{aligned}
\ee
which implies 
\be \label{eq:E3eq}
\begin{aligned}
\hat E_3= \frac{1}{2 \sqrt{s_1}} (s_1 + m_3^2 - m_2^2) \, .
\end{aligned}
\ee
The expression for $\hat E_1$ can also be obtained in a similar way, resulting in
\be \label{eq:E1eq}
\begin{aligned}
\hat E_1= \frac{1}{2 \sqrt{s_1}} (s - s_1 - m_1^2)  \, .
\end{aligned}
\ee
Finally, returning to equation~\eqref{eq:s2eq}, and using the results found in eqs.~\eqref{eq:p1eq},\eqref{eq:p23eq},~\eqref{eq:E3eq}, \eqref{eq:E1eq}, we can see that, for a fixed $s_1$, the expression only depends on $\cos{\theta_{13}}$. This indicates that when $\theta_{13} = \pi$, we reach $s_2 = s_2^{+}$, and when $\theta_{13} = 0$, we obtain $s_2 = s_2^{-}$. By using the final expressions for the energies and momenta in terms of $s_1$, we can express the limits of $s_2$ as follows
\be \label{eq:s2ilim}
 s_2^{\pm} =  m_1^2 +m_3^2 + \frac{1}{2 s_1} \qty[(s - s_1 - m_1^2) (s_1 + m_3^2 - m_2^2) \pm \lambda^{1/2}(s_1, s,m_1^2) \lambda^{1/2}(s_1, m_2^2,m_3^2)] \, .
\ee

Therefore, to compute the three-body decays of $\chi_2$, we used the eq.~\eqref{eq:GammaF}, with the integration limits provided by eqs.~\eqref{eq:s1ilim} and~\eqref{eq:s2ilim}. For instance, in the case of decays into fermions,  $\chi_2 \to \chi_1 \, \bar f f$, we consider the following replacements: $M=m_2$, $m_1=m_1$, $m_2 = m_3 = m_f$, and the squared matrix element is given by 
\begin{alignat*}{3}
  \abs{\overline{\mathcal{M}}_{2 \to 1 \, \bar f f }}^2 &=
  \; \; \frac12 && \abs{\mathcal{M}_{2 \to 1 \, \bar f f}}^2 \\
  & = \, C^f \; \big[ && 16 \,m_1^2 s_2 - 16 \, s_2^2 - 16 \, m_1^2 m_2^2 + 16 \, m_2^2 \, s_2 + 8 \, m_1^2 \, s_1 - 16 \, m_1 m_2 s_1  \\
  & && -16 \, s_2 s_1 + 8 \, m_2^2 \, s_1 - 8 \, s_1^2 + 32 \, s_2 m_f^2 - 32 \, m_1 m_2 m_f^2  -16 \,  m_f^4 \big] \, ,
\end{alignat*}
where in the first line we average over the two spin states of the particle $\chi_2$, and $C^f=(3,1,1/2)$ is the same as the one that appears in eq.~\eqref{eq:GZff}~\footnote{The factor of 3 for quarks accounts for the colors, while the factor of $1/2$ is a consequence of coupling exclusively to left-handed neutrinos.}. We verified that our amplitudes match with the ones computed in Ref.~\cite{Jodlowski:2019ycu}. 

The formula described above for fermions was applied to leptonic decays. In addition to leptonic degrees of freedom, in the low-energy regime we are focusing on, the final physical states include hadrons rather than individual quarks. Therefore, particularly for $B$-like models, the correct computation of the hadronic contributions is essential. For decays into hadronic final states, $\chi_2 \to \chi_1 + \mathcal{H}$, we evaluated the most significant contributions, from $\mathcal{H}= \pi \gamma , \pi \pi , K K$, by using the same decay width expression. For the amplitudes, we followed the calculations presented in~Ref.~\cite{Foguel:2022ppx}. 

\begin{figure}[h!]
\begin{center}
\includegraphics[width=0.6\textwidth]
{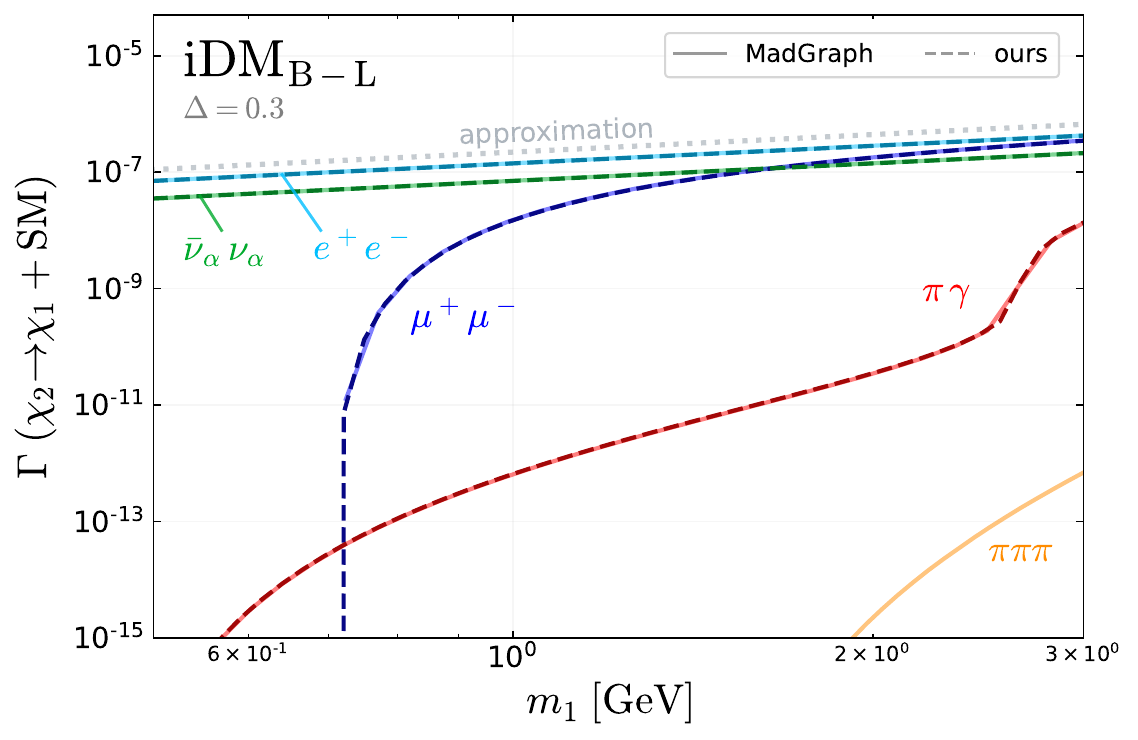}
\end{center}
\vglue -0.8 cm
\caption{\label{fig:wmad}  Comparison between analytical expressions (dashed) for the dark fermion $\chi_2$ three-body decays, given by eq.~\eqref{eq:GammaF}, with the numerical computation simulated with MadGraph (solid). The colors represent different final states. The dotted gray line corresponds to the approximate decay width expression of eq.~\eqref{eq:chi2decay}. For this plot we considered the iDM$_{B-L}$ model with $\Delta=0.3$, $\alpha_D = 1$ and $g_{B-L}=1$.
}
\end{figure}

To validate these analytic formulas, we also simulated the decays of $\chi_2$ with the use of Madgraph~\cite{Alwall:2014hca} along with the iDM$_Q$ FeynRules model file that we developed. Figure~\ref{fig:wmad} presents a comparison between the decay widths $\Gamma (\chi_2 \to \chi_1 + {\rm SM})$ for various SM final states, indicated by different colors. The solid lines represent the numerical computation of MadGraph, while the dashed lines show our computed widths using the analytic expression of eq.~\eqref{eq:GammaF}. For illustration, the iDM$_{B-L}$ parameters were fixed to $\Delta=0.3$, $\alpha_D = 1$, and $g_{B-L}=1$. From the figure, we can see that the analytic computation matches very well with the simulation results. The orange line depict the numerical computation for the four-body decay $\chi_2 \to \chi_1 \pi^0 \pi^+ \pi^-$, demonstrating that this contribution is negligible.  Additionally, the dotted gray line represents the decay width into fermions using the approximate expression from eq.~\eqref{eq:chi2decay}. As one can see, the approximation tends to overestimate the actual decay width and is only valid when $m_2 - m_1 \gg 2 m_f$. An interesting remark is that if ones multiplies the approximate expression by the correction factor $K=0.64$, as suggested by Ref.~\cite{NA64:2021acr}, the curve aligns with the correct result for the electron decay (light blue).

To conclude the observations regarding $\chi_2$ decays, let us elaborate on the computation of the other hadronic channels represented in orange in figure~\ref{fig:brchi2d04}. Although we considered the relevant hadronic contributions from pions and kaons, several other hadronic channels could also be included in this analysis (see, for example, Ref.~\cite{Foguel:2022ppx}). To estimate the impact of all inclusive other hadronic contributions, we subtracted the individual fermionic and hadronic channels computed analytically from the approximate expression for the total decay width $\Gamma_2$ of $\chi_2$, as given in eq. (B7) of Ref.~\cite{Jodlowski:2019ycu}. For iDM$_Q$ models that also couple with neutrinos, we considered $Z_Q \to \nu \nu$ in eq. (B6) and (B7)  instead of electrons. This approach provides a rough estimate of additional hadronic contributions that might have been overlooked.

\section{Details on the Relic Densities}
\label{app:rddetails}

In this appendix, we will provide further details on the relic density freeze-out curves for the iDM$_Q$ models. We begin by exploring the impact of varying the mass splitting $\Delta$ on the abundance yields and cosmological evolution.

\begin{figure}[h!]
\begin{center}
\includegraphics[width=0.48\textwidth]{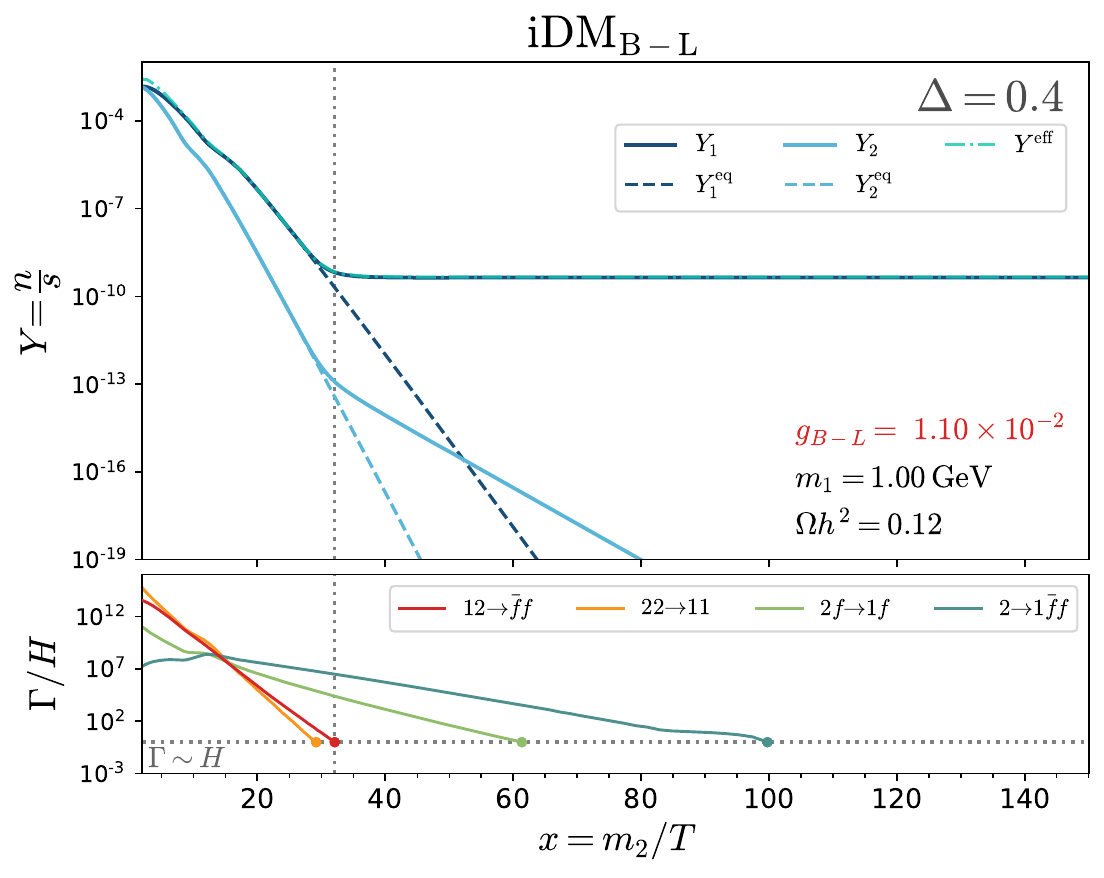}
\includegraphics[width=0.48\textwidth]{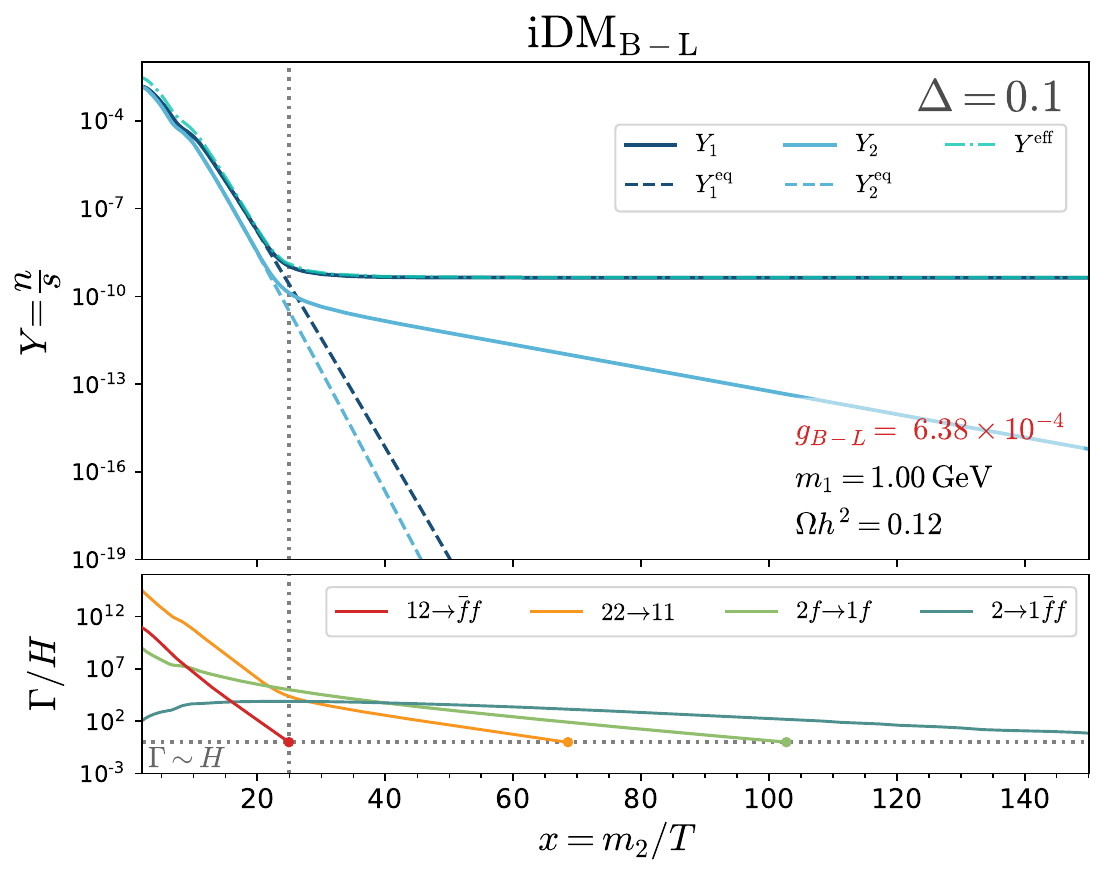}
\includegraphics[width=0.49\textwidth]{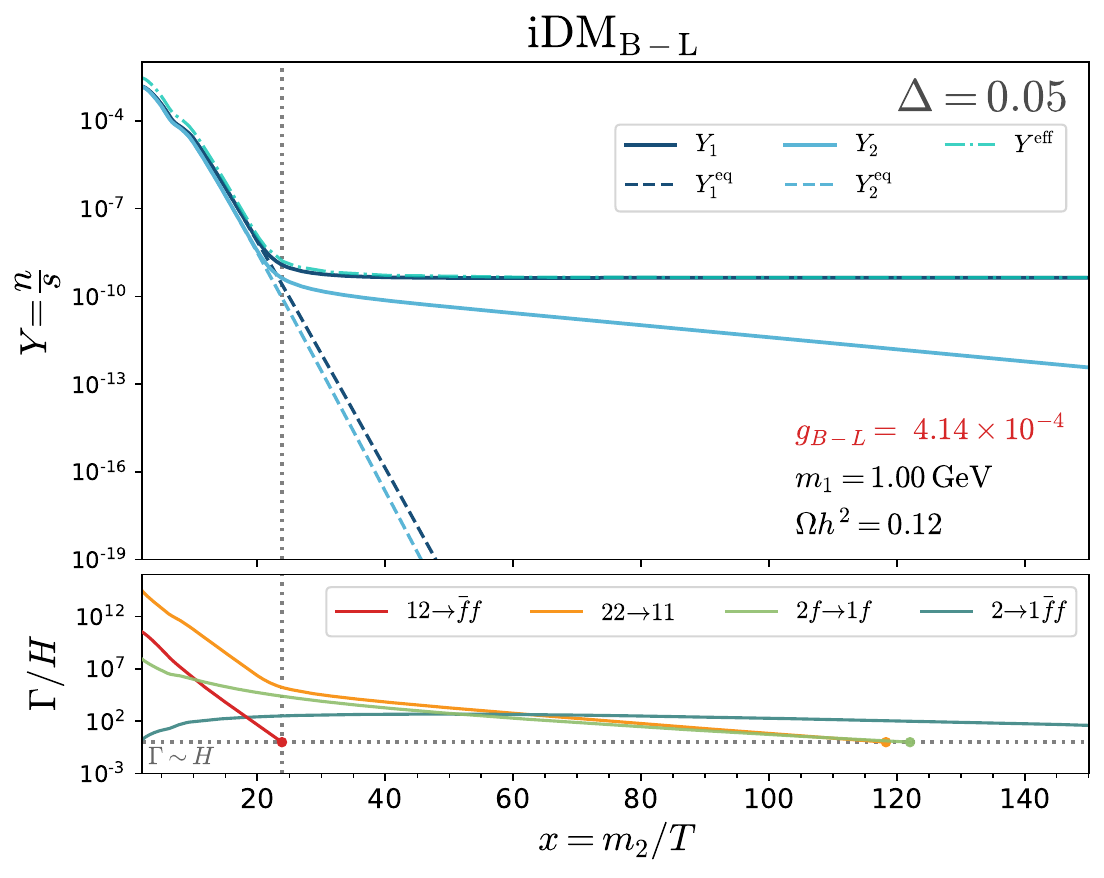}
\includegraphics[width=0.49\textwidth]{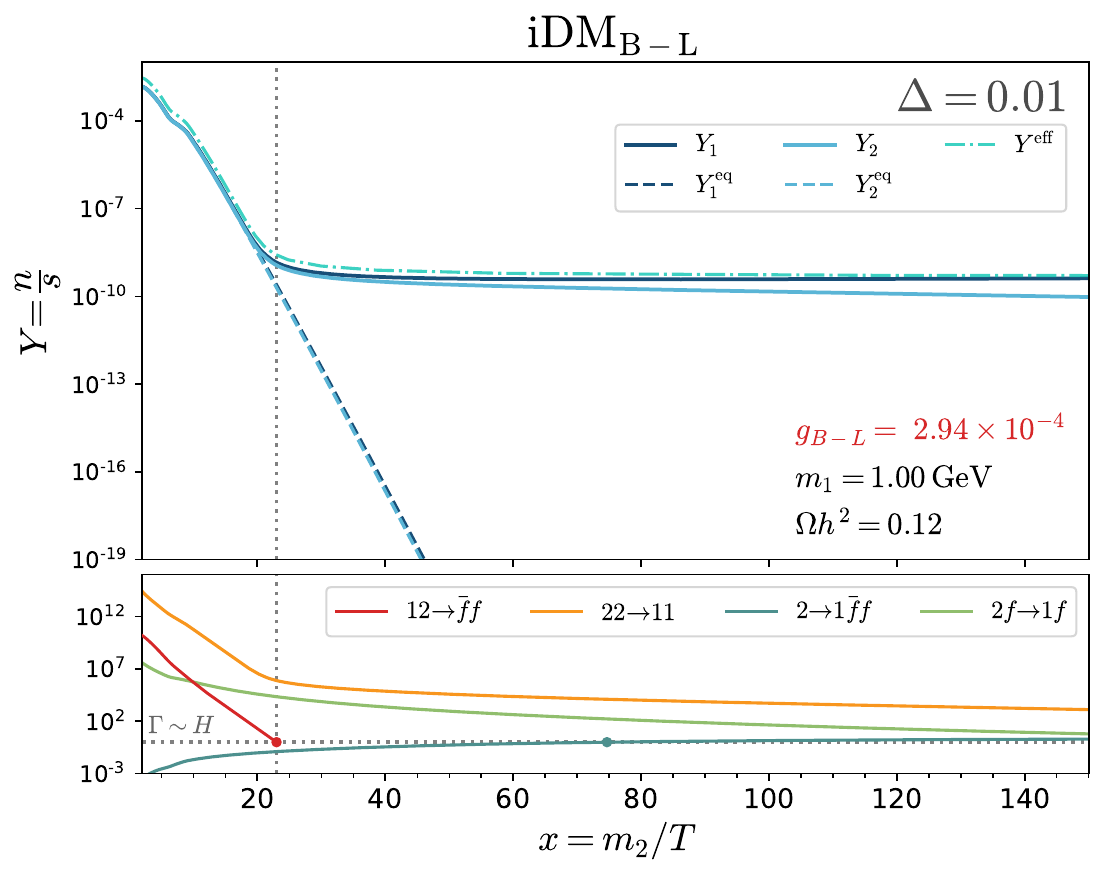}
\end{center}
\vglue -0.8 cm
\caption{\label{fig:fo_curve_BL} Same as figure~\ref{fig:fo_curve_models}, but focusing on the effects of varying mass splittings: $\Delta=0.4$ (upper left),  $\Delta=0.1$ (upper right), $\Delta=0.05$ (lower left), and $\Delta=0.01$ (lower right) on the evolution of particle yields and rates for the iDM$_{B-L}$ model. The other parameters were fixed at $\alpha_D=0.1$, $R=3$, $m_1 = 1 \, {\rm GeV}$. The $g_Q$ value corresponds to the coupling that correctly reproduces the DM observed abundance.}
\end{figure}

Figure~\ref{fig:fo_curve_BL} illustrates the behavior of the comoving number densities and reaction rates for the iDM$_{\rm B-L}$ model with decreasing values of the mass splitting: $\Delta = 0.4$, $\Delta = 0.1$, $\Delta = 0.05$, and $\Delta = 0.01$. Several key observations can be made. First, as the mass splitting approaches zero, the yields of the two dark fermions begin to track each other more closely. This is expected, as the decay rate scales with $\Delta^5$ according to eq.~\eqref{eq:chi2decay}. As a result, lowering $\Delta$ substantially impacts the decay rate $\Gamma_{2}$, which is reflected by the decrease in the blue curve in the lower panels. Similarly, the effectiveness of the coannihilation approximation, as described by eq.~\eqref{eq:boltzcoann} and represented by  $Y^{\rm eff}$, becomes less accurate as $\Delta$ decreases, since the $\chi_2$ abundance is no longer negligible. Another notable feature is that, as the splitting decreases, the $\chi_2 - \chi_1$ scattering (orange line) becomes more efficient, eventually surpassing the $\chi_2 -$fermion kinetic decoupling (green line). This behavior is the same as the one observed for the dark photon iDM models in Ref.~\cite{Berlin:2023qco}.

\begin{figure}[h!]
\begin{center}
\includegraphics[width=0.65\textwidth]{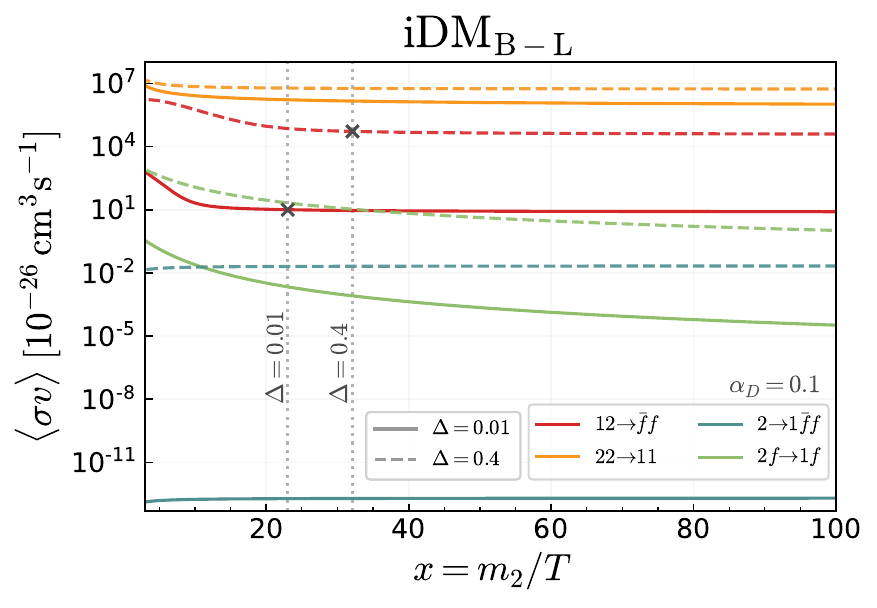}
\end{center}
\vglue -0.8 cm
\caption{\label{fig:sigmav} Thermally averaged cross-section times velocity of the four number-changing $\chi_1$ channels as a function of $x$ for the iDM$_{B-L}$, with $\alpha_D =0.1$,  $m_1= 1$ GeV, and considering $\Delta=0.01$ (solid lines) and $\Delta=0.4$ (dashed lines). The channels are labeled by the following colors: coannihilation (red), $\chi_2 - \chi_1$ scattering (orange), $\chi_2 -$fermion scattering (green) and $\chi_2$ decays (blue). The $g_{B-L}$ values are fixed to reproduce the correct DM relic density (see figure~\ref{fig:fo_curve_BL}). For $\chi_2$ decays, $\ang{\sigma v}$ represents $\ang{\Gamma}_{2  \to 1 ff}$, and for fermion scatterings, we use $\ang{\Gamma}_{2 \to 1 ff}/n_{f}^{\rm eq}$. The vertical dotted lines mark the freeze-out $x_{\rm fo}$ values for $\Delta=0.01$ ($x_{\rm fo}$=23) and $\Delta=0.4$ ($x_{\rm fo}$=32). The crosses indicate the corresponding value of $\ang{\sigma v}_{12 \to ff}$  at freeze-out.}
\end{figure}

The choice of a larger (smaller) $\alpha_D$ does not change the overall behavior of the curves; it only decreases (increases) the required value of $g_{B-L}$ to achieve the correct DM relic density at freeze-out. Additionally, $g_{B-L}$ decreases when the mass splitting $\Delta$ becomes smaller, as illustrated in the figure. These statements can be verified by referring to the analytic formula in eq.~\eqref{eq:sigvcoann}, and they are also evident in figure~\ref{fig:relicB3lt}. The correct DM relic density is typically associated with $\ang{\sigma v}_{12 \to ff} \sim 10^{-25} \; {\rm cm^3/s}$, which is the characteristic WIMP-like value. Therefore, when $\alpha_D$ increases, we need to reduce $g_Q$ to maintain the same thermal average. In a similar manner, since $\ang{\sigma v}_{12 \to ff}$ is inversely proportional to $\Delta$, reducing the mass splitting leads to a correspondingly lower value of $g_Q$. However, as the number density of $\chi_2$ decreases for larger $\Delta$, achieving the correct decoupling temperature for the coannihilation rate requires higher $\ang{\sigma v}_{12 \to ff}$ values to compensate for the reduced number density.

This last observation can be better visualized in figure~\ref{fig:sigmav}, which shows the values of $\ang{\sigma v}$ for each $\chi_1$ number-changing process, labeled by colors, in the iDM$_{B-L}$  for $\alpha_D=0.1$ and two mass splitting values:  $\Delta=0.01$ (solid lines) and $\Delta=0.4$ (dashed lines). The dotted vertical lines indicate the freeze-out temperature $T_{\rm fo}$  for each mass splitting, i.e., the value $x_{\rm fo}$ when the coannihilation rate $\Gamma_{12} \sim H$. For $\Delta=0.01$, we have $\ang{\sigma v}_{12 \to ff} \simeq 10^{-25} \; {\rm cm^3/s} $ at $x_{\rm fo}= 23$, and for $\Delta=0.4$, we have $\ang{\sigma v}_{12 \to ff} \simeq 5.2 \times 10^{-22} \; {\rm cm^3/s} $ at $x_{\rm fo} = 32$. As previously mentioned, increasing $\Delta$ results in a smaller number density $n_2$, which implies a larger $\ang{\sigma v}_{12 \to ff}$. Consequently, as we decrease $\Delta$, the red curves in the figure also moves to smaller $\ang{\sigma v}$, approaching the canonical WIMP-like value. This trend is also visible in figure~\ref{fig:fo_curve_BL}, where the red lines $\Gamma_{12}/H$ in the lower plots intersect the horizontal line at smaller $x_{\rm fo}$ values as $\Delta$ decreases, indicating that lower $\Delta$ correspond to higher $T_{\rm fo}$.

\begin{figure}[h!]
\begin{center}
\includegraphics[width=0.62\textwidth]{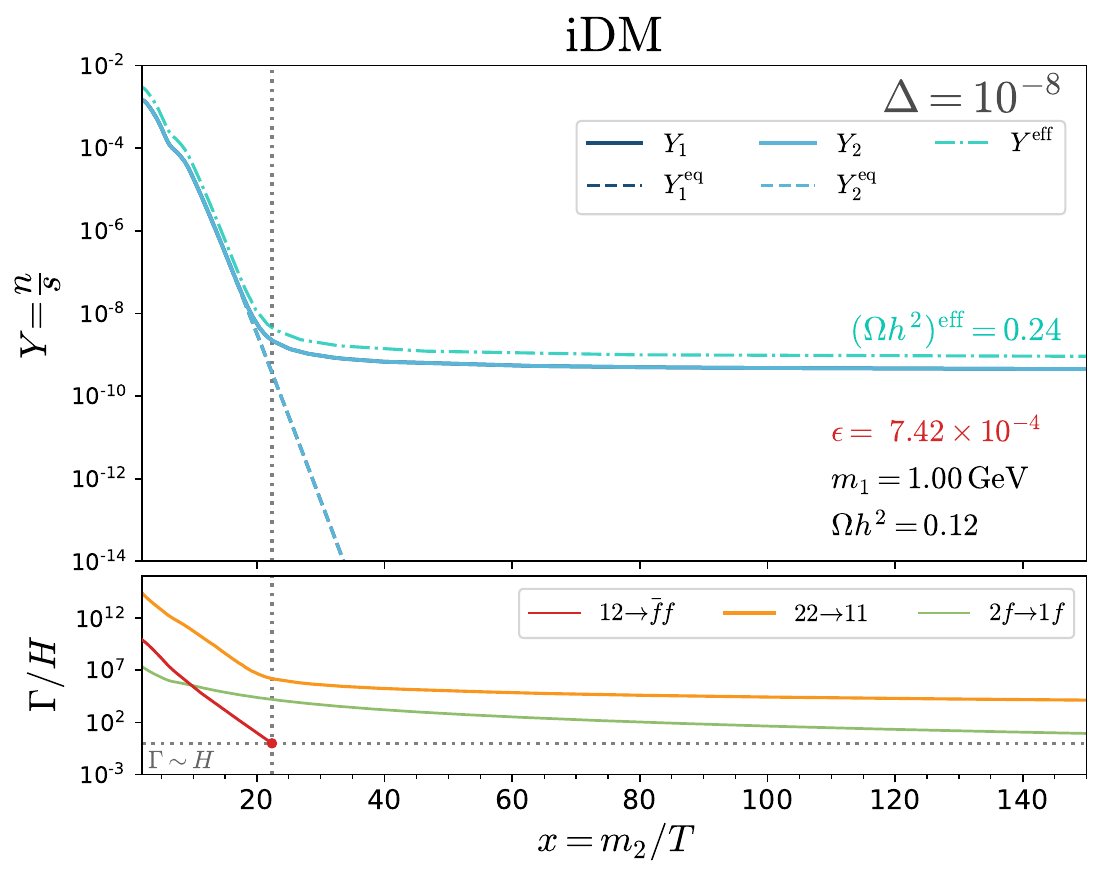}
\end{center}
\vglue -0.8 cm
\caption{\label{fig:fo_curve_DP} Freeze-out curve for the \textit{vanilla} dark photon iDM model in the extreme case of very small mass splitting, $\Delta = 10^{-8}$. We fix $\alpha_D=0.1$, $R=3$, and $m_1 = 1 \, {\rm GeV}$. The effective relic abundance $(\Omega h^2)^{\rm eff}$, depicted in dash-dotted light blue, represents the density value for the effective $Y^{\rm eff}$ curve, resulting in this case in twice the correct DM abundance.}
\end{figure}

Now, to illustrate a case with a very small inelastic mass splitting, we present in figure~\ref{fig:fo_curve_DP} an extreme case where $\Delta=10^{-8}$, corresponding to $\delta=10$~eV, for the \textit{vanilla} dark photon iDM model, with fixed $\alpha_D=0.1$, $R=3$, and $m_1 = 1 \, {\rm GeV}$. In this scenario, the decays of the heavier fermion are kinematically forbidden, resulting in an inefficient depletion. As a result, the effective abundance $Y^{\rm eff}$ is twice the observed relic value, \textit{i.e.}, $(\Omega h^2)^{\rm eff} \sim 0.24$. These findings qualitatively align with the analysis in Ref.~\cite{Berlin:2023qco} regarding the order and strength of all processes in the absence of the decay mode.

\begin{figure}[h!]
\begin{center}
\includegraphics[width=0.45\textwidth]{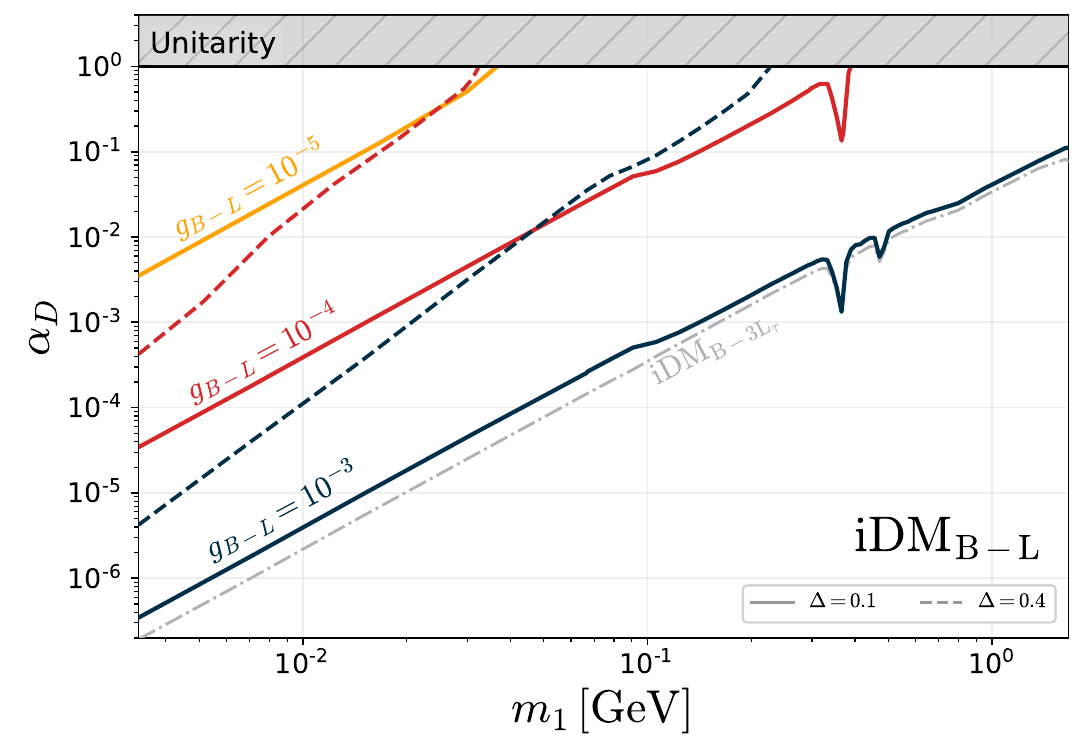}
\includegraphics[width=0.45\textwidth]{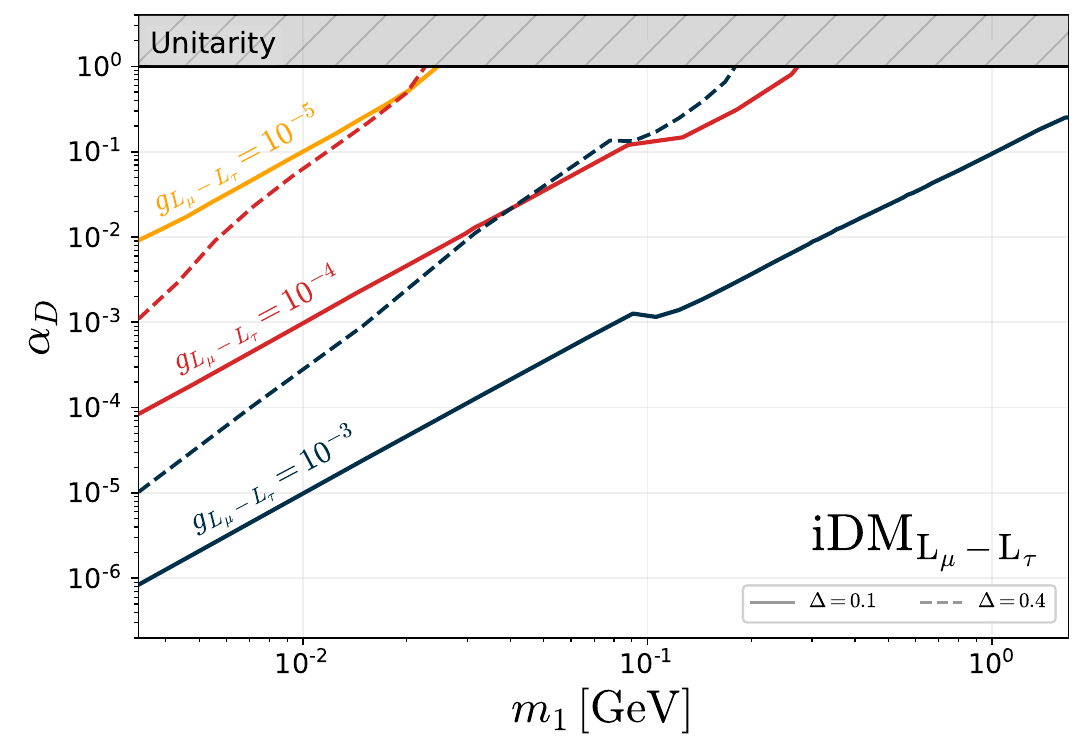}
\includegraphics[width=0.45\textwidth]{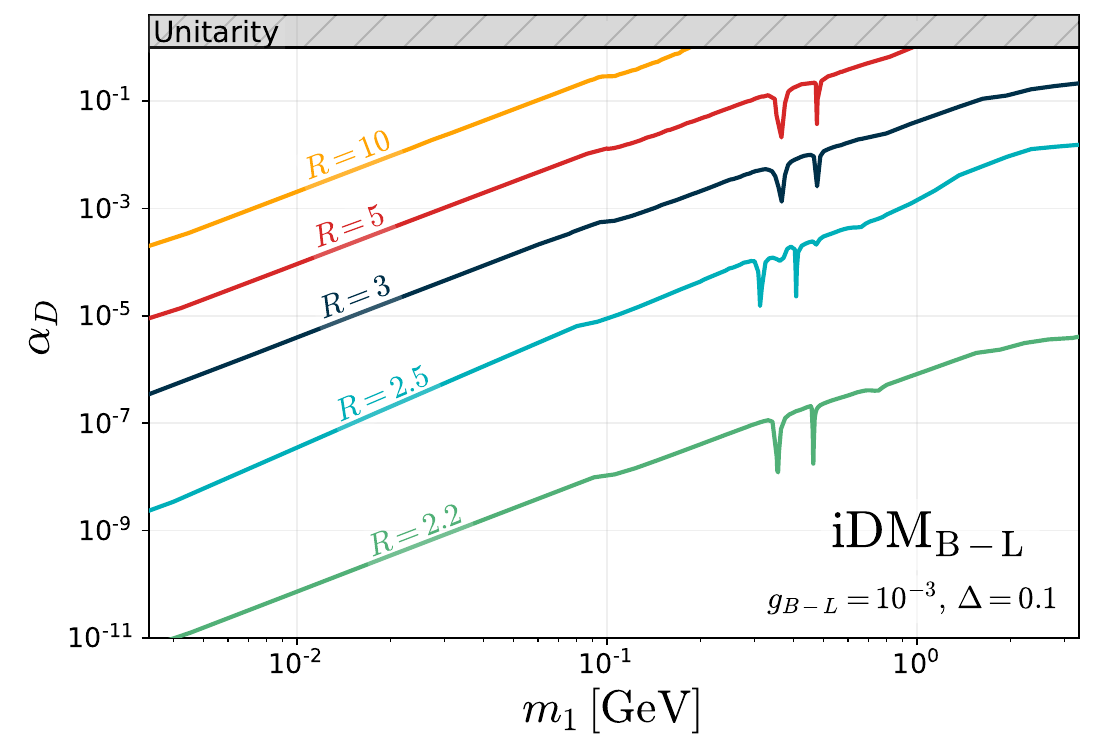}
\end{center}
\vglue -0.8 cm
\caption{\label{fig:tt_dark} Thermal target curves in the $\alpha_D \times m_1$ parameter space for different parameter and model choices. The upper left (right) panel shows results for the iDM$_{B-L}$ (iDM$_{L_{\mu}-L_{\tau}}$) model with a fixed mass ratio of $R = 3$, while varying gauge coupling $g_Q$ (colors) and iDM mass splitting $\Delta$ (line styles). The lower panel presents the curves for the iDM$_{B-L}$ model for fixed coupling $g_{B-L} = 10^{-3}$ and $\Delta = 0.1$, but for different mass ratios $R = m_{Z_{B-L}}/m_1$, as indicated by the colors. The hatched region shows the \textit{naive} perturbative unitarity bound.}
\end{figure}

Finally, in figure~\ref{fig:tt_dark}, we present the curves in the $\alpha_D \times m_1$ parameter space that yield the correct dark matter relic density for different parameter choices. In the upper left panel, we consider the iDM$_{B-L}$ model and show the thermal targets for three different values of the gauge coupling, distinguished by color, as well as for two different iDM splittings: $\Delta = 0.1$ (solid) and $\Delta = 0.4$ (dashed). The behavior of other iDM baryophilic models is very similar, as illustrated by the comparison between the solid blue line and the gray dash-dotted line, which represents the thermal target of iDM$_{B-3L_{\tau}}$ for fixed $g_{B-3L_{\tau}} = 10^{-3}$ and $\Delta = 0.1$. The upper right panel shows the same curves but for the iDM$_{L_{\mu}-L_{\tau}}$ model. Both upper panels assume a fixed mass ratio of $R=3$. In contrast, the lower panel shows the thermal targets for the iDM$_{B-L}$ model with fixed $g_{B-L} = 10^{-3}$ and $\Delta = 0.1$, while varying the mass ratio $R = m_{Z_{B-L}}/m_1$ between the mediator and dark matter masses, as indicated by the different colors. As the ratio approaches the resonance regime ($R \sim 2$), the cross-section is significantly enhanced, shifting the curve to even lower coupling values.

\section{ReD-DeLiVeR code}
\label{app:rdcode}

In the following appendix, a brief description of the structure of the \textsc{ReD-DeLiVeR} Python package is provided, along with some user instructions.

The \textsc{ReD-DeLiVeR} code is an updated version of the previous \textsc{DeLiVeR} code, which was designed to compute decay widths and branching ratios of general light vector mediators, including decays into various hadronic channels. In this new version, users can now choose to include a dark matter extension, which can be either a simplified DM model or the iDM$_Q$ scenario. The code can be utilized to numerically compute the relic abundance of the chosen dark matter candidate, using the free parameters of the model as input. Additionally, users can calculate process rates and identify thermal targets in the coupling-mass parameter space of the model. The code is publicly available at~\cite{ReD-DeLiVeR}, along with an user-friendly tutorial in Jupyter Notebook format.

\vspace{15pt}

The \textsc{ReD-DeLiVeR} package structure can be synthesized in different python files and folders, as displayed below:
\vspace{-3pt}
\begin{leftbar}[15cm]
\vspace{-8pt}
\begin{verbatim}
    ReD-DeLiVeR/
    
        > src/
            > vecdecays.py
            > width_inel.py
            > utilities.py
            > pars.py
            > chan.py
            > form_factors/
            > functions/
            > data/
        > relic_density.py
        > txt_targets/
        > models/   
        
\end{verbatim}
\end{leftbar}

Let us now discuss the functionalities of each module.

\vspace{5pt}
\noindent\colorbox{gray!30}{\texttt{> src/}}
\vspace{5pt}

The source code folder contains modules and directories from the original \textsc{DeLiVeR} code used for computing the $Z_Q$ mediator's hadronic decay widths. This includes the \verb+form_factors/+ and \verb+functions/+ folders: the former stores the vector meson dominance (VMD) calculations for several hadronic channels, while the latter provides relevant functions used for such computations. Additionally, the \verb+chan.py+ and \verb+pars.py+ files provide a list of defined hadronic channels and relevant parameters. For more information about these files, we refer the reader to Ref.~\cite{Foguel:2022ppx} and \cite{DeLiVeR}.

In addition to these hadronic-related tools, this folder also contains the Python scripts used to compute the widths and branching ratios of generic vector mediators (\verb+vecdecays.py+) and the heavier dark fermion state $\chi_2$ (\verb+width_inel.py+). Let us first describe the dark vector decays.

\vspace{8pt}
\hspace{2pt} \colorbox{teal!20}{\texttt{> vecdecays.py} }
\vspace{8pt}

This module is divided into three classes, which we list below.

\begin{itemize}
    \item \texttt{Processes()}: in this class we define the analytic formulas for the vector decay widths into fermions and DM final states.

    \item \texttt{decayZp(Processes)}: This class is responsible for computing the $Z_Q$ decay widths based on the provided input, which includes the model name, couplings ($Q$-charges), DM type, mass splitting $\Delta$ (for inelastic DM), and the mass ratio $R$. Optionally, the user can specify a hadron-quark transition energy value, \texttt{mhad}, although it is not required. In addition to decay widths, the class calculates branching ratios, lifetimes, and decay lengths. A complete list of SM final states is available in the Jupyter tutorial. For DM, the user can select between complex scalar, Majorana fermion, Dirac fermion, inelastic, or no DM insertion.
    
    \item \texttt{Model(decayZp)}: Finally, using this class, the user can save the computed widths and/or branching ratios into text files inside the \texttt{models/} folder, while scanning a range of $m_{Z_Q}$ masses.

\end{itemize}

For instance, let us compute the decay widths for the case of the iDM$_{B-L}$ model for fixed $\Delta=0.4$ and $R=3$. First, we define the couplings and call the \texttt{Model} class:
\begin{lstlisting}[language=Python]
>>> from src import vecdecays as vd
>>> coupBL = [1./3,1./3,1./3,1./3,1./3,1./3,-1,-1,-1,-1,-1,-1]
>>> BLmodel = vd.Model(name= "BL_boson", coup= coupBL, DMtype= "inelastic", split=0.4, Rrat=3, mhad= 1.737, folder = "models")
\end{lstlisting}
where the coupling assignment order is $[q_Q^d,q_Q^u,q_Q^s,q_Q^c,q_Q^b,q_Q^t,q_Q^e,q_Q^\mu,q_Q^\tau,q_Q^{\nu_e},q_Q^{\nu_\mu},q_Q^{\nu_\tau}]$.
Then we can compute the normalized widths via
\begin{lstlisting}[language=Python]
>>> BLmodel.calcnormwid(mmin=1e-3 ,mmax=10.0, step=10000)
\end{lstlisting}
where we show the default values of the minimum (\texttt{mmin}), maximum (\texttt{mmax}) and total number of mass values (\texttt{step}). During this computation, the code generates the .txt files for the normalized widths. In the \verb+models/BL_boson+ folder, it saves three different files corresponding to the fermionic and hadronic widths organized by channel, along with a final file that compiles the widths into four categories: QCD (quarks and hadrons), charged leptons, neutrinos, and total. It will also create a file in the \verb+models/DM_model+ directory that contains the normalized widths into the selected DM candidate. Besides, this method generates some instance attributes that contain the computed widths.

The normalized width files generated in the step before will be used for the computation of the decay widths and branching ratios for user-defined couplings, which can be done with the commands
\begin{lstlisting}[language=Python]
>>> BLmodel.calcwid(gQ= 1e-3, gDM= 1)
>>> BLmodel.calcbr(gQ=1,gDM=1)
\end{lstlisting}

Now that we described the basic usage of the vector mediator script, we can move on to the dark fermion $\chi_2$ decays.

\vspace{8pt}
\hspace{2pt} \colorbox{teal!20}{\texttt{> width\_inel.py}}
\vspace{8pt}

This script contains only one class, \texttt{decayChi2()}, which receives as input the coupling array and the $R$ ratio. This class posses the methods necessary for the computation of the widths, and branching ratios, as described in the appendix~\ref{app:chi2dec}. 

\vspace{5pt}
\noindent\colorbox{gray!30}{\texttt{> relic\_density.py}}
\vspace{5pt}

The script \verb+relic_density.py+ serves as the main module of the package and includes three distinct classes:

\begin{itemize}
    
     \item \texttt{Cosmology()}: This class defines basic cosmology-related functions described in section~\ref{sec:relic}, including the number of relativistic degrees of freedom $g_\rho$, the Hubble rate $H$, and the equilibrium number density $n_i^{\rm eq}$, all expressed as functions of $x$ or the temperature.

    \item \texttt{CrossSections(Model, Cosmology)}: this class inherits the methods of the \verb+Model+ class (from \verb+vecdecays.py+) as well as the \verb+Cosmology+ class. It is initialized by the same parameters as \verb+Model+, together with an additional attribute for the number of degrees of freedom of the DM candidate. The class provides the cross-section functions and decay rates of eq.~\eqref{eq:boltziDM}, along with methods for computing thermal averages.

    \item \texttt{Boltzmann(CrossSections)}: Finally, the \verb+Boltzmann+ class is responsible for solving the differential Boltzmann equations and computing the relic density, given the initialized model parameters. It can also evaluate the rates and plot the freeze-out curves. Additionally, it includes a method to compute the thermal target curve in the $g_Q \times m_{Z_Q}$ plane for a specified range of $m_{Z_Q}$ values. This class also features methods that utilize parallelization to optimize the computation of the thermal target. The resulting thermal target curves are saved as .txt files in the \verb+txt_targets/+ directory.
\end{itemize}

Let us provide an example of how to use this module. We will again consider the iDM$_{B-L}$ model with fixed values of $\Delta=0.4$ and $R=3$. First, we instantiate the \verb+Boltzmann+ class:

\begin{lstlisting}[language=Python]
>>> import relic_density as rd
>>> boltzBL = rd.Boltzmann("BL_boson", coupBL, DMtype= "inelastic", split=0.4, Rrat=3, mhad= 1.737, dof = 2)
\end{lstlisting}

Next, we can compute the relic density for a specified choice of dark matter mass and couplings:

\begin{lstlisting}[language=Python]
>>> boltzBL.set_DM(mDM= 1, gQ= 1e-3, gDM = 1.1)
>>> boltzBL.relic_density_idm()
Out[]: relic =  2.5122727752688716 for mV= 3  and gQ= 0.001
\end{lstlisting}
In this case, the computed relic density is relatively high, suggesting that an increase in the coupling is necessary to decrease it to the correct value. The freeze-out curves can be plotted using the following command:
\begin{lstlisting}[language=Python]
>>> boltzBL.plot_FO(name = "FOplot")
\end{lstlisting}
If the \verb+name+ option is specified, the plot will be saved in the \verb+models/<modelname>/relic+ folder, along with a .txt file containing the computed yields for each $x$ value.

Finally, the parameter space can be scanned to produce the thermal target curve using the following method:
\begin{lstlisting}[language=Python]
>>> boltzBL.compute_target(mVarr= [0.01,10,20])
\end{lstlisting}
where we choose to specify the variable \verb+mVarr+, which is an array indicating the initial, final and number of values for the mediator mass. The output will be saved in a file located in the \verb+txt_targets/+ directory, containing the $(g_Q, m_{Z_Q})$ values that reproduce the observed DM relic density $\Omega h^2 = 0.12$.

Now, since this computation may take some time, it is useful to parallelize the process using multiprocessing methods. We utilize the \verb+pathos+ library for this purpose, which can be installed via:
\begin{lstlisting}[language=bash]
$ pip install pathos
\end{lstlisting}
The thermal target curve can then be computed more efficiently by using the method:
\begin{lstlisting}[language=Python]
>>> boltzBL.parallel_target(mVarr=[0.01,10,30], nump = 15, fileN= "file")
\end{lstlisting}
In this command, the variable \verb+nump+ specifies the number of processes to run concurrently, while the results will be saved at the \verb+txt_targets/+ folder with the provided \verb+fileN+ string appended to the end of the file names.

\section{Details on the Experimental Limits}
\label{app:bounds}

To recast or compute the various experimental bounds discussed in section~\ref{sec:pheno}, we applied distinct strategies depending on the type of search. In this appendix, we first provide additional details on the methods used for these calculations. We then elaborate on the bounds for the regime in which the dark charges $q_D$ are fixed while $\alpha_D$ fluctuates. Let us begin by categorizing the different searches.

\vspace{4pt}
\paragraph*{Invisible searches} For the cases of BaBar, Belle II, LDMX, NA62 and NA64-$e$, we recast the original bounds using a modified version of the formula presented in Ref.~\cite{Ilten:2018crw}. The original searches considered a dark photon $Z_\gamma$ decaying predominantly into invisible dark matter states. To recast this scenario for an iDM$_Q$ search, we solved the following expression for each value of mass and coupling
\be \label{eq:invrecast}
\frac{\sigma_{Z_Q}}{\sigma_{Z_\gamma}} \qty{ \rm{BR}(Z_Q \to \chi_1 \chi_2) \qty[ \rm{BR}(\chi_2 \to \chi_1  \bar \nu \nu) + P_{\rm out}^{\chi_2} \, \rm{BR}(\chi_2 \to \chi_1 + {\rm vis}) ] +  \rm{BR}(Z_Q \to \bar \nu \nu ) } - 1 = 0 \, ,
\ee
where $\sigma_{Z_Q}$ represents the $Z_Q$ production cross-section, and $P_{\rm out}^{\chi_2}$ is the probability that $\chi_2$ decays outside the detector, given by
\be 
 \rm P_{\rm out}^{\chi_2} = e^{- L_{\rm out}/\lambda_{\chi_2}} \, ,
\ee
with $\lambda_{\chi_2}$ denoting the decay length of the heavier dark fermion and $L_{\rm out}$ being the size of the detector volume. For the cross-section ratio in eq.~\eqref{eq:invrecast}, we applied the same scalings described in Ref.~\cite{Ilten:2018crw}. 

For the other relevant invisible limits,  slight modifications to the recasting procedure were necessary. In the case of the NA64-$\mu$ invisible search~\cite{NA64:2024klw}, the original limit was computed for $L_\mu - L_\tau$ mediators. Therefore, when recasting to iDM${L\mu - L_\tau}$, the cross-section ratio in eq.~\eqref{eq:invrecast} simplifies to the gauge coupling ratio. For the CDF invisible limit, they provided a limit on the quantity $g_Q \sqrt{\rm{BR}(Z_Q \to \rm DM )}$ for each mass value, which we rescaled accordingly for the iDM$_B$ scenario.

Equation~\eqref{eq:invrecast} accounts for all possible invisible signatures in an iDM$_Q$ search:  either the mediator decays into dark states, which subsequently decay invisibly into neutrinos or escape the detector, or the mediator decays directly into neutrinos—though this is usually highly suppressed unless $g_Q \sim g_D$. We also verified that the scenario in which the mediator itself escapes the detector volume is negligible for the values of $\alpha_D$ considered here.

\begin{figure}[h!]
\begin{center}
\includegraphics[width=0.6\textwidth]{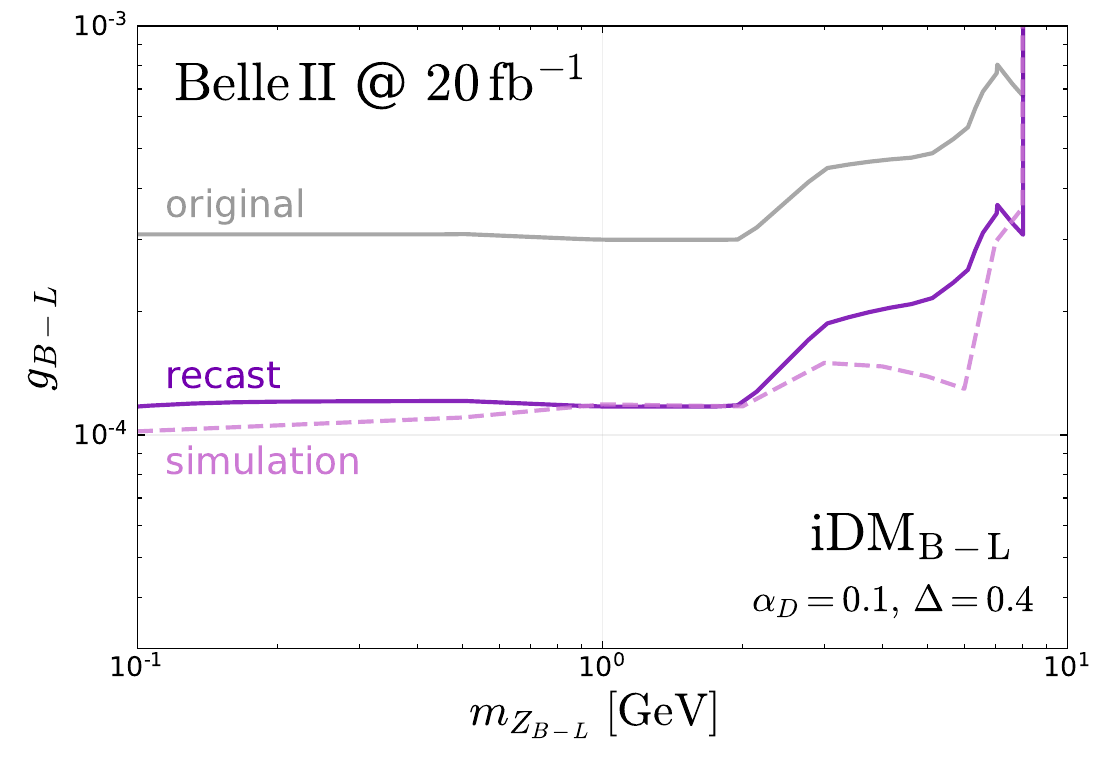}
\end{center}
\vglue -0.8 cm
\caption{\label{fig:maddump} Comparison between the iDM$_{B-L}$ recast calculation (dark purple) of the Belle II invisible bound (gray) with the result obtained from the MadDump simulation (light purple). For the comparison, we fix $\alpha_D = 0.1$, $R=3$, and $\Delta=0.4$. }
\end{figure}

To validate the recasting procedure employed here, we simulated a search for iDM$_{B-L}$ in the Belle II detector using our iDM$_Q$ FeynRules model file along with the software's MadGraph~\cite{Alwall:2014hca} and MadDump~\cite{Buonocore:2018xjk}. First, we generated the dark mediators via $e^+e^- \to Z_Q \gamma$ and decayed $Z_Q \to \chi_1 \chi_2$ inside MadGraph, verifying that the kinematical distributions matched those shown in figure 3 of Ref.~\cite{Duerr:2019dmv}. The cross-section for $e^+e^- \to Z_Q \gamma$ also matched the result in figure 205 of Ref.~\cite{Belle-II:2018jsg}. Next, we modified the MadDump source code to implement the Belle II detector geometry and computed the decay probabilities of $\chi_2$ within MadDump. Figure~\ref{fig:maddump} compares the recast Belle II bound (which corresponds to the limit that appears in the iDM$_{B-L}$ panel for $\Delta=0.4$ in figure~\ref{fig:limits1}) with the limit obtained from the MadDump simulation along with the collaboration's background estimates from Table 143 of Ref.~\cite{Belle-II:2018jsg}. The gray line represents the original predicted bound for a dark photon search presented by Belle II, from which our recast was derived. As shown, the simulation agrees well with our recast computation.

\vspace{4pt}
\paragraph*{Semi-visible searches}  For the E137, CHARM, NuCal and LSND semi-visible searches, we recast from the original iDM limits by solving the following recast formula
\be \label{eq:visrecast}
\frac{\sigma_{Z_Q}}{\sigma_{Z_\gamma}} 
\qty(\frac{\rm{BR}(Z_Q \to \chi_1 \chi_2) }{\rm{BR}(Z_\gamma \to \chi_1 \chi_2) })
\qty(\frac{\rm{BR}(\chi_2 \to Z_Q^*  \to \chi_1 e^+ e^-)}{\rm{BR}(\chi_2 \to Z_\gamma^* \to  \chi_1 e^+ e^-)})  
\qty(\frac{ \varepsilon_Q(\tau_{\chi_2})}{\varepsilon_\gamma(\tau_{\chi_2})}) - 1 = 0 \, ,
\ee
where the variables related to $Z_\gamma$ are computed at the fixed $(m_{Z_\gamma}, \epsilon)$ values from the experimental limit. At each mass $m_{Z_\gamma} = m_{Z_Q}$, we solve the equation for $g_Q$. The detector efficiency ratio $\varepsilon_Q/ \varepsilon_\gamma$ accounts for the probability of decaying inside the detector and depends on the lifetime of the heavier dark fermion. This was calculated using the effective proper-time fiducial decay region technique described in Ref.~\cite{Ilten:2018crw}, but in the iDM case,  adapted for the propagation of the $\chi_2$ state. 

For the $4\mu$ searches from BaBar and CMS, which constrained $L_\mu - L_\tau$ mediators, we recast the limits to the iDM$_{L_\mu - L_\tau}$ model by using the equation
\be \label{eq:visrecast}
\frac{g_Q}{g^{\rm exp}_{L_\mu - L_\tau}}
 \qty[\frac{\rm{BR}(Z_Q \to \mu^+ \mu^-)+ \rm{BR}(Z_Q \to \chi_1 \chi_2) \rm{BR}(\chi_2 \to \chi_1 \mu^+ \mu^-) P_{\rm in}^{\chi_2}}{\rm{BR}(Z_{L_\mu - L_\tau} \to \mu^+ \mu^-)}] - 1 = 0 \, ,
\ee
which we solve for $g_Q$ for each fixed value of $(m_{Z_{L_\mu - L_\tau}}, g^{\rm exp}_{L_\mu - L_\tau})$. The branching ratio $\rm{BR}(Z_{L_\mu - L_\tau} \to \mu^+ \mu^-)$  is computed in the original bound model, \textit{i.e.}, the minimal $U(1)_{L_\mu - L_\tau}$ scenario. The other terms, which depend on $g_Q$ and $Z_Q$, represent the recast parameters for the iDM$_{L_\mu - L_\tau}$ model. The probability that $\chi_2$ decays inside the detector, $P_{\rm in}^{\chi_2}$, is given by
\be 
 \rm P_{\rm in}^{\chi_2} = e^{- L_{\rm in}/\lambda_{\chi_2}} - e^{- L_{\rm out}/\lambda_{\chi_2}} \, ,
\ee
where $L_{\rm in}$ is the propagation distance to enter the detector, and $L_{\rm out}$ is the distance to exit it.

\vspace{4pt}
\paragraph*{Meson decays} Finally, for the $J/\Psi$ and $\Upsilon$ invisible decay searches, we explicitly computed the excluded regions by considering the analytic expression for the branching ratio. For instance, the expression for $J/\Psi$ meson decay into pseudo-Dirac fermion DM, $\rm{BR}(J/\Psi \to \bar \chi \chi)$, can be found in Ref.~\cite{Schuster:2021mlr}. We adapted this formula to the iDM scenario, finding that
\be 
\begin{aligned}
    \frac{{\rm BR}(J/\Psi \to \chi_1 \chi_2)}{{\rm BR}(J/\Psi \to e^+ e^-)} =& \, \frac{\alpha_D}{\alpha_e} \qty(\frac{q_Q^c g_Q}{q_{\rm em}^c e})^2 \frac{m_V^4}{(m_{Z_Q}^2 - m_V^2)^2 + m_{Z_Q}^2 \Gamma_{Z_Q}^2} \\
    & \times \qty(1- \frac{\Delta^2}{R^2} m_r^2)^{3/2}  \qty( 1 + \frac{m_r^2 (\Delta +2)^2}{2 R^2}) \sqrt{1 -\frac{m_r^2 (\Delta+2)^2}{R^2} } \, ,
\end{aligned}
\ee
where $\alpha_e \equiv e^2/(4 \pi)$, $m_V$  is the vector meson mass and $m_r \equiv m_{Z_Q}/m_V$. A similar expression can be derived for the $\Upsilon$ decay, with the appropriate modifications for the $b$ quark couplings.

Now, in order to compute the bound, it is essential to remember that, in the iDM$_Q$ scenario, the $\chi_2$ state can decay to visible modes, spoiling the invisible limit. Therefore, the effective invisible branching ratio must account for this possibility, as well as any potential decays of $Z_Q$ into neutrinos. The final formula for the iDM$_Q$ contribution to the invisible vector meson $V$ branching ratio can be expressed as
\be 
\begin{aligned}
    {\rm BR}(V \to {\rm inv})|_{\rm iDM_Q}= & \, {\rm BR}(V \to \chi_1 \chi_2) \, \qty[\rm{BR}(\chi_2 \to \chi_1  \bar \nu \nu) + P_{\rm out}^{\chi_2} \, \rm{BR}(\chi_2 \to \chi_1 + {\rm vis})] \\
    & +  {\rm BR}(V \to Z_Q^* \to \bar \nu \nu ) \, .
\end{aligned}
\ee

Hence, limits can be established on the $m_{Z_Q} \times g_Q$ parameter space by requiring that the above expression for the invisible vector meson branching ratio does not exceed the measured experimental limits
\be 
{\rm BR}(V \to {\rm inv})|_{\rm iDM_Q} < {\rm BR}(V \to {\rm inv})|_{\rm exp} \, .
\ee

\vspace{15pt}

Now that we have outlined the methodology behind the different bound computations, let us turn to the study of the scenario where the dark charge $q_D$ is fixed. As discussed previously, we focus on the regime where $g_D \gg g_Q$, as it allows for a successful freeze-out of DM candidates within an experimentally viable parameter space. 
Since $g_D \equiv q_D \, g_Q$, this condition often requires large $q_D$ values. Indeed, for smaller charge values, the thermal target curves tend to shift upwards, potentially conflicting with the bounds. Theoretically, there is no restriction against arbitrarily large values of $q_D$. Nevertheless, it is interesting to examine at which charge this scenario starts to open up available parameter space.

\begin{figure}[h!]
\begin{center}
\includegraphics[width=0.48\textwidth]{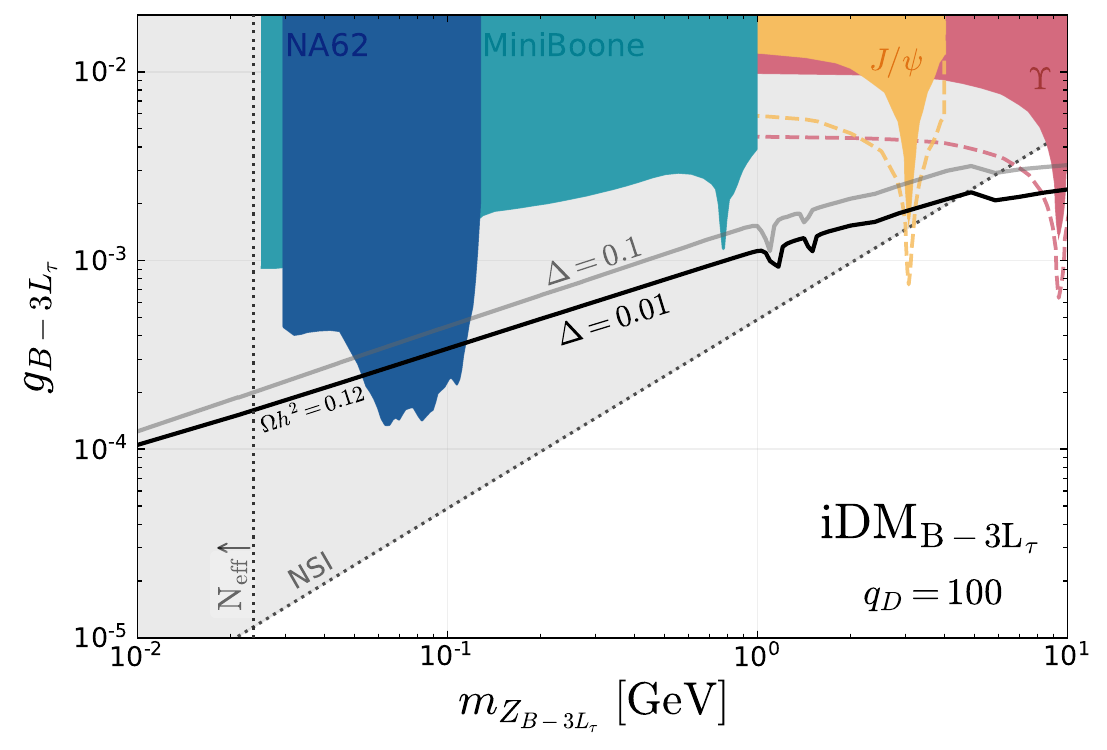}
\includegraphics[width=0.48\textwidth]{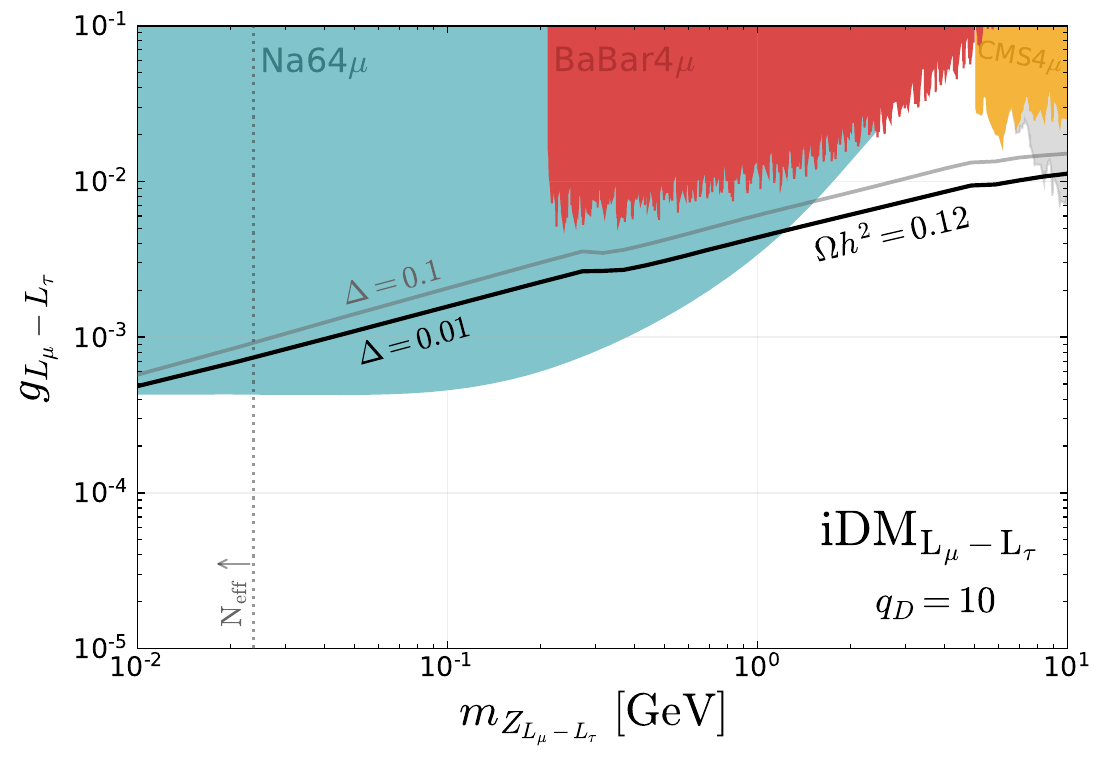}
\end{center}
\vglue -0.8 cm
\caption{\label{fig:bounds_qD} Same as in figure~\ref{fig:limits2}, but for a fixed dark charge value and varying $\alpha_D$. The left (right) panel considers the case of iDM$_{B-3L_\tau}$ (iDM$_{L_\mu-L_\tau}$) for fixed $q_D=100$ ($q_D=10$). }
\end{figure}

Figure~\ref{fig:bounds_qD} presents the bound plots for iDM$_{B-3L_\tau}$ (left) and iDM$_{L_\mu-L_\tau}$ (right), with fixed dark charge values of $q_D = 100$ (left) and $q_D=10$ (right). The solid black line shows the thermal target curve for the mass splitting $\Delta=0.01$, while the gray line represents $\Delta=0.1$. For iDM$_{B-3L_\tau}$, there in no effective changes due to these two mass splitting. In the case of iDM$_{L_\mu-L_\tau}$, there is a slight increase in the CMS-4$\mu$ bound, as shown by the gray region on the right plot. Regarding the thermal targets, increasing $q_D$ lowers the thermal target curves, and similarly, decreasing $\Delta$ has the same effect. Larger dark charge values also tend to enlarge the invisible bounds while reducing the visible ones, as seen for MiniBooNE, meson bounds, BaBar, and CMS. For the case of NA62 and NA64$\mu$, there is no change with respect to $q_D$, as the reduction in the mediator branching ratio to decays into iDM is compensated by decays into neutrinos for the models considered in the figure.

In summary, the figure shows that for the case of iDM$_{B-3L_\tau}$, there is already available parameter space starting from charges of order $q_D \sim \mathcal{O}(100)$, while for iDM$_{L_\mu-L_\tau}$, new regions can already be unlocked for $q_D \sim \mathcal{O}(10)$.

\clearpage
\bibliographystyle{utphys}
\bibliography{InelasticDM}  

\end{document}